\documentclass[journal]{IEEEtran}
\usepackage{cite}
\usepackage{url}

\usepackage[dvipsnames,svgnames]{xcolor}
\usepackage{graphicx}
\usepackage{subfig}
\usepackage{psfrag}
\DeclareGraphicsExtensions{.xps}
\usepackage{epstopdf} 
\usepackage{overpic}
\usepackage{enumerate}
\usepackage{enumitem}
\usepackage[cmex10]{amsmath}
\usepackage{amssymb}
\usepackage{amsthm}

\def\tr{\mathrm{tr}}

\theoremstyle{theorem}
\newtheorem{theorem}{Theorem}
\newtheorem{definition}{Definition}
\newtheorem{lemma}{Lemma}
\newtheorem{corollary}{Corollary}
\newtheorem{example}{Example}
\newtheorem{remark}{Remark}

\usepackage[textwidth=30mm]{todonotes}
\newcommand{\mathacr}[1]{\mathsf{#1}}

\newcommand{\vect}[1]{\mathbf{#1}}

\def\Psiv{\vect{Q}}

\def\tr{\mathrm{tr}}
\def\rank{\mathrm{rank}}

\def\Htran{\mbox{\tiny $\mathrm{H}$}}
\def\Ttran{\mbox{\tiny $\mathrm{T}$}}
\def\CN{\mathcal{N}_{\mathbb{C}}} 
\def\taupu{\tau_{p}} 
\def\bphiu{\boldsymbol{\phi}} 

\def\Pu{\mathcal{P}} 
\def\sigmaUL{\sigma^2_{\mathrm{ul}}} 
\def\Cyj{\vect{A}_{jk}}

\begin{document}

\IEEEoverridecommandlockouts

\title{Towards Massive MIMO 2.0:\\{ 
Understanding spatial correlation, interference suppression, and pilot contamination}}
\author{
\IEEEauthorblockN{Luca Sanguinetti, \emph{Senior Member, IEEE}, Emil Bj{\"o}rnson, \emph{Senior Member, IEEE}, Jakob Hoydis, \emph{Member, IEEE}
\thanks{
\newline \indent L.~Sanguinetti is with the Dipartimento di Ingegneria dell'Informazione, University of Pisa, 56122 Pisa, Italy. E.~Bj\"ornson is with the Department of Electrical Engineering (ISY), Link\"{o}ping University, 58183 Link\"{o}ping, Sweden. J.~Hoydis is with Nokia Bell Labs, Paris-Saclay, 91620 Nozay, France.}
}}
\maketitle

\begin{abstract}
Since the seminal paper by Marzetta from 2010, Massive MIMO has changed from being a theoretical concept with an infinite number of antennas to a practical technology. The key concepts are adopted in 5G and base stations (BSs) with $M=64$ full-digital transceivers have been commercially deployed {in sub-6\,GHz bands}. The fast progress was enabled by many solid research contributions {of which} the vast majority assume spatially uncorrelated channels and signal processing schemes developed for single-cell operation. These assumptions make the performance analysis and optimization of Massive MIMO tractable but have three major caveats: 1) practical channels are spatially correlated; 2) large performance gains can be obtained by multicell processing, without BS cooperation; 3) the interference caused by pilot contamination creates a finite capacity limit, as $M\to\infty$. There is a thin line of papers that avoided these caveats, but the results are easily missed. Hence, this tutorial article explains the importance of considering spatial channel correlation and using signal processing schemes designed for multicell networks. We present recent results on the fundamental limits of Massive MIMO, which are not determined by pilot contamination but the ability to acquire channel statistics. These results will guide the journey towards the next level of Massive MIMO, which we call ``Massive MIMO 2.0''.
\end{abstract}



\section{Introduction}

The data traffic in cellular networks has grown at an exponential pace for decades, thanks to the continuous evolution of the wireless technology. The cellular throughput is determined by three key factors \cite{massivemimobook}:
\begin{equation}
\textrm{Throughput [bit/s/km}^2\textrm{]} = \frac{ \textrm{Spectrum [Hz]} \times \textrm{SE [bit/s/Hz/cell]} }{\textrm{Cell size [km}^2\textrm{/cell]}}
\end{equation}
where SE stands for spectral efficiency. The traditional way to manage the traffic growth is to allocate more frequency spectrum and to reduce the cell sizes by deploying more base stations (BSs).
Consequently, contemporary cellular networks are already densely deployed in the urban parts of many countries and there is little bandwidth left in the sub-6\,GHz bands that are attractive for wide-area coverage  \cite{Andrews2017a}. In contrast, the growth in SE, measured in bit/s/Hz/cell, has been rather modest over the past decades. 
To continue the network technology evolution, it is thus vital to make major improvements in the SE. 

The SE represents the number of bits of information that can be reliably communicated per sample (channel use) and is an increasing function of the signal-to-interference-and-noise ratios (SINRs) of the communication links. Hence, the SE between a BS and a user equipment (UE) is limited by the channel's signal attenuation and interference from other transmissions that take place at the same time and frequency. The signal attenuation can be reduced by using high-gain antennas at the BS to form fixed beams toward the cell center, while interference can be managed by scheduling UEs orthogonally in the time-frequency domain. These approaches have served us well, but are now being changed to allow for higher SE in future 5G networks~\cite{Parkvall2017a}.

The most promising wireless technology to improve the SE is Massive MIMO (multiple-input multiple-output) \cite{Larsson2014a}. This physical-layer technology equips each BS with an array of many active antennas, which are used to spatially multiplex many UEs; that is, to communicate with them on the same time-frequency resource. By combatting the signal attenuation and interference through spatial signal processing, such as uplink (UL) receive combining and downlink (DL) transmit precoding, the SE per cell can be improved by orders-of-magnitude over classical cellular networks.
Massive MIMO is essentially a scaled-up version of the space-division multiple access (SDMA) concept from the 1990s \cite{Swales1990a,Anderson1999a}, 
with many more antennas and more aggressive spatial multiplexing than we could imagine before T.~Marzetta's seminal paper from 2010 \cite{marzetta2010noncooperative}.
Massive MIMO has since then gradually changed from being a controversial theoretical concept with an extremely large number of BS antennas to a mainstream technology that has found its way into the 5G New Radio standard \cite{Parkvall2017a}. The first 64-antenna 
Massive MIMO BSs have been added to the Ericsson  AIR, Huawei  AAU, and Nokia  AirScale product lines and commercially deployed \cite{Sprint2018feb}. 
 This manifests that Massive MIMO is no longer a promising concept but a reality for cellular networks (below 6 GHz). 

There is a rich body of scientific papers that analyze Massive MIMO and a handful of well-cited overview articles \cite{Rusek2013a,Larsson2014a,Swindlehurst-JSTSP-14,Bjornson2016b}, but the vast majority make two simplifying assumptions:
\begin{enumerate}
\item The propagation channels to the multiantenna BSs are spatially uncorrelated;
\item The signal processing schemes originally conceived for single-cell operation are heuristically applied to multicell scenarios.
\end{enumerate}
These assumptions make the SE analysis and optimization analytically tractable \cite{Marzetta2016a}, but there are three major caveats:
\begin{enumerate}
\item Practical {Massive MIMO} channels are spatially correlated, as seen from measurement campaigns \cite{Kermoal2002a, McNamara2002a, Gao2015a,Gao2015b} and physical arguments  \cite[Sec.~2]{massivemimobook}, \cite{Ertel1998a};
\item Huge gains can be obtained by developing signal processing schemes for multicell operation 
\cite{ashikhmin2012pilot,Adhikary2017a,BjornsonHS17};
\item The inter-cell interference, particularly the phenomenon of pilot contamination, becomes a critical limiting factor due the aforementioned simplifying assumptions.
\end{enumerate}
There is a thin, but solid, line of theoretical research that has avoided some or all of these caveats \cite{Huh2012a,Hoydis2013,Yin2013a,EmilEURASIP17,BjornsonHS17,Neumann-TSP17}, but its development has been relatively slow due to the less tractable analysis. The important insights and key messages from this line of research, which complements and sometimes contradicts the common views, are easily missed since they constitute a minor fraction of the vast literature on Massive MIMO. Nevertheless, this  research has now become mature, thanks to fundamental works such as \cite{Ngo2012b,Huh2012a,Hoydis2013,Yin2013a,EmilEURASIP17} and the recent developments in \cite{BjornsonHS17,Neumann-TSP17,Adhikary2017a} that have provided the \emph{last important pieces to the puzzle}. That is why {it is a good time} to summarize these results and insights.

This tutorial article reviews the basics of Massive MIMO in Section~\ref{sec:what-is-massive-mimo}. Then, in Section~\ref{sec:spatially-correlated} we explain the importance of considering spatially correlated channels and designing signal processing schemes that take the spatial correlation and multicell interference into account.  We then demonstrate the way to quantify the SE and finally present recent results on the fundamental SE limits of Massive MIMO, which are not determined by  pilot contamination (as is the case for spatially uncorrelated channels \cite{marzetta2010noncooperative,Marzetta2016a}) but the ability to acquire accurate channel statistics.
The lack of these insights has not prevented the first deployments of Massive MIMO, but will guide the evolution of the technology towards what we call Massive MIMO 2.0.


\section{What is Massive MIMO in this article?}
\label{sec:what-is-massive-mimo}

The term ``Massive MIMO'' lacks a concise and universal definition.
Marzetta's original paper \cite{marzetta2010noncooperative} demonstrates that the acquisition of channel state information (CSI) is the limiting factor in communication systems with many antennas. The paper further shows that a system with an unlimited number of antennas should operate in time-division duplex (TDD) mode and exploit channel reciprocity to acquire all the necessary CSI from a finite number of UL pilot signals. However, the paper does not call this ``Massive MIMO'' and provides no definition to be used in practical systems. Massive MIMO has anyway become a buzzword and the telecom industry prefers to treat it as synonymous with SDMA (or multiuser MIMO) with more than ten BS antennas, in order to quickly reach the market with Massive MIMO branded products. To demonstrate how the technology can be taken to the next level, in this article we will consider a stricter definition:

%
%
%
%
%

\begin{definition} A Massive MIMO cellular network consists of:
\begin{itemize}
\item $L \geq 2$ cells operating according to a synchronous TDD protocol;
\item BSs equipped with $M \geq 64$ antennas with fully digital transceiver chains;
\item Linear combining and precoding schemes capable of spatially multiplexing $K\geq 8$ UEs per cell;
\item More BS antennas than active UEs: $M/K > 1$.
\end{itemize}
\end{definition}

This definition is in line with the canonical form of Massive MIMO for sub-6\,GHz bands in \cite{massivemimobook} and includes Marzetta's original setup from \cite{marzetta2010noncooperative} as a special case as $M\to \infty$.
It is also in line with real-time Massive MIMO testbeds \cite{Chen2017b,Malkowsky2017}, field trials \cite{Liu2017a}, and recent commercial deployments \cite{Sprint2018feb}. 
{In fact, we selected the ranges $M\geq 64$ and $K \geq 8$ since the existing sub-6\,GHz products have 64 antennas and support multiplexing of 8 streams, while the testbeds support even larger $M$ and $K$. These products constitute the first technology generation, which we refer to as Massive MIMO 1.0.} 
Note that our definition does not specify a particular frequency range, but only the use of fully digital transceivers, which are currently used in sub-6\,GHz bands and will be used at higher frequencies in the next few years from now \cite{Tawa2018a}. We refer the reader to \cite{Adhikary2013,Choi2014a} for alternative MIMO concepts tailored to frequency-division duplex (FDD) mode\footnote{
{The conclusions and insights of this paper are valid in the uplink of both TDD and FDD. The main challenge with FDD is the estimation/feedback overhead, which becomes prohibitive in mobile scenarios. This is why Massive MIMO operates in TDD mode, but if the overhead can be managed, the same conclusion would appear in the downlink of FDD.}} and to \cite{heath2016overview,7959169} for overviews of MIMO communications with analog or hybrid analog-digital transceivers. These research topics are important for mmWave communications, where the channels might contain a sparse set of spatial components and fully digital transceivers will take some more time to develop. 
{Nevertheless, they are outside the scope of this article, which considers TDD and fully digital transceivers that can be used at sub-6\,GHz as well as mmWave bands.}

{For brevity, we only consider single-antenna UEs, but the results can be readily applied to UEs with multiple antennas by viewing them as virtual UEs that transmit separate signals, representing different data streams \cite[Sec.~1.3]{massivemimobook}.}

\subsection{The transmission protocol and spatially uncorrelated channel model}

\begin{figure}[t]
     \begin{center}
            \includegraphics[width=0.5\textwidth]{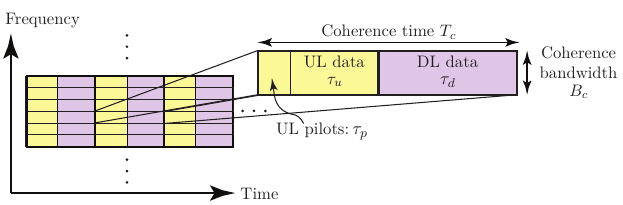}
    \end{center}
    \caption{The time-frequency plane is divided into coherence blocks over which the
channel response is time-invariant and frequency-flat. The $\tau_c$ samples of each block are used in a TDD fashion for UL pilots, UL data, and
DL data.}
  \label{fig1}
\end{figure}

The BS can only make efficient use of its multiple antennas if it knows the wireless channels of the UEs. Since they change over time and frequency, a TDD protocol is used where the UL and DL transmissions fit into the \emph{coherence block} of the channel; see Fig.~\ref{fig1}. This represents the time-frequency block in which the channels can be approximated as time-invariant and flat-fading. 
In doing so, each channel is the same in the UL and DL directions 
 and can be estimated at the BS using only UL pilots, which are known deterministic signals sent by the UEs. The number of complex-valued samples $\tau_c = T_c B_c$ in a coherence block is, according to the Nyquist-Shannon sampling theorem, determined by the coherence time $T_c$ and coherence bandwidth $B_c$ of propagation channels. Both depend on several factors such as the delay spread of the propagation environment (i.e., how frequency-selective the channel is), UE mobility, and carrier frequency \cite{Marzetta2016a}. Typical values for $\tau_c$ ranges from hundreds (with high-mobility and high-channel dispersion) to thousands of samples (with low-mobility and low-channel
dispersion). As shown in Fig.~\ref{fig1}, the $\tau_c$ samples are used for three purposes: 1) $\tau_p$ samples for UL pilots; 2) $\tau_u$ samples for UL data; 3) $\tau_d$ samples for DL data.
The pilots and data are transmitted at different times and the TDD protocol is synchronized across the cells. There are other variations of the TDD protocol, for which we refer the interested reader to \cite{Fernandes2013aa,Upadhya2017a,Verenzuela2017a}.

\begin{figure*}[t!]
\begin{center}
\begin{overpic}[unit=1mm,width=.5\columnwidth]{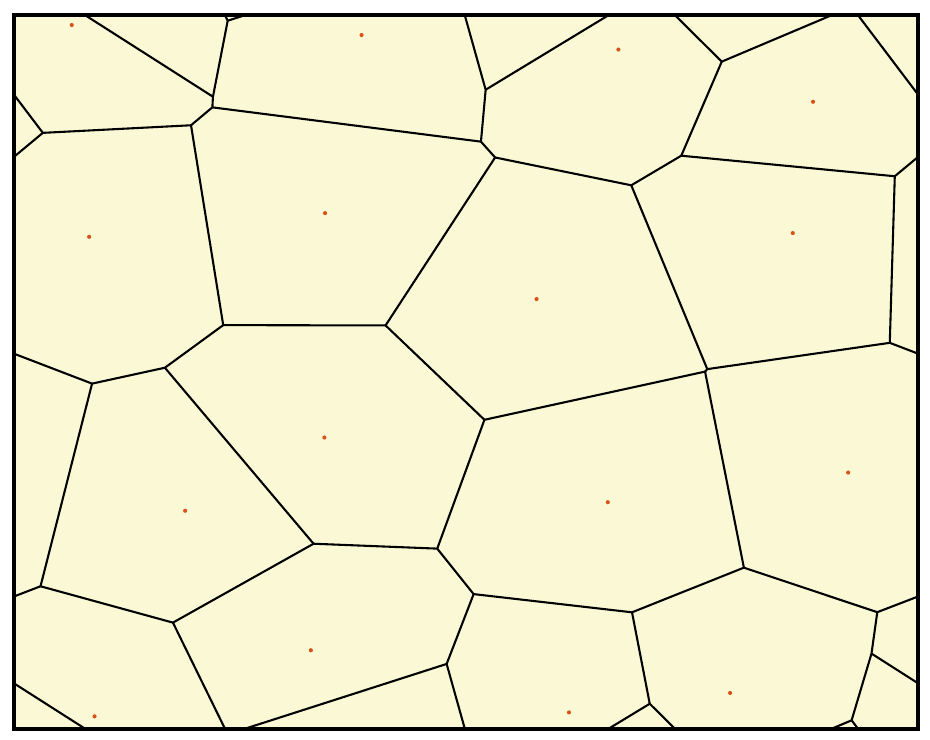}
\put(40,-10){\footnotesize $f=1$}
\end{overpic} 
\begin{overpic}[unit=1mm,width=.5\columnwidth]{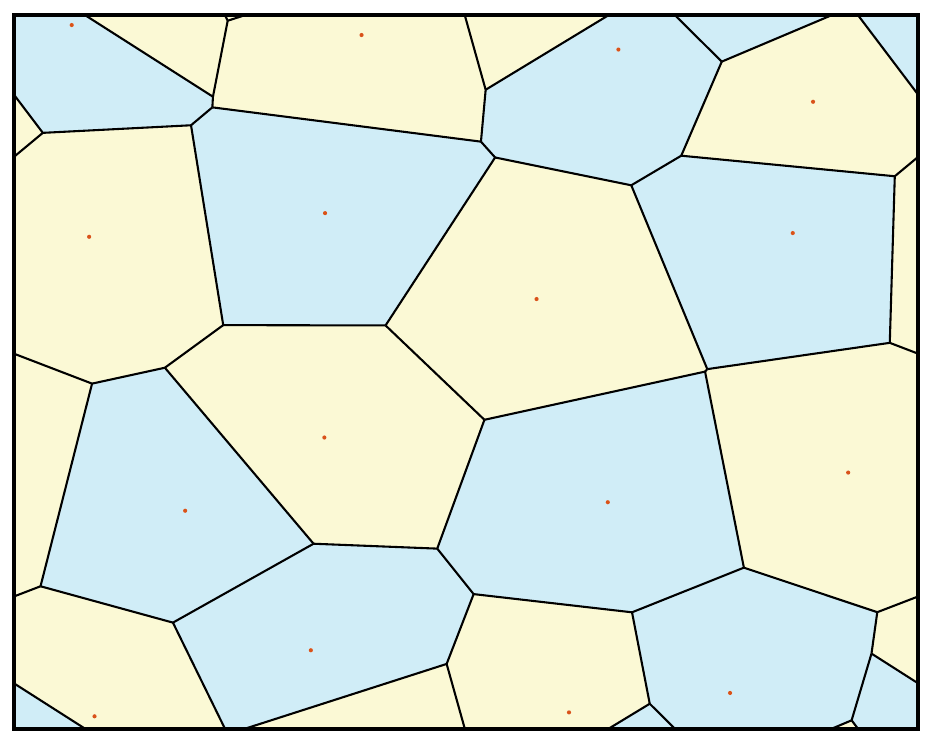}
\put(40,-10){\footnotesize $f=2$}
\end{overpic} 
\begin{overpic}[unit=1mm,width=.5\columnwidth]{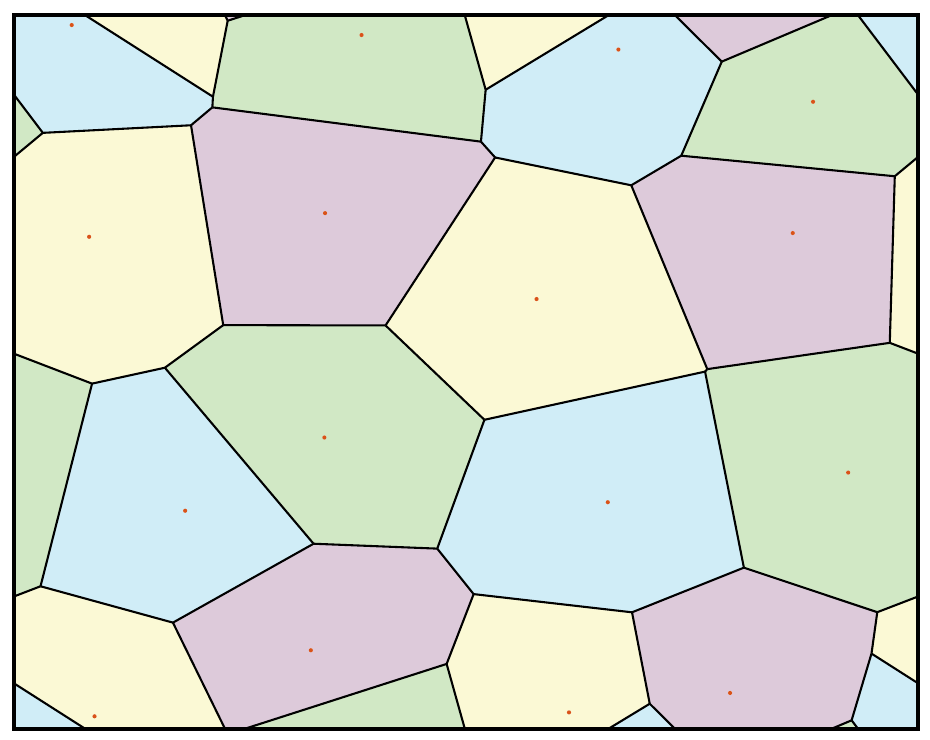}
\put(40,-10){\footnotesize $f=4$}
\end{overpic} 
\end{center} \vspace{5mm}
\caption{Illustration of the investigated cellular network with three pilot reuse factors. Each group is indicated with a distinct color and uses a disjunct set of pilots. The data transmission is carried out without a reuse factor.}\label{figure_network_layout} 
\end{figure*}

A common assumption is $\taupu = K f \leq \tau_c$, so that we can create a pilot book ${\mathbf{\Phi}} \in \mathbb{C}^{\taupu\times \taupu}$ of $Kf$ mutually orthogonal pilot sequences of length $\tau_p$; for example, using a Walsh-Hadamard matrix \cite{massivemimobook}. These are divided into $f$ disjunct groups with $K=\taupu/f$ pilots each, where the integer $f$ is called
the \emph{pilot reuse factor}. Each cell is associated with one of the $f$  groups, according to a predefined pilot reuse pattern, as illustrated in Fig.~\ref{figure_network_layout}.
Hence, the $K$ UEs in a cell use mutually orthogonal pilots and these pilots are reused in a fraction $1/f$ of the $L$ cells. For simplicity, the UEs are numbered in such a way that UE $k$ in two pilot-sharing cells uses the same pilot.

We denote $\bphiu_{jk} \in \mathbb{C}^{\taupu}$ the pilot sequence used by UE~$k$ in cell~$j$ and assume that it is transmitted with power $\rho_{\rm{ul}}$. 
This implies $ \bphiu_{jk}^{\Ttran} \bphiu_{jk}^{*}  = \taupu\rho_{\rm{ul}}$. We call $\Pu_{j} \subset \{1,\ldots, L\}$ the group of cells that utilize the same pilot set as cell~$j$, such that 
\begin{equation}
\bphiu_{jk}^{\Ttran} \bphiu_{li}^{*} = \begin{cases}
\taupu\rho_{\rm{ul}} & \textrm{if } l \in \Pu_{j} \textrm{ and } i = k \\
0 & \textrm{otherwise}.
\end{cases}
\end{equation}
Hence, the pilot sequence $\bphiu_{jk}$ is orthogonal to all other UEs' pilots, except for UE $k$ in cell $l \in \Pu_{j} \setminus \{ j \}$.

We denote by $\vect{h}_{lk}^{j} \in \mathbb{C}^{M}$ the channel between UE~$k$ in cell~$l$ and BS~$j$. This channel has traditionally often been modeled as 
\begin{align}\label{eq:iid-Rayleigh fading}
\vect{h}_{lk}^{j} \sim \CN( \vect{0}_M, \beta_{lk}^{j} \vect{I}_M)
\end{align}
where the Gaussian distribution accounts for the random small-scale fading realization in each coherence block, while $\beta_{lk}^{j}$ describes the macroscopic large-scale fading (e.g., geometric attenuation and shadow fading). The channel model in \eqref{eq:iid-Rayleigh fading} is called \emph{uncorrelated Rayleigh fading}, since the elements of $\vect{h}_{lk}^{j}$ are uncorrelated (and also independent) and have Rayleigh distributed magnitudes. This implies that $\vect{h}_{lk}^{j}$ has no dominant spatial directivity---the signals are equally likely to arrive to the BS from any direction---which is practically questionable. This model has been (and still is) very popular in the Massive MIMO literature since it is analytically tractable and leads to neat, understandable closed-form SE expressions \cite{Marzetta2016a}.


As said earlier, to make efficient use of its antennas, it is crucial for a BS to learn channels. It is particularly important for BS~$j$ to have
estimates of the channels $\{{\vect{h}}_{jk}^{j}: k=1,\ldots,K\}$ from the UEs in its own cell. 
These estimates are obtained from the received pilot signal $\vect{Y}_j^{p} \in \mathbb{C}^{M \times \taupu}$ at BS~$j$, which is given by 
\begin{align} \label{eq:uplink-pilot-model}
\vect{Y}_j^{p} = \underbrace{ \sum_{i=1}^{K} \sqrt{\rho_{\rm {ul}}} \vect{h}_{ji}^{j} \bphiu_{ji}^{\Ttran}  }_{\textrm{Desired pilots}} + \underbrace{\sum_{l=1,l \neq j}^{L} \sum_{i=1}^{K}  \sqrt{\rho_{\rm {ul}}} \vect{h}_{li}^{j} \bphiu_{li}^{\Ttran}  }_{\textrm{Inter-cell pilots}} + \underbrace{ \vphantom{\sum_{l=1,l \neq j}^{L} } \vect{N}_{j}^{p}}_{\textrm{Noise}}
\end{align}
where the first term accounts for the desired pilots from within the cell, the second term is interference, and $\vect{N}_{j}^{p} \in \mathbb{C}^{M \times \taupu}$ is independent receiver noise with i.i.d.\ elements distributed as $\CN(0,\sigma^{2}_{\rm{ul}})$. 
Since the channels 
are realizations of random variables, Bayesian estimators are desirable whenever the large-scale fading coefficients $\{\beta_{lk}^{j}\}$ are known. The linear minimum mean-squared error (MMSE) estimate of the channel $\vect{h}_{jk}^{j}$ 
is
\begin{align}\notag
\widehat{\vect{h}}_{jk}^{j}  &={\beta}_{jk}^{j} \psi_{jk}^{j}   \left(\frac{1}{\taupu\sqrt{\rho_{\rm {ul}}}}\vect{Y}_j^{p}\bphiu_{jk}^{*}  \right)\\&= {\beta}_{jk}^{j} \psi_{jk}^{j} \left(\vect{h}_{jk}^{j} + \sum_{l \in \Pu_{j}\setminus j} \vect{h}_{lk}^{j}  +  \frac{1}{\taupu\sqrt{\rho_{\rm {ul}}}}\vect{N}_{j}^{p}\bphiu_{jk} ^{*} \right)\label{eq:MMSEestimator_h_jli_uncorr}
\end{align}
with 
$\psi_{jk}^{j}= \left( \sum_{l \in \Pu_{j} }  \beta_{lk}^{j} +  \frac{1}{\taupu}\frac{\sigma^2_{\rm {ul}}}{\rho_{\rm {ul}}}  \right)^{-1}$.
The normalized mean-squared error (NMSE) given by
\begin{equation}\label{eq:NMSE_Uncorr}
\frac{\mathbb{E}\{ \| \vect{h}_{jk}^{j} - \widehat{\vect{h}}_{jk}^{j} \|^2 \}}{\mathbb{E}\{ \| \vect{h}_{jk}^{j} \|^2 \}}  = 1- \frac{{\beta_{jk}^{j}}}{\beta_{jk}^{j} + \hspace{-15mm} \underbrace{\sum_{l \in \Pu_{j}\setminus\{j\} }  \beta_{lk}^{j}}_{\text{Interference from UEs using the same pilot}} \hspace{-12mm}+\frac{1}{\taupu}\frac{\sigma^2_{\rm {ul}}}{\rho_{\rm {ul}}}}.
\end{equation}
Since the interference generated by the pilot-sharing UEs increases the NMSE in \eqref{eq:NMSE_Uncorr}, it also reduces the channel estimation quality. This ``pilot interference'' is often called \emph{pilot contamination} and behaves differently from the noise;
it not only reduces the estimation quality, but has an additional impact on the SE since \eqref{eq:MMSEestimator_h_jli_uncorr} contains the channels of pilot-sharing UEs. This effect will later be discussed in detail.

\subsection{Basic signal processing}

We analyze the achievable SE, focusing for now on the UL. Similar to \eqref{eq:uplink-pilot-model}, 
the received signal $\vect{y}_j \in \mathbb{C}^{M}$ at BS~$j$ during UL data transmission is modeled as
\begin{align} \label{eq:uplink-signal-model}
\vect{y}_j = \underbrace{ \sum_{i=1}^{K} \vect{h}_{ji}^{j} s_{ji}}_{\textrm{Desired signals}} + \underbrace{\sum_{l=1,l \neq j}^{L} \sum_{i=1}^{K} \vect{h}_{li}^{j} s_{li}}_{\textrm{Intra- and inter-cell interference}} + \underbrace{\vphantom{\sum_{i=1,i\ne k}^{K} } \vect{n}_{j}}_{\textrm{Noise}}
\end{align}
where $\vect{n}_{j} \sim \CN(\vect{0}_{M}, \sigma^2_{\rm {ul}}\vect{I}_{M})$ is independent noise. The data signal from UE~$i$ in cell~$l$ is denoted by $s_{li} \sim \CN({0}, \rho_{\rm {ul}})$, with $\rho_{\rm {ul}}$ being the transmit power. {For the sake of discussion, all UEs transmit with the same power. Although better performance can be achieved by using some power allocation strategy, this does not have any fundamental impact on the scaling behaviors. We refer the interested reader to \cite[Sec. 7.1]{massivemimobook}.}

To detect the signal $s_{jk}$ from UE~$k$, BS~$j$ selects the receive combining vector $\vect{v}_{jk} \in \mathbb{C}^{M}$ to separate the desired signal from the interference and noise. This vector is multiplied with \eqref{eq:uplink-signal-model} to obtain
\begin{align} \notag
\!\!\vect{v}_{jk}^{\Htran} \vect{y}_{j} = \vect{v}_{jk}^{\Htran}\vect{h}_{jk}^{j} s_{jk} &+ \underbrace{ \sum_{i=1,i\ne k}^{K} \vect{v}_{jk}^{\Htran}\vect{h}_{ji}^{j} s_{ji}}_{\textrm{Intra-cell interference}} \\&+ \underbrace{\sum_{l=1,l \neq j}^{L} \sum_{i=1}^{K} \vect{v}_{jk}^{\Htran}\vect{h}_{li}^{j} s_{li}}_{\textrm{Inter-cell interference}} + \underbrace{\vphantom{\sum_{i=1,i\ne k}^{K} } \vect{v}_{jk}^{\Htran}\vect{n}_{j}}_{\textrm{Noise}}.\label{eq:vector-channel-UL-processed}
\end{align}
Two popular choices for $\vect{V}_{j} = \left[ \vect{v}_{j 1} \, \ldots \, \vect{v}_{j K}  \right]$ are maximum-ratio (MR) and zero-forcing (ZF) combining \cite{Marzetta2016a}:
\begin{equation} \label{eq:combining-schemes}
\!\!\!\vect{V}_{j} = \begin{cases}
\vect{V}_{j}^{\rm{MR}}  = \widehat{\vect{H}}_{j}^{j} & \textrm{with MR combining} \\ 
\vect{V}_{j}^{\rm{ZF}}  =\widehat{\vect{H}}_{j}^{j} \left(
 (  \widehat{\vect{H}}_{j}^{j})^{\Htran} \widehat{\vect{H}}_{j}^{j}  \right)^{-1} & \textrm{with ZF combining}\!\!\!
\end{cases}
\end{equation}
with $\widehat{\vect{H}}_{j}^{j}  = [   \widehat{\vect{h}}_{j 1}^{j} \, \ldots \, \widehat{\vect{h}}_{j K}^{j} ]\in \mathbb{C}^{M \times K}$ containing the estimates of the intra-cell channels in cell $j$.  MR and ZF are both suboptimal but have been considered since the beginning of SDMA and applied to Massive MIMO since the early works. MR  has low computational complexity and maximizes the power of the desired signal, but neglects the existence of interference. ZF has higher complexity (due to the inversion of a $K \times K$ matrix) but can partially suppress intra-cell interference. 
In a single-cell scenario with perfect CSI, MR and ZF are asymptotically optimal at low and high SNRs, respectively \cite{Bjornson2013d}.
In fact, both schemes were conceived for single-cell operation; that is, they rely only on the channel estimates from UEs in the own cell and ignore the existence of other cells. We will see that this is a major drawback.

\begin{table*}[t]
\renewcommand{\arraystretch}{1.}
\centering
\caption{System parameters of the running example. The asymmetric network (with wrap-around) in Fig.~\ref{figure_network_layout} is used.}
\label{table:system_parameters_running_example}
\begin{tabular}{|c|c|}
\hline \bfseries $\!\!\!\!\!$ Parameter $\!\!\!\!\!$ & \bfseries Value\\
\hline\hline

      Network area &  $1$\,km $ \times\,  1$\,km \\
  Number of cells and UEs per cell&  $L = 16, K=10$ \\

UL noise power and UL transmit power & $\sigma_{\rm{ul}}^2 = -94$\,dBm, $\rho_{\rm{ul}}= 20$\,dBm \\
%

Samples per coherence block & $\tau_c = 200$ \\

Pilot reuse factor & $f=1,2$ or $4$ \\

Distance between UE $k$ in cell $l$ and BS~$j$ & $d_{lk}^{\,j}$ \\

\begin{tabular}{@{}c@{}} Large-scale fading coefficient for \\the channel between UE $k$ in cell $l$ and BS~$j$\end{tabular}
& $\beta_{lk}^{j} =  -148.1 - 37.6 \, \log_{10} \left( \frac{d_{lk}^{j}}{1\,\textrm{km}} \right) + F_{lk}^{j}$\,dB\\
 
Shadow fading between UE $k$ in cell $l$ and BS~$j$ & $F_{lk}^{j} \sim \mathcal{N}(0,10)$ \\




\hline
\end{tabular}
\end{table*}

The SE that a UE can achieve is upper bounded by the channel capacity. Thus, an \emph{achievable SE} is any number that is below the capacity. While the classical ``Shannon formula'' cannot be applied when the receiver has imperfect CSI, there are well-established capacity lower bounds that can be used. One popular choice is the \emph{use-and-then-forget (UatF) bound} \cite{massivemimobook,Marzetta2016a}, whose name comes from the fact that the channel estimates are used for designing the receive combining vectors and then effectively ``forgotten'' before the signal detection.

\begin{theorem} \label{theorem:uplink-capacity-forgetbound} 
An UL SE of UE~$k$ in cell~$j$ is 
\begin{equation} \label{eq:uplink-SE-expression-forgetbound}
\underline{\mathacr{SE}}^{\mathrm{ul}}_{jk}= \frac{\tau_u}{\tau_c} \log_2  ( 1 + \underline{\gamma}^{\mathrm{ul}}_{jk} ) \quad \textnormal{[bit/s/Hz]} 
\end{equation}  
where $\underline{\gamma}^{\mathrm{ul}}_{jk}$ can be interpreted as the SINR and is given by
\begin{equation} \label{eq:uplink-SINR-expression-forgetbound}
\begin{split}
\underline{\gamma}^{\mathrm{ul}}_{jk} = \frac{  | \mathbb{E}\{ \vect{v}_{jk}^{\Htran} \vect{h}_{jk}^{j}  \} |^2  }{ 
\sum\limits_{l=1}^{L} \sum\limits_{i=1}^{K_l}   \mathbb{E} \{  | \vect{v}_{jk}^{\Htran} \vect{h}_{li}^{j} |^2 \}
-  | \mathbb{E}\{ \vect{v}_{jk}^{\Htran} \vect{h}_{jk}^{j}  \} |^2 + \frac{\sigma^2_{\rm {ul}}}{\rho_{\rm {ul}}} \mathbb{E}\{ \|  \vect{v}_{jk}   \|^2 \}  }.
\end{split}
\end{equation}
The expectations are computed with respect to all sources of randomness and the pre-log factor $\frac{\tau_u}{\tau_c}$ accounts for the fraction of samples per
coherence block used for UL data.
\end{theorem}
The achievable SE above has been widely used in Massive MIMO for the following main reasons. Firstly, it can be applied with any combining scheme and channel estimator. Secondly, it is analytically tractable, in the sense that the expectations can be computed in closed form in some special cases \cite{Marzetta2016a}, as shown next.

\begin{corollary} \label{cor:closed-form-MR_uncorr}If MR combining is used with uncorrelated Rayleigh fading, the SINR in \eqref{eq:uplink-SINR-expression-forgetbound} of Theorem~\ref{theorem:uplink-capacity-forgetbound} becomes \begin{equation} \label{eq:uplink-rate-expression-forgetbound-MR-uncorr}
\underline{\gamma}_{jk}^{\mathrm{ul}} =  \frac{ (\beta_{jk}^{j})^2 \psi_{jk}^{\,j} M }{ 
\underbrace{  \vphantom{ \sum\limits_{(l,i) \in \Pu_{jk}  \setminus  \{ j \} } }  \sum\limits_{l=1}^{L} \sum\limits_{i=1}^{K}  \beta_{li}^{j}}_{\textrm{Non-coherent interference}}
+ \underbrace{\sum\limits_{l \in \Pu_{j}  \setminus  \{j\} }  (\beta_{lk}^{j} )^2 \psi_{jk}^{\,j}  M}_{\textrm{Coherent interference}}
+ \frac{\sigma^2_{\rm {ul}}}{\rho_{\rm {ul}}}  }
\end{equation}
with $\psi_{jk}^{j}= \left( \sum_{l \in \Pu_{j} }  \beta_{lk}^{j} +  \frac{1}{\taupu}\frac{\sigma^2_{\rm {ul}}}{\rho_{\rm {ul}}}  \right)^{-1}$.
\end{corollary}

The SINR in \eqref{eq:uplink-rate-expression-forgetbound-MR-uncorr} has a simple interpretation. The numerator represents the desired signal power, which increases linearly with $M$, known as the \emph{array gain}. The first term in the denominator is a summation over all UEs in the network and is referred to as \emph{non-coherent interference}, because it is independent of $M$. The second term only involves the UEs in the other cells using the same pilot and is a consequence of pilot contamination. 
This term is called \emph{coherent interference} since it increases linearly with $M$, just as the signal term. This phenomenon occurs  since the channel estimate in \eqref{eq:MMSEestimator_h_jli_uncorr} is a linear combination of pilot-sharing UEs' channels.

 A similar closed-form SINR expression as \eqref{eq:uplink-rate-expression-forgetbound-MR-uncorr} can be obtained with ZF (see \cite{Marzetta2016a}) with the  difference that some of the non-coherent interference terms are reduced and the signal and coherent interference terms scale with $M-K$ (rather than with $M$). This is because ZF sacrifices $K$ spatial dimensions to suppress intra-cell interference. Since the SINR has the same simple structure in these well-studied cases, one might suspect that this is always the case in Massive MIMO, but we will later show that it is a deceiving generalization.



\subsection{Basic spectral efficiency analysis}

To quantify the UL SE in a consistent way throughout this article, we consider a running example with 16 asymmetric cells (and wrap around) as illustrated in Fig.~\ref{figure_network_layout}. The key parameters are given in Table \ref{table:system_parameters_running_example}. 
The UEs are uniformly and independently distributed in
each cell, at distances larger than $35$\,m from the BS. The results are averaged over $100$ UE distributions, with $50$ channel realizations each. 
There are $fK$ pilots in each coherence block and the remaining samples are used for UL data transmission, i.e., $\tau_u = \tau_c - f K$. 

\begin{figure}[t!]
\begin{center}
\includegraphics[width=1\columnwidth]{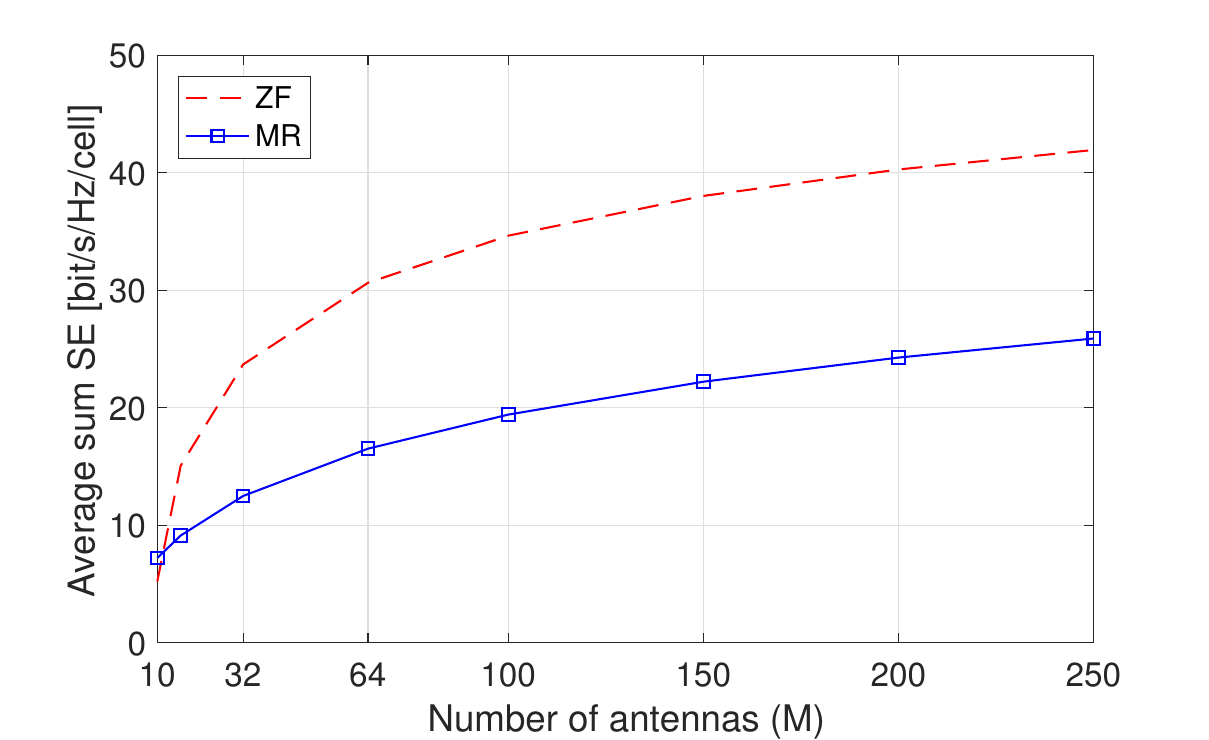}
\end{center}
\caption{Average UL sum SE as a function of the number of BS antennas for
MR and ZF combining with uncorrelated Rayleigh fading, $K=10$ UEs per cell, and pilot reuse factor $f=1$.}\label{figure:ul_se_vs_antennas_Rayleigh} 
\end{figure}

Fig.~\ref{figure:ul_se_vs_antennas_Rayleigh} shows the UL sum SE as a function of $M$ with MR and ZF combining, and $f=1$. By going from $M = 10$ to $M = 250$, the SE improves substantially with both schemes. With ZF, it passes from $5.21$ to $41.94$ bit/s/Hz, and from $7.21$ to $25.89$ bit/s/Hz with MR. 
{Although an exact comparison is not possible, we note that these numbers are substantially higher than the sum UL SE of $2.8$ bit/s/Hz/cell achieved by basic LTE systems; see \cite[Remark 4.1]{massivemimobook}.} Both MR and ZF provide higher SE per cell than that already for $M\ge16$. The SE is increased by more than $10\times$ with ZF when $M\ge64$, which reduces to $5\times$ with MR. This provides evidence that the Massive MIMO technology is  capable of improving the SE by an order of magnitude. 


\begin{figure}[t!]
\begin{center}
\includegraphics[width=1\columnwidth]{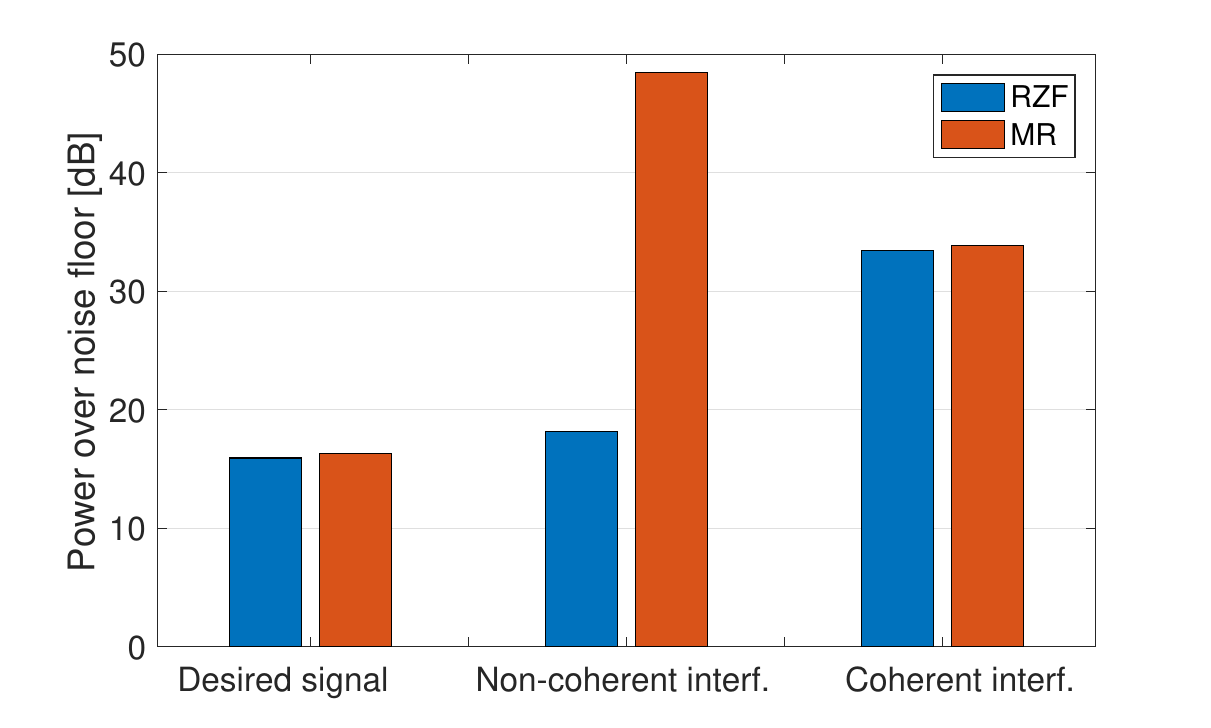}
\end{center} 
\caption{Average UL power of the desired signal, non-coherent interference, and coherent interference for the weakest UE in the cell with $M=100$ and $f=1$. Uncorrelated Rayleigh fading is considered.}\label{figure:ul_average_power_weakest} 
\end{figure}

To understand the difference between MR and ZF, Fig.~\ref{figure:ul_average_power_weakest} considers the weakest UE in an arbitrary cell, which is the one with the lowest received signal power. The average signal and interference powers are shown, normalized with respect to the noise power. We consider $M=100$ and $f=1$. The non-coherent interference is the largest term in the SINR when using MR. In contrast, it is the coherent interference caused by the pilot-sharing UEs that dominates  when using ZF. The reason is not that the coherent interference has grown---it is roughly the same as with MR---but ZF is able to effectively suppress the non-coherent interference. With MR, the non-coherent interference power dominates since it does not suppress any interference. ZF sacrifices a few dB of signal power to suppress the non-coherent interference by 30\,dB, which explains the large SE difference in Fig.~\ref{figure:ul_se_vs_antennas_Rayleigh} and demonstrates that signal processing schemes that target interference suppression are crucial to get the most out of Massive MIMO. Moreover, we notice that a new limiting factor can appear when a previous one has been resolved. It has been claimed that coherent interference can be avoided by increasing the pilot reuse factor $f$. On the contrary, in this simulation, the
 SE actually reduces as $f$ increases, since the improved estimation quality and reduced coherent interference are not compensating for the fact that the  pre-log factor $(\tau_c - f K)/\tau_c$ in \eqref{eq:uplink-SE-expression-forgetbound} decreases with $f$. 
As we will show later on, signal processing schemes developed for multicell operation are required to truly benefit from having $f>1$.

\subsection{Is pilot contamination the final frontier?}

Massive MIMO was originally characterized as a multiuser MIMO system operating in the ``Marzetta limit'' where $M\to\infty$ while the number $K$ of UEs is fixed \cite{marzetta2010noncooperative}. Unlike the traditional ``large-system limit'' that had been studied before and in which $M,K\to\infty$ with a fixed ratio, the Marzetta limit has the practical benefit that the $fK$ pilot resources required for channel estimation remain finite even in the asymptotic limit. As discussed in Section~\ref{ref:Asymptotic_discussion} in more detail, this asymptotic regime has received much attention in the literature, not because $M$ will be nearly infinite in practice, but to understand the scaling behavior and ultimate performance limits.
With MR and uncorrelated Rayleigh fading, the following result was proved in \cite{marzetta2010noncooperative} and  follows directly from Corollary~\ref{cor:closed-form-MR_uncorr}.

\begin{corollary} \label{cor:asymptotic_MR_uncorr} 

With uncorrelated Rayleigh fading, as $M \to \infty $ the SINR with MR is upper limited by
\begin{equation} 
\begin{split}
\underline{\gamma}_{jk}^{\mathrm{ul}} \to \frac{ (\beta_{jk}^{j})^2 }{ 
\underbrace{\sum\limits_{l \in \Pu_{j}  \setminus  \{j\} }  (\beta_{lk}^{j} )^2}_{\textrm{Coherent interference}}}. 
\end{split}
\end{equation}
\end{corollary}
This corollary shows that the SINR converges to a limit where the impact of noise and non-coherent interference vanishes, making the coherent interference caused by pilot contamination the only remaining performance-limiting factor. The same limit is achieved by ZF and sophisticated variations thereof \cite{Hoydis2013,Ngo2012b}, and also when studying the DL with a similar methodology.

Based on these results, the research community has speculated, or taken for granted, that pilot contamination constitutes a fundamental SE limitation. 
More specifically, the following claims have been repeated many times in the literature (including by the authors of this article):
\begin{itemize}
\item Due to pilot contamination, the SE saturates as $M\to \infty$;
\item MR is asymptotically optimal;
\item More sophisticated schemes than MR can only improve the SE for finite values of $M$.
\end{itemize}
There are two good reasons to question these claims.
Firstly, they are supported by a single anecdotal model, namely uncorrelated Rayleigh fading. Practical channels are spatially correlated \cite{Paulraj97}, while uncorrelated channels appear under extremely strict physical requirements \cite{massivemimobook}, which are certainly not satisfied in practice. 
Secondly, the signal processing schemes that have dominated the literature (such as MR and ZF), were initially designed for single-cell operation, and then used without questioning in multicell contexts. A few early works (e.g., \cite{Hoydis2013}) discussed the possibility for a BS to  use channel estimates from UEs in other cells to develop ``multicell'' interference-suppression schemes, but conjectured that this would not be helpful because of their low estimation quality. The multicell approach was first analyzed for Massive MIMO in \cite{Ngo2012b,EmilEURASIP17}, but only for uncorrelated Rayleigh fading where a finite SE limit was observed as $M\to \infty$.
Two important questions are thus:
\begin{enumerate}
\item What is the role of spatial correlation in Massive MIMO?
\item How is the SE affected by spatial correlation and the use of multicell interference-suppression schemes?
\end{enumerate}
Both questions will be addressed in this tutorial and, interestingly, will allow us to conclude that coherent interference behaves very differently once we move away from the uncorrelated fading assumption.

\section{Spatially Correlated Channels}
\label{sec:spatially-correlated}

As any $M$-dimensional vector, the channel  $\vect{h}_{lk}^{j} \in \mathbb{C}^{M}$ is characterized by its norm $\|\vect{h}_{lk}^{j}\|$ and its direction $\vect{h}_{lk}^{j} / \|\vect{h}_{lk}^{j}\|$ in the vector space. Both are modeled as random variables in a fading channel.
With the uncorrelated Rayleigh fading model in \eqref{eq:iid-Rayleigh fading}, $\|\vect{h}_{lk}^{j}\|^2$ has a (scaled) chi-squared distribution and is independent of $\vect{h}_{lk}^{j} / \|\vect{h}_{lk}^{j}\|$, which is uniformly distributed over the unit sphere in $\mathbb{C}^M$. This characterizes a spatially uncorrelated channel.

\begin{definition} \index{spatial channel correlation} \label{def:spatial-correlation}
A fading channel $\vect{h} \in \mathbb{C}^M$ is \emph{spatially uncorrelated} if the channel gain $\| \vect{h} \|^2$ and the channel direction $\vect{h}/\| \vect{h} \|$ are independent random variables, and the channel direction is uniformly distributed over the unit-sphere. The channel is otherwise \emph{spatially correlated}.
\end{definition}

As seen from this definition, the conditions for a channel to be spatially uncorrelated are very strict. This explains why all practical channels are spatially correlated, even if some are ``more'' correlated than others. There are (at least) {three} physical explanations for spatial correlation:
\begin{enumerate}
\item The propagation environment produces more multipath components to the BS from some spatial directions than from others;
\item The BS antennas have spatially dependent antenna patterns and varying polarization.
\item {The array geometry is creating spatially under- or over-sampling.}
\end{enumerate}
{These} factors contribute to making practical channels spatially correlated. 

\subsection{Correlated Rayleigh fading}

A tractable way to model spatially correlated channels with no line-of-sight path is the \emph{correlated Rayleigh fading} model: 
\begin{equation} \label{eq:correlated-Rayleigh-model}
\vect{h}_{lk}^{j} \sim \CN \left( \vect{0}_{M}, \vect{R}_{lk}^{j}  \right)
\end{equation}
where $\vect{R}_{lk}^{j} \in \mathbb{C}^{M \times M}$ is the spatial correlation matrix. The normalized trace $\beta_{lk}^{j} = \frac{1}{M} \tr \big( \vect{R}_{lk}^{j} \big)$
is the average channel gain from an antenna at BS~$j$ to UE~$k$ in cell~$l$. As with uncorrelated Rayleigh fading, the Gaussian distribution is used to model the small-scale fading variations, while $\vect{R}_{lk}^{j}$ describes the macroscopic propagation characteristics.  In theory, the correlation matrix is fixed, while in practice it is known to change on a time-scale much larger than the coherence time of the small-scale fading. We assume that $\vect{R}_{lk}^{j}$ is known at the BS. Its estimation will be discussed in Section~\ref{SectionVI}. The eigenstructure of $\vect{R}_{lk}^{j}$ determines the spatial correlation properties of the channel $\vect{h}_{lk}^{j}$; that is, which spatial directions are statistically more likely to contain signal components than others. High spatial correlation is characterized by large eigenvalues variations, {but this property can generally not be utilized to conclude that one channel is more correlated than another \cite{Jorswieck2007a}.} Models for generation of $\vect{R}_{lk}^{j}$ can be found in \cite[Sec. 7.3]{massivemimobook} {and these utilize the angular spread to quantify the amount of spatial correlation.}

Uncorrelated Rayleigh fading, as in \eqref{eq:iid-Rayleigh fading}, occurs in the special case when $\vect{R}_{lk}^{j} = \beta_{lk}^{j}  \vect{I}_{M}$ is a scaled identity matrix, which implies that all the diagonal elements are equal and all off-diagonal are zero. Practical channels are different in both respects, as proved by numerous measurements campaigns. This will be further explained in Section \ref{sec:channel_measurements}.
\begin{remark}[{Karhunen-Lo\`eve} representation]
Let the eigendecomposition of $\vect{R}_{lk}^{j}$ be given by 
\begin{align}\label{eq:eigendecomposition}
\vect{R}_{lk}^{j}={\bf U}_{lk}^{j}{\bf \Lambda}_{lk}^{j}({\bf U}_{lk}^{j})^{\Htran}
\end{align} 
where ${\bf \Lambda}_{lk}^{j}\in \mathbb{C}^{r\times r}$ is a diagonal matrix containing the $r = \rank(\vect{R}_{lk}^{j})$ non-zero eigenvalues of $\vect{R}_{lk}^{j}$ and ${\bf U}_{lk}^{j}\in \mathbb{C}^{N\times r}$ is the tall {semi-unitary} matrix containing the eigenvectors of $\vect{R}_{lk}^{j}$ corresponding to the non-zero eigenvalues. This matrix satisfies ${\big(\vect{U}_{lk}^{j}\big)}^{\Htran} \vect{U}_{lk}^{j} = \vect{I}_{r}$. Using the Karhunen-Lo\`eve representation, we can write the channel vector $\vect{h}_{lk}^{j}$ in \eqref{eq:correlated-Rayleigh-model} as
\begin{equation} \label{eq:correlated-Rayleigh-model_1}
\vect{h}_{lk}^{j} = {\bf U}_{lk}^{j}\big({\bf \Lambda}_{lk}^{j}\big)^{1/2}
\vect{e}_{lk}^j 
\end{equation}
where $\vect{e}_{lk}^j \sim \CN \left( \vect{0}_{r},  \vect{I}_{r} \right)$. This is a convenient way to generate fading realizations of a spatially correlated channel.
\end{remark}

\subsection{Uniform linear arrays and angular representation}\label{sec:ULA}
Although antenna arrays come in arbitrary shapes and sizes---depending on the use case, site geometry, and carrier frequency---the most widely used model in the MIMO literature is the uniform linear array (ULA). Consider UE $k$ in cell $l$ and suppose that its signal is received at an arbitrary BS $j$ as the superposition of $N$ physical signal paths, each reaching the ULA as a planar wave from a particular angle $\varphi_{lk}^j (n)\in[0,2\pi]$ for $n=1,\ldots,N$. The channel vector is then obtained as
\begin{equation} \label{eq:ULA}
\vect{h}_{lk}^j = \sum\limits_{n=1}^{N} g_{lk}^j(n){{\bf a}}( \varphi_{lk}^j(n))
\end{equation}
where $g_{lk}^j(n) \in \mathbb{C}$ accounts for the gain and phase rotation of the $n$th physical path and ${{\bf a}}( \varphi_{lk}^j(n))\in \mathbb{C}^M$ is the array response vector of the ULA, given by
\begin{equation} \label{eq:array_response_vector}
{{\bf a}}\big( \varphi\big) = \big[1 \;\; e^{j2\pi \Delta  \cos(\varphi)} \, \ldots \,  e^{j2\pi \Delta(M-1)  \cos(\varphi)}\big]^{\Ttran}
\end{equation}
where $\Delta$ denotes the spacing between adjacent antennas, normalized by the wavelength. Suppose the angles are i.i.d.~random variables with angular probability density function $f(\bar\varphi) $ and $g_{lk}^j (n)$ are i.i.d. random variables with zero mean and variance $\mathbb{E}\{|g_{lk}^j (n)|^2\}$.
Then, the correlation matrix $\vect{R}_{lk}^j= \mathbb{E}\{\vect{h}_{lk}^j{(\vect{h}_{lk}^j)}^{\Htran}\}$
has a Toeplitz form with the $(m_1,m_2)$th element given by 
\begin{equation} \label{eq:ULA_covariance_element}
\left[\vect{R}_{lk}^j\right]_{m_1,m_2}= \beta_{lk}^j \int e^{j2\pi \Delta (m_1-m_2)  \cos(\bar\varphi)}f(\bar\varphi)d\bar\varphi
\end{equation}
where $\beta_{lk}^j = \sum_{n=1}^N \mathbb{E}\{|g_{lk}^j (n)|^2\}$ represents the total average gain of the physical paths. {Notice that the Toeplitz form arises with other array geometries.\footnote{The wide sense stationary uncorrelated scattering (WSSUS) assumption is only required for the Toeplitz form to hold \cite[Ch. 2.4]{PaulrajBook}.}} By using the Szeg\H{o}'s theorem \cite{Gray2006}, it can be proved that the Toeplitz matrix ${\bf R}_{lk}^j$ becomes asymptotically equivalent to a circulant matrix as $M\to \infty$. A well-known property of circulant matrices \cite{Gray2006} is that their eigenvector matrix equals the unitary discrete Fourier transform (DFT) matrix $\vect{F} \in \mathbb{C}^{M\times M}$ with elements 
\begin{equation}\label{eq:DFT}
\left[\vect{F}\right]_{l,n}= \frac{1}{\sqrt{M}}e^{j\frac{2\pi}{M} (l-1)(n-1)} 
\end{equation}
for  $l,n\in\{1,2,\ldots,M\}$. The consequence is that, in the regime of large $M$ where the Toeplitz matrix ${\bf R}_{lk}^j$ is well approximated by a circulant matrix, we can approximate ${\bf U}_{lk}^j$ in \eqref{eq:eigendecomposition} with a submatrix $\bar{\bf F}\in \mathbb{C}^{N\times r}$, formed by a selection of $r$ columns of $\bf F$. Therefore, \eqref{eq:eigendecomposition} becomes 
\begin{align}\label{eq:eigendecomposition_approx}
\vect{R}_{lk}^{j}\approx\bar{\bf F}{\bf \Lambda}_{lk}^{j}\bar{\bf F}^{\Htran}.
\end{align} 
Another consequence is that the computationally efficient fast Fourier transform (instead of the Karhunen-Lo\`eve representation in \eqref{eq:correlated-Rayleigh-model_1}) can be used for the generation of ${\bf{h}}_{lk}^j$.

Notice that the inverse DFT transformation of ${\bf{h}}_{lk}^j$ given by $\bar{\bf{h}}_{lk}^j = \bar{\bf F}^{\Htran}{\bf{h}}_{lk}^j$ is nothing but its angular domain representation \cite[Ch. 7.3.3]{Tse:2005}. 
If the antennas in the ULA are \emph{critically spaced} at half the wavelength (i.e., $\Delta=1/2$), then the inverse DFT transformation decomposes the channel from any spatial physical direction $\varphi_{lk}^j(n)$ along one particular angle $\pm \arccos (k/L)$ for $k=0,\ldots,M-1$, where $L=M\Delta$ is the normalized length of the ULA. The parameter $1/L$ is a measure of the array's angular resolvability. Note that this resolution is not affected by a change of the number of antennas as long as the length $L$ remains the same. 

{Notice that the DFT approximation requires the macroscopic large-scale fading to maintain constant over the antenna array. Therefore, it cannot be used to capture near-field scattering and other phenomena that may lead to such variations. These have been observed by several measurement campaigns as shown next.}

\subsection{A look at channel measurements}\label{sec:channel_measurements}
\begin{figure}[t!]
\begin{center}
\subfloat[Diagonal elements of the  correlation matrices.]{\label{figure:measurements_fhh_diagonals}
\begin{overpic}[unit=1mm,width=1\columnwidth]{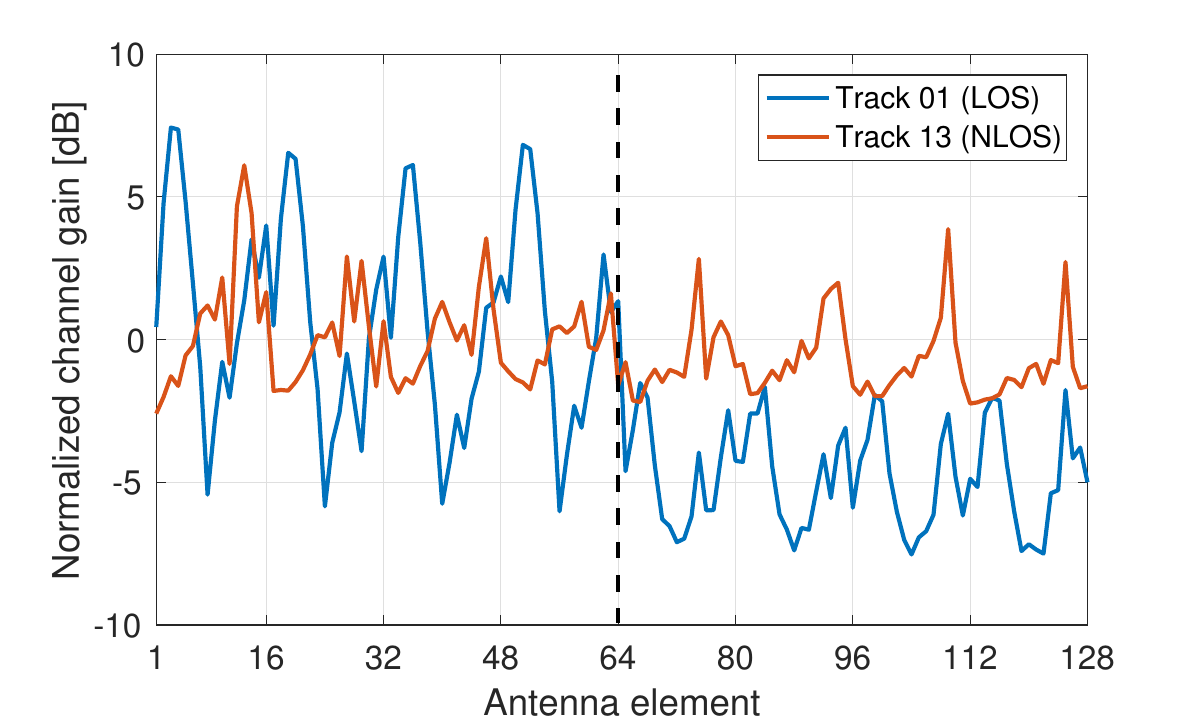}
\put(29,10){\scriptsize  Vertical polarized}
\put(50,13){\vector(-1,0){7}}
\put(53,13){\vector(1,0){7}}
\put(54,10){\scriptsize  Horizontal polarized}
\end{overpic}} \\
\subfloat[Eigenvalues of the  correlation matrices, in decreasing order.]{\label{figure:measurements_fhh_eigenvalues}
\begin{overpic}[unit=1mm,width=1\columnwidth]{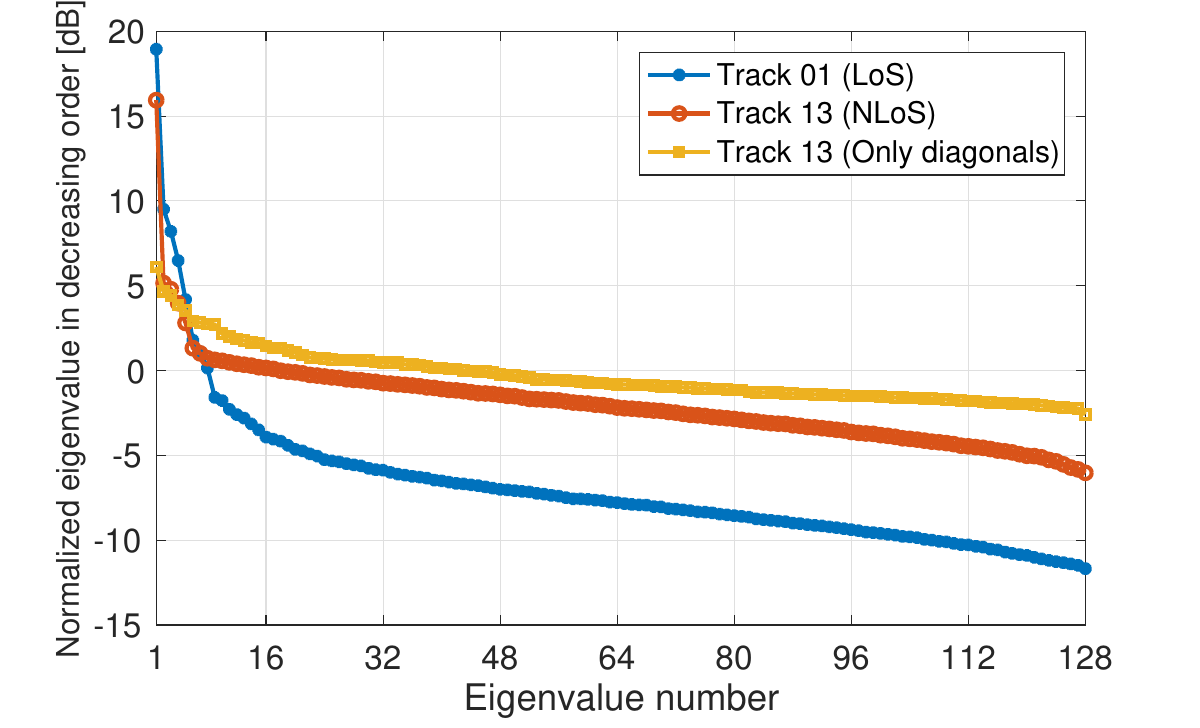}
\end{overpic}} 
\end{center}
\caption{Measurements results obtained at Fraunhofer Heinrich Hertz Institute in a scenario where a transmitting UE is received by a $128$-antenna uniform cylindrical array that is located outdoor an height of approximately $30$\;m, in both LoS and NLoS. Each spatial correlation matrix is estimated by averaging 1000 samples over time and frequency. The matrices are normalized to have trace equal to $M$. The diagonal elements of the correlation matrices are shown in (a) and demonstrate the non-uniform channel gains for the different antennas. The eigenvalues of the correlation matrices are shown in (b) and demonstrate the non-uniform distribution over multipath components from different spatial directions. The yellow curve shows what would have been achieved if the off-diagonal elements were zero in the NLoS case. Note that every UE has a unique spatial correlation structure.}\label{figure:measurements_fhh}
\end{figure}

Fig.~\ref{figure:measurements_fhh} shows the key properties of the measured correlation matrices of two UEs selected from a measurement campaign at Fraunhofer Heinrich Hertz Institute \cite{Thiele19}. The measurements were conducted at a 3.75\,GHz carrier frequency using a bandwidth of 250\,MHz.  A uniform cylindrical array was used with $16$ columns, $4$ dual-polarized patch elements with an antenna spacing of $52$\,mm in the vertical direction. The total number of antennas is $128$. The BS was located outdoor at {a height} of approximately $30$\,m whereas the transmitting UE moved over $13$ different tracks, each of $40$\,m length.\footnote{We refer to Lars Thiele (email: lars.thiele@hhi.fraunhofer.de) for more details on the measurement campaign.}

Fig.~\ref{figure:measurements_fhh}(a) shows the diagonal elements of the normalized (such that the trace is equal to $M=128$) correlation matrices, each representing the average normalized channel gain between an individual BS antenna and the UE transmitting from ``Track 01'' and ``Track 13''. We used 1000 measurement samples taken over a 10\,m distance to create correlation matrices.
The differences between the two UE are the spatial locations and that the propagation occurs respectively in line-of-sight (LoS) and non-line-of-sight (NLoS). The power varies substantially around the average channel gain according to a random-like pattern that is unique for every UE, and the periodic behaviors are created by the cylindric array geometry. The standard deviation of the gain variations (measured in dB scale) lies between 2 and 3 dB for the different cases. 
An explanation for this effect is that different multipath components are seen by different parts of the array.

Fig.~\ref{figure:measurements_fhh}(b) shows the decreasingly ordered eigenvalues of the measured correlation matrices. In both LoS and NLoS cases, the eight largest eigenvalues contribute to more than 99\% of the channel gains.
Hence, the measured channel vectors are approximately spanned by the corresponding eight eigenvectors, which are different for each case, although the shape of the eigenvalue curves are similar. An explanation for this effect is that the vast majority of the multipath components originate from a relatively small number of spatial directions.
If the off-diagonal elements were zero in the NLoS case (corresponding to uncorrelated fading), then the yellow curve would be created by the gain variations of Fig.~\ref{figure:measurements_fhh}(a), which has a quite similar behavior.

Analogous observations can be made from the measurements reported in \cite{Gao2015a,Gao2015b}. Hence, we conclude that practical spatial correlation matrices have three main properties:
\begin{itemize}
\item Non-identical diagonal elements;
\item Non-zero off-diagonal elements;
\item A unique structure for every position of UE and BS.
\end{itemize}




\subsection{Basic impact of spatial correlation in Massive MIMO}

{Spatial correlation was initially viewed as detrimental in the early papers on MIMO communications, since this is the case for high-data rate communications over single-user MIMO channels \cite{Paulraj97,Loyka2001a}. However, the situation is quite different in multiuser communications where it is known that spatial correlation can be both beneficial and detrimental \cite{Clerckx2008b}. More precisely,}
the relation between the dominant eigenvectors of the UEs play an important role, in addition to the eigenvalue distribution. We will demonstrate this with an example, where we assume for simplicity that only cell $j$ is active. Moreover, we artificially assume that each correlation matrix has $r=M/K$ eigenvalues equal to one, while the remaining eigenvalues are zero. The correlation matrix of UE $k$ can  thus be represented as
\begin{equation}\label{eq:orthogonal-correlation}
	\vect{R}_{jk}^{j} =  \vect{U}_{k} \vect{U}_{k}^{\Htran}
	\end{equation}
where the columns of $\vect{U}_{k} \in\mathbb{C}^{M\times r}$ are the unit-norm eigenvectors corresponding to the $r$ non-zero eigenvalues. 
We will investigate two extreme cases: when $ \vect{U}_{k} $ is different for every UE and when it is the same.

In the first case, we assume that $\vect{U}_{k}^{\Htran} \vect{U}_{i} = \vect{0}_{r \times r}$ $\forall k\neq i$, which implies that the individual correlation matrices are all mutually orthogonal. Multiplying $\vect{y}_j$ in \eqref{eq:uplink-signal-model} by the eigenvector matrix $\vect{U}_{k}^{\Htran}$ of UE $k$, we obtain
\begin{align}
\vect{U}_{k}^{\Htran}\vect{y}_j =  \vect{e}_{jk}^j s_{jk} + \vect{U}_{k}^{\Htran}\vect{n}_j \label{eq:individual-correlation}
\end{align}
where $\vect{e}_{jk}^j = \vect{U}_{k}^{\Htran} \vect{h}_{jk}^j  \sim \CN( \vect{0}_{r}, \vect{I}_{r})$. Notice that \eqref{eq:orthogonal-correlation} is the system model of an $r$-antenna single-user channel with $\vect{e}_{jk}^j$ being the uncorrelated Rayleigh fading channel vector and $\vect{U}_{k}^{\Htran}\vect{n}_j$ is the noise. 
Hence, thanks to the structure of the spatial correlation matrices, the multiuser channel is divided into $K$ interference-free single-user channels, each having $r$  ``antennas'' instead of $M$. The interference removal makes spatial correlation very beneficial in this case.

In the second case, we assume that all UEs share the same correlation matrix $\vect{R}_{jk}^{j} = \vect{U}\vect{U}^{\Htran}$ for $k=1,\ldots,K$, with $\vect{U}\in\mathbb{C}^{M\times r}$. Under these circumstances, we have
\begin{align}
 \vect{U}^{\Htran}\vect{y}_j =   \sum_{i=1}^K \vect{e}_{ji}^j s_{ji} + \vect{U}^{\Htran}\vect{n}_j.
\end{align}
Unlike \eqref{eq:individual-correlation}, this is a $K$-user channel with $r$  uncorrelated ``antennas''. Hence, due to the common spatial correlation matrix, we effectively lose many of the antennas without gaining anything back in terms of interference removal, which is clearly an undesirable effect of spatial correlation.

\subsection{Linear independence and orthogonality of matrices}

Spatial correlation can apparently be beneficial in Massive MIMO if the UEs have sufficiently different spatial correlation matrices. Two difference measures  play a key role in this article and will now be defined.

A set of linearly independent vectors is such that no vector in the set can be written as a linear combination of the others. When applied to correlation matrices, this concept generalizes to the following definition.

\begin{definition}
\label{linear-independence}
Consider the correlation matrix $\vect{R} \in \mathbb{C}^{M \times M}$. \index{linearly independent}
This matrix is \emph{linearly independent} of the correlation matrices $\vect{R}_1,\ldots,\vect{R}_N \in \mathbb{C}^{M \times M}$ if
\begin{equation} \label{eq:linearly-independent-covariance-matrices-def}
\left\| \vect{R} - \sum_{i=1}^{N} c_i \vect{R}_i \right\|^2_F >0
\end{equation}
for all $c_1,\ldots,c_N \in \mathbb{R}$, {with $\left\| \vect{A} \right\|_F = \sqrt{\tr({\vect{A}^{\Htran}\vect{A}})}$ being the Frobenius norm}.
We further say that $\vect{R}$ is  \emph{asymptotically linearly independent}\index{asymptotically linearly independent} of $\vect{R}_1,\ldots,\vect{R}_N$ if
\begin{equation} \label{eq:linearly-independent-covariance-matrices-def-asymp}
\underset{M}{\mathrm{lim \, inf}} \, \frac{1}{M} \left\| \vect{R} - \sum_{i=1}^{N} c_i \vect{R}_i \right\|^2_F >0
\end{equation}
for all $c_1,\ldots,c_N \in \mathbb{R}$.
\end{definition}

The linear independence condition in \eqref{eq:linearly-independent-covariance-matrices-def} implies that $\vect{R}$ cannot be written as a linear combination of $\vect{R}_1,\ldots,\vect{R}_N$. All these matrices can have full rank, but different eigenvalues and/or eigenvectors. The asymptotically linear independence condition in \eqref{eq:linearly-independent-covariance-matrices-def-asymp} is more restrictive since it both requires linear independence and that the subspace in which the matrices differ has a norm that grows at least linearly with $M$.


Another measure of the difference between vectors is orthogonality. When applied to correlation matrices, this concept generalizes as follows.

\begin{definition}
\label{def:orthogonal-matrices}
Two correlation matrices $\vect{R}_1,\vect{R}_2 \in \mathbb{C}^{M \times M}$ are \emph{spatially orthogonal}\index{spatially orthogonal} if  
\begin{equation} \label{eq:orthogonal-matrices-def}
\tr \left( \vect{R}_1 \vect{R}_2 \right)= 0.
\end{equation}
We further say that $\vect{R}_1$ and $\vect{R}_2$  \emph{asymptotically spatially orthogonal}\index{asymptotically spatially orthogonal} if
\begin{equation} \label{eq:orthogonal-matrices-def-asymp}
 \frac{1}{M} \tr \left( \vect{R}_1 \vect{R}_2 \right) \to 0 \quad \textrm{as }M \to \infty.
\end{equation}
\end{definition} 

The orthogonality condition in \eqref{eq:orthogonal-matrices-def} implies $\vect{R}_1 \vect{R}_2 = \vect{0}_{M \times M}$, which means that the eigenspaces are non-overlapping. Hence, $\vect{R}_1$ and $\vect{R}_2$ can only be orthogonal if they are rank-deficient. 

To understand the difference between these two measures defined above, we provide the following two examples.

\begin{example} \label{example:linear}
Consider two UEs with spatial correlation matrices
\begin{equation} \label{eq:simple-covariance-matrices}
\vect{R}_1 = \begin{bmatrix} a & c \\ c^\star & b \end{bmatrix}, \quad \vect{R}_2 = \begin{bmatrix} d & f \\ f^\star & e \end{bmatrix}.
\end{equation}
In practice, these matrices are determined by the propagation environment, antenna patterns, and UEs' locations. To model that, suppose the values of $a,b,c,d,e,f$ are drawn as realizations from continuous random variables. The matrices $\vect{R}_1$ and $\vect{R}_2$ are linearly independent if and only if the vectors $[ a \,\,b\,\, c]^{\Ttran}$ and $[d\,\, e \,\,f]^{\Ttran}$ are non-parallel. Recall that the probability to get one specific realization from a continuous distribution is always zero, while there can be a non-zero probability to obtain a realization in a given interval. Consequently, the probability that the two vectors are parallel is zero. 

The two matrices are spatially orthogonal if every element of $\vect{R}_1 \vect{R}_2 $ is zero. The first element is $ad+cf^\star$ and, due to the continuous distributions, the realization $ad+cf^\star=0$ occurs with zero probability.

The bottomline is that linear independence is very likely to occur in practice, while spatial orthogonality is very unlikely. 
\end{example}
The next example considers the asymptotic variants of linear independence and spatial orthogonality.

\begin{example} \label{example:asymptotic-linear}
Consider two UEs and assume that their spatial correlation matrices can be well approximated as $\vect{R}_1 = {\vect{F}} {\vect{F}}^{\Htran}$ and 
\begin{equation} \label{eq:matrix-example-independence2}
\vect{R} = \vect{F} \left[ {\begin{array}{*{20}{c}}
{{\lambda_1}}&0& \cdots \\
0& \ddots &0\\
 \vdots &0&{{\lambda_M}}
\end{array}} \right] \vect{F}^{\Htran}
\end{equation}
where $\vect{F} \in \mathbb{C}^{M\times M}$ is the unitary DFT matrix defined in \eqref{eq:DFT}. Recall from Section \ref{sec:ULA} that such correlation matrices arise in the large antenna regime when using ULAs. Each column of the unitary matrix $\vect{F} \in \mathbb{C}^{M \times M}$ represents a multipath component arriving from one of the  angular directions $\{\pm \arccos(k/L) : k =0,\ldots,M-1\}$ with $L$ being the normalized length of the ULA. These angular directions are equally strong in $\vect{R}_1$, while $\lambda_{1},\ldots,\lambda_{M}$ are i.i.d.~positive random variables with non-zero variance that determines the strength of the physical directions in $\vect{R}$. 
This difference between $\vect{R}_1$ and $\vect{R}$ appears naturally when UEs are at different locations, since the propagation paths to the various scattering clusters are then different.


From Definition~\ref{linear-independence} and the law-of-large numbers, we have that
\begin{align} \notag
\!\!\!\mathop {\liminf}\limits_M \frac{\| \vect{R} - c_1 \vect{R}_{1} \|_F^2}{M}
&\mathop {\geq} \mathop {\lim}\limits_M \sum_{m=1}^{M} \frac{\left( \lambda_m-\frac{1}{M}\sum\limits_{n=1}^{M}\lambda_n\right)^2}{M} \notag\\&\mathop {=} \; \mathbb{E} \left\{ (\lambda_m-\mathbb{E}\{ \lambda_m \})^2 \right\}
 \end{align}
where the lower bound is obtained by setting $c_1 = \frac{1}{M}\sum_{n=1}^{M} {\lambda_n}$. Since $\mathbb{E}\{ (\lambda_m-\mathbb{E}\{ \lambda_m \})^2\}$ is the non-zero variance of $ \lambda_m$, we conclude that $\vect{R}$ and $\vect{R}_1$ are asymptotically linearly independent.

To determine if the correlation matrices are asymptotically spatially orthogonal, we follow Definition~\ref{def:orthogonal-matrices}:
\begin{align}
 \!\!\!\!\frac{1}{M} \tr \left( \vect{R} \vect{R}_1 \right)  = \frac{1}{M} \sum_{m=1}^{M} \lambda_m \to \mathbb{E} \left\{ \lambda_m \right\}> 0 \;\; \textrm{as }M \to \infty
 \end{align}
 since the random variables $\lambda_m$ are positive and have non-zero variance. Hence, the matrices $\vect{R}$ and $\vect{R}_1$ are not asymptotically spatially orthogonal.
\end{example}

 \begin{figure*}
\setcounter{equation}{35}
\begin{align}\label{eq:uplink-rate-expression-forgetbound-MR}
\underline{\gamma}_{jk}^{\mathrm{ul}} =  \frac{  \tr \! \left( \vect{R}_{jk}^{j} {(\Psiv_{jk}^{j})}^{-1}\vect{R}_{jk}^{j} \right)  }{ 
\underbrace{ \vphantom{\sum\limits_{(l,i) \in \Pu_{jk}  \setminus  \{j\} }}  \sum\limits_{l=1}^{L} \sum\limits_{i=1}^{K}  \frac{ \tr \!\left( \vect{R}_{li}^{j}  \vect{R}_{jk}^{j} {(\Psiv_{jk}^{j})}^{-1}   \vect{R}_{jk}^{j} \right) \! }{\tr \!\left( \vect{R}_{jk}^{j} {(\Psiv_{jk}^{j})}^{-1}  \vect{R}_{jk}^{j} \right) }}_{\textrm{Non-coherent interference}}
+ \underbrace{\sum\limits_{l \in \Pu_{j}  \setminus  \{j\} }  \frac{ \left| \tr  \left( \vect{R}_{lk}^{j} {(\Psiv_{jk}^{j})}^{-1}  \vect{R}_{jk}^{j} \right)  \right|^2  }{\tr  \left( \vect{R}_{jk}^{j} {(\Psiv_{jk}^{j})}^{-1}  \vect{R}_{jk}^{j} \right) }}_{\textrm{Coherent interference}}+ \frac{\sigmaUL}{\rho_{\rm{ul}}}  } .
\end{align}
\hrule
\end{figure*}

Example~\ref{example:asymptotic-linear} indicates that asymptotic linear independence likely occurs, while asymptotic spatial orthogonality is unlikely. Although the channel models were artificial in the above examples, linear independence will always appear in practice due to the natural irregularities of propagation channels, which were observed in Fig.~\ref{figure:measurements_fhh}b and modeled randomly in Example~\ref{example:asymptotic-linear}.
When it comes to spatial orthogonality, \cite{Adhikary2013,Yin2013a} used the angular representation for ULAs to prove that two correlation matrices become asymptotically spatially orthogonal if the channels from the UEs have non-overlapping angular supports.
The measurement results in Fig.~\ref{figure:measurements_fhh} have large eigenvalue variations, but the asymptotic orthogonality metric in \eqref{eq:orthogonal-matrices-def-asymp} is still far from zero and does not decrease when going from 32 to 64 antennas. 
A likely explanation is that the multipath components are not confined to an angular interval, but spread over the angular domain in a random-like pattern, as shown by the measurements reported in \cite[Fig.~4b]{Gao2015a}.



In conclusion, we can rely on the asymptotic linear independence when designing signal processing schemes for Massive MIMO, while spatial orthogonality only appears in special cases and, thus, we cannot rely on it in the signal processing.

 \begin{figure*}
 \setcounter{equation}{37}
\begin{align}
\gamma_{jk}^{\rm {ul}} & =  \frac{ |  \vect{v}_{jk}^{\Htran} \hat{\vect{h}}_{jk}^j |^2  }{{\mathbb{E}}\left\{ 
\!\sum\limits_{l=1,l\ne j}^L\sum\limits_{i=1}^K | \vect{v}_{jk}^{\Htran} {\vect{h}}_{li}^j |^2 + \sum\limits_{i=1,i\ne k}^K | \vect{v}_{jk}^{\Htran} {\vect{h}}_{ji}^j |^2
+| \vect{v}_{jk}^{\Htran} \tilde{\vect{h}}_{jk}^j |^2+  \frac{\sigma_{\rm{ul}}^2}{\rho_{\rm{ul}}}\vect{v}_{jk}^{\Htran} \vect{v}_{jk} 
\Big| \{\widehat{\bf{H}}_l^{j} \} \right\}} \notag\\&= \frac{ |  \vect{v}_{jk}^{\Htran} \hat{\vect{h}}_{jk}^j |^2  }{ 
 \vect{v}_{jk}^{\Htran}  \left(   \sum\limits_{l=1,l\ne j}^L\sum\limits_{i=1}^K \hat{\vect{h}}_{li}^j {(\hat{\vect{h}}_{li}^j)}^{\Htran}+\sum\limits_{i=1,i\ne k}^K \hat{\vect{h}}_{jk}^j {(\hat{\vect{h}}_{jk}^j)}^{\Htran} +   \vect{Z}_j\right) \vect{v}_{jk}  
}   \label{eq:uplink-instant-SINR}
\end{align}
\hrule
\end{figure*}


\section{Signal Processing for Massive MIMO 2.0}
This section provides the core signal processing methods that underpins Massive MIMO 2.0, with particular emphasis on the exploitation of spatial channel correlation. We begin by revising the MMSE channel estimator to take such correlation into account. The optimal receive combining is derived and then simplified to obtain suboptimal single-cell schemes. Then, the DL precoding is dealt with and shown to resemble the UL results.

\subsection{MMSE channel estimation with {spatial} correlation}

Since the pilots are transmitted synchronously in all cells, BS~$j$ can use the $\tau_p$ pilots to estimate both the channels from its own UEs and the channels from UEs in other cells. The latter estimates can be used for inter-cell interference suppression. Assuming that the correlation matrices are known at the BSs (see Section \ref{SectionVI} for further details), the MMSE estimate of $\vect{h}_{li}^{j}$ at BS~$j$ under correlated Rayleigh fading is
\setcounter{equation}{32}
\begin{equation} \label{eq:MMSEestimator_h_jli}
\hat{\vect{h}}_{li}^{j}  = \vect{R}_{li}^{j} \big(\Psiv_{li}^{j}\big)^{-1}  \left(\frac{1}{\tau_p\sqrt{\rho_{\rm{ul}}}}\vect{Y}_j^{p}\bphiu_{li}\right)
\end{equation}
where $\vect{Y}_j^{p}$ is given in \eqref{eq:uplink-pilot-model} and
\begin{equation} \label{eq:Psiv-definition}
\Psiv_{li}^{j} =  \sum_{l' \in \Pu_{l} }\vect{R}_{l'i}^{j}  +  \frac{1}{\tau_p}\frac{\sigma_{\rm{ul}}^2}{\rho_{\rm{ul}}}  \vect{I}_{M}.
\end{equation}
The estimation error $\tilde{\vect{h}}_{li}^{j} = \vect{h}_{li}^{j} - \hat{\vect{h}}_{li}^{j}$ is independent of $\hat{\vect{h}}_{li}^j$ and has correlation matrix $
\vect{C}_{li}^{j} = \mathbb{E} \{ \tilde{\vect{h}}_{li}^{j} ( \tilde{\vect{h}}_{li}^{j} )^{\Htran} \}  = \vect{R}_{li}^{j} - \vect{\Phi}_{li}^j
$ with $\vect{\Phi}_{li}^j =\vect{R}_{li}^{j}
{(\Psiv_{li}^{j})}^{-1}  \vect{R}_{li}^{j}$.
The NMSE with spatially correlated channels takes the form
\begin{equation}\label{eq:NMSE_corr}
\frac{\mathbb{E}\{ \| \vect{h}_{jk}^{j} - \hat{\vect{h}}_{jk}^{j} \|^2 \}}{\mathbb{E}\{ \| \vect{h}_{jk}^{j} \|^2 \}} = \frac{\tr{\big(\vect{C}_{jk}^{j}\big)}}{\tr{\big(\vect{R}_{jk}^{j}\big)}} = 1 -\frac{\tr{\big(\vect{R}_{jk}^{j}
\big(\Psiv_{li}^{j}\big)^{-1}  \vect{R}_{jk}^{j}\big)}}{\tr{\big(\vect{R}_{jk}^{j}\big)}}
\end{equation}
and is increased by the interference from pilot-sharing UEs, which enters into $\Psiv_{li}^{j}$ in \eqref{eq:Psiv-definition}. Unlike the NMSE in \eqref{eq:NMSE_Uncorr} for uncorrelated fading, \eqref{eq:NMSE_corr} depends not only on the average channel gains but also on the full spatial correlation matrices $\vect{R}_{jk}^{j}$ and $\vect{R}_{lk}^{j}$ for $l\in \mathcal{P}_j$. Intuitively, the NMSE should be better when the interfering UEs' have very different spatial correlation properties (e.g., different dominant eigenspaces) and this intuition can be confirmed by inspecting the expression in \eqref{eq:NMSE_corr}. 
In the extreme case of $\vect{R}_{jk}^{j}\vect{R}_{lk}^{j}=\vect{0}_M$, one can show that the NMSE  in \eqref{eq:NMSE_corr} is completely unaffected by the interfering UE $k$ in cell $l$. Therefore, in theory, it is possible to completely avoid pilot contamination between two UEs if their correlation matrices are spatially orthogonal. While this condition is unlikely to hold in practice (as explained earlier) a good rule-of-thumb is to assign pilots to the UEs such that $\tr(\vect{R}_{jk}^{j}\vect{R}_{lk}^{j})$ is rather small for all pilot-sharing UEs \cite{Yin2013a}.


One can show that the NMSE in \eqref{eq:NMSE_corr} reduces with $M$ \cite[Sec.~3]{massivemimobook}, which is different from the NMSE  for uncorrelated fading in \eqref{eq:NMSE_Uncorr} that is independent of $M$. This is a logical result since the spatial correlation matrix imposes a structure that effectively reduces the randomness;
 when estimating a particular element of the channel vector, the received signals on all antennas provide useful information about that element.

\subsection{Uplink spectral efficiency and optimal receive combining}
\label{SectionIV.B}
Before looking into the optimal design of receive combining, we first generalize
Corollary \ref{cor:closed-form-MR_uncorr} for MR combining to the case of correlated Rayleigh fading \cite[Sec.~4]{massivemimobook}. This will be instrumental to understand how the UL SE is affected by spatial correlation. 

\begin{corollary} \label{cor:uplink-capacity-forgetbound-MR} 
If MR combining is used with correlated Rayleigh fading, then the SINR expression in Theorem~\ref{theorem:uplink-capacity-forgetbound} becomes \eqref{eq:uplink-rate-expression-forgetbound-MR} at the top of this page. 
\end{corollary}

The above SINR expression resembles that of \eqref{eq:uplink-rate-expression-forgetbound-MR-uncorr} in Corollary \ref{cor:closed-form-MR_uncorr}. Since the trace of an $M \times M$ matrix is the sum of the $M$ diagonal elements, the signal term in the numerator increases linearly with $M$. For the same reason, the interference terms ${ \tr \big( \vect{R}_{li}^{j}  \vect{R}_{jk}^{j} {(\Psiv_{li}^{j})}^{-1}  \vect{R}_{jk}^{j} \big) \! }/{\tr \!\big( \vect{R}_{jk}^{j} {(\Psiv_{li}^{j})}^{-1}  \vect{R}_{jk}^{j} \big) }$ remain (roughly) constant when changing $M$ and thus their sum constitutes the non-coherent interference. On the other hand, the coherent interference is obtained by summing up the terms ${ | \tr  \big( \vect{R}_{lk}^{j} {(\Psiv_{li}^{j})}^{-1}  \vect{R}_{jk}^{j} \big) |^2  }/{\tr  \big( \vect{R}_{jk}^{j} {(\Psiv_{li}^{j})}^{-1}  \vect{R}_{jk}^{j} \big) }$ that scale linearly with $M$. Unlike \eqref{eq:uplink-rate-expression-forgetbound-MR-uncorr}, the strength of the coherent interference is determined by how similar the spatial correlation matrices $\vect{R}_{lk}^{j}$ and $\vect{R}_{jk}^{j} $ are. It is small when the spatial correlation properties are very different, and zero in the special case when $\vect{R}_{lk}^{j} \vect{R}_{jk}^{j}=\vect{0}$. This is in line with the observations made before for channel estimation and shows that the SE in Massive MIMO can benefit by spatially correlated channels.

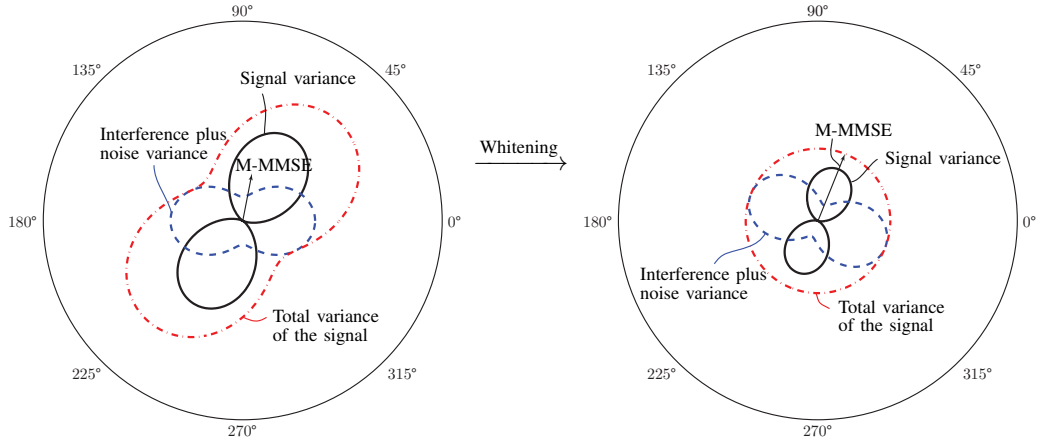
\begin{figure*}[t!]
\begin{center}
\subfloat{\label{figure_MMSEprewhitening1}
\begin{overpic}[unit=1mm,width=.7\columnwidth]{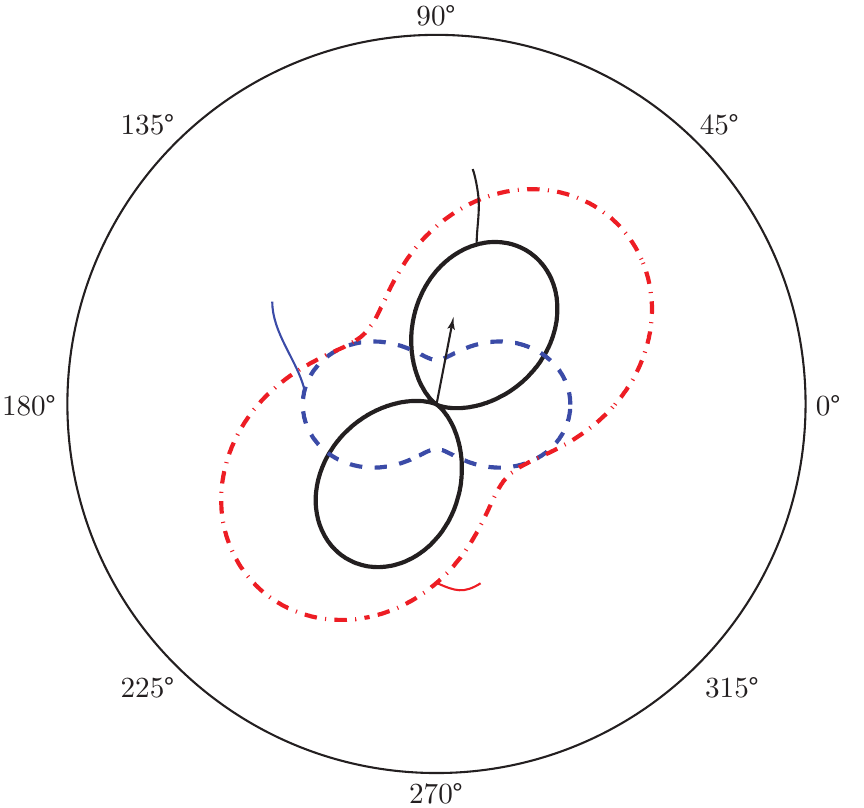}
\put(49,57.5){\scriptsize M-MMSE}
\put(50,76){\scriptsize Signal variance}
\put(20,64){\scriptsize Interference plus}
\put(20,60){\scriptsize noise variance}
\put(57,25){\scriptsize Total variance}
\put(57,21){\scriptsize  of the signal}
\end{overpic}} 
\begin{minipage}[t]{1.2cm}
\vspace{-4cm}
$\xrightarrow{\textrm{{Whitening}}}$
\end{minipage}
\subfloat{\label{figure_MMSEprewhitening2}
\begin{overpic}[unit=1mm,width=.7\columnwidth]{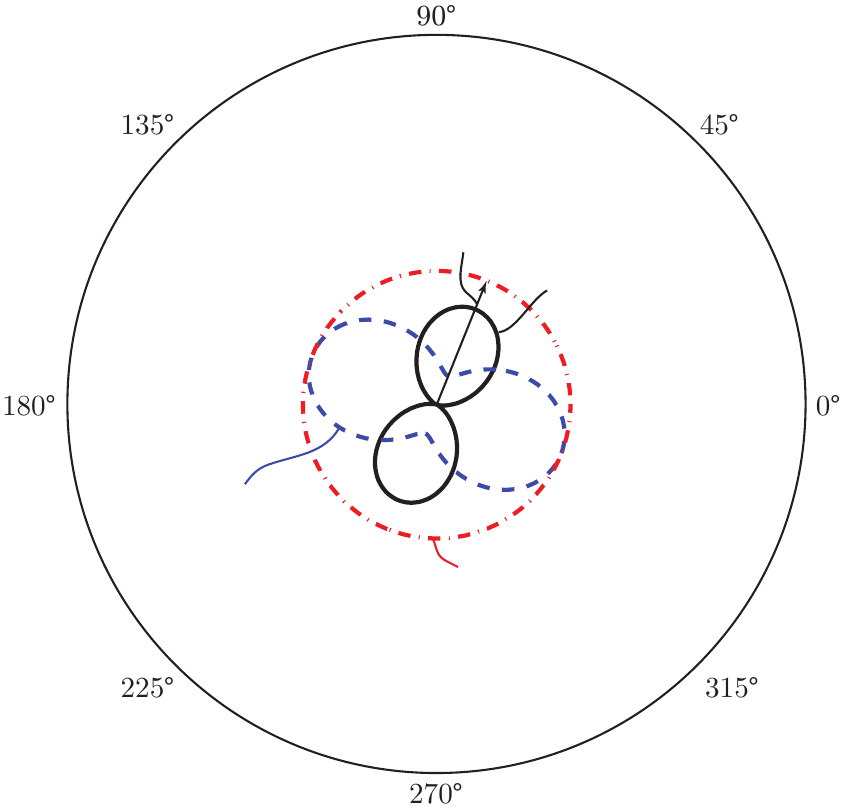}
\put(50,65){\scriptsize M-MMSE}
\put(65,59){\scriptsize Signal variance}
\put(12.5,34){\scriptsize Interference plus}
\put(12.5,30){\scriptsize noise variance}
\put(55,27){\scriptsize Total variance}
\put(55,23){\scriptsize of the signal}
\end{overpic}} 
\end{center}
\caption{Illustration of how the variance of the received signal is distributed over different spatial directions (in terms of angles) and how M-MMSE combining finds the optimal direction for signal reception. Originally, M-MMSE combining maximizes the SINR by finding a non-trivial spatial direction with the best tradeoff between high signal variance and low interference/noise variance. After whitening the received signal, M-MMSE combining corresponds to using the spatial direction that jointly maximizes the signal variance and minimizes the interference/noise variance.}\label{figure:MMSEprewhitening}
\end{figure*}

%

So far, an UL SE was computed in Theorem~\ref{theorem:uplink-capacity-forgetbound} on the basis of the UatF capacity lower bound. As mentioned earlier, this has been the most popular bound in Massive MIMO because it leads to tractable SINR expressions and also because it is valid for any combining vector and channel estimator. However, when MMSE channel estimation is used, an alternative capacity bound can be used \cite{Hoydis2013,Marzetta2016a}. 

\begin{theorem}\label{theorem:uplink-SE}
If the MMSE estimator is used to compute channel estimates for all UEs, an UL SE of UE $k$ in cell $j$ is
\setcounter{equation}{36}
\begin{equation} \label{eq:uplink-rate-expression-general}
\begin{split}
\mathsf{SE}_{jk}^{\rm {ul}} = \frac{\tau_u}{\tau_c} \mathbb{E} \left\{ \log_2  \left( 1 + \gamma_{jk}^{\rm {ul}}  \right) \right\} \quad \textrm{[bit/s/Hz] }
\end{split}
\end{equation}
with the instantaneous effective SINR given in \eqref{eq:uplink-instant-SINR} at the top of this page where ${\mathbb{E}}\{\cdot|\{{{\widehat{\bf H}}_l^{j}}\}\}$ denotes the conditional expectation given the {MMSE} channel estimates $\{{{\widehat{\bf H}}_l^{j}}: l=1,\ldots,L\}$ available at BS~$j$, and 
\setcounter{equation}{38}
\begin{align}
\vect{Z}_j = \sum\limits_{l=1}^{L} \sum\limits_{i=1}^{K} (\vect{R}_{li}^{j} - \vect{\Phi}_{li}^j) +  \frac{\sigma_{\rm{ul}}^2}{\rho_{\rm{ul}}}  \vect{I}_{M}. 
\end{align}
\end{theorem}
This alternative bound is intuitively tighter than the UatF bound in Theorem \ref{theorem:uplink-capacity-forgetbound}, since the channel estimates are used also in the signal detection, and not only in the selection of combining vectors. Unlike the UatF bound, \eqref{eq:uplink-rate-expression-general} can only be applied when the estimate $\hat{\vect{h}}_{li}^{j}$ and the estimation error $\tilde{\vect{h}}_{li}^{j} = \vect{h}_{li}^{j} - \hat{\vect{h}}_{li}^{j}$ are independent random variables---a condition that is satisfied by the MMSE estimator. The bound in \eqref{eq:uplink-rate-expression-general} has been analyzed in a number of articles that consider MR and other heuristic schemes that appear to perform well on uncorrelated Rayleigh fading channels.
Instead of resorting to heuristics, we notice that the SINR in \eqref{eq:uplink-instant-SINR}  only depends on one combining vector, $\vect{v}_{j k}$, and the SINR has the form of a generalized Rayleigh quotient. Hence, the combining vector that maximizes it can be obtained as follows \cite{Ngo2012b,EmilEURASIP17,BjornsonHS17}.

\begin{theorem} The instantaneous effective {SINRs} in \eqref{eq:uplink-instant-SINR} for the {UEs}  in cell $j$ are jointly maximized by the receive combining
\begin{equation} \label{eq:MMSE-combining}
\vect{V}_{j}^{\rm{M-MMSE}} =   \Bigg(  \sum\limits_{l=1}^L \widehat{\vect{H}}_{l}^j {(\widehat{\vect{H}}_{l}^j)}^{\Htran} + \vect{Z}_j  \Bigg)^{\!-1}    \widehat{\vect{H}}_{j}^j.
\end{equation}
\end{theorem}
\begin{table*}[t]
            \caption{Number of complex multiplications per coherence block of different combining schemes. The MMSE channel estimator is assumed with all schemes except for M-MMSE-EW and OBE, for which the EW-MMSE and LS estimators are used, respectively.}  \label{tab:cost_linear_processing}
\centering
    \begin{tabular}{|c|c|c|c|} 
    \hline
     {Scheme} & {Combining vector computation} & {Channel estimation}& {Precomputation based on channel statistics}  \\      \hline\hline 
    
    	 M-MMSE  &  $LK\frac{M^2 + M}{2} + \frac{M^3-M}{3} + KM^2$ &$M\tau_p^2 + LKM^2$& $\frac{M^3-M}{3}\tau_p + LKM^3$\\   \hline
    	S-MMSE   &   $K\frac{M^2 + M}{2} +\frac{M^3-M}{3} + KM^2 $ &$MK\tau_p + KM^2$&$\frac{M^3-M}{3}K + KM^3$\\ \hline
    	ZF 	&   $M\frac{K^2+K}{2}+\frac{K^3-K}{3} +MK^2$ &$MK\tau_p + KM^2$&-\\ \hline
	MR 		& $-$  &$MK\tau_p + KM^2$&-\\   \hline
	M-MMSE-EW	& $LK\frac{M^2 + M}{2} + \frac{M^3-M}{3} + KM^2$  &$M\tau_p + KM$&$LKM$\\   \hline
        OBE  	&   $ KM^2$&$MK\tau_p$&$K\frac{7M^3-M}{3}+\frac{M^3-M}{3} + 3KM^3(\frac{L}{f}\frac{\frac{L}{f}+1}{2})+K\frac{(\frac{L}{f})^3 - \frac{L}{f}}{3}$\\ \hline

    \end{tabular}
\end{table*}

This optimal combining scheme also minimizes the conditional MSE $\mathbb{E} \{|s_{jk} - \vect{v}_{jk}^{\Htran}\vect{y}_j|^2\,|\, \{ \widehat{\vect{H}}_{l}^{j} \}\}$ between the data signal $s_{jk}$ and the received signal $\vect{v}_{jk}^{\Htran}\vect{y}_j$ after receive combining. Hence, it would be appropriate to call it MMSE combining, but that can easily lead to confusion since the MSE-minimizing scheme takes a different form in single-cell and multicell scenarios. 
For that reason, \cite{EmilEURASIP17} proposed to call \eqref{eq:MMSE-combining} the 
\emph{multicell MMSE (M-MMSE) combining} scheme.
The `multicell' notion refers to the fact that $\vect{V}_{j}^{\rm{M-MMSE}}$ is computed by utilizing both the intra- and inter-cell channel estimates that can be computed locally at BS~$j$, using the existing pilot signaling. Intuitively, M-MMSE suppresses the strong interfering signals from wherever they may originate, in contrast to ZF that only considers interference from within the own cell. 
No cooperation between the cells is needed to implement M-MMSE, even if the name might give that impression.


The structure of $\vect{V}_{j}^{\rm{M-MMSE}} $ is quite intuitive. The inverted matrix in \eqref{eq:MMSE-combining} is the conditional correlation matrix $\Cyj = \mathbb{E} \{ \vect{y}_j \vect{y}_j^{\Htran}  \,|\, \{\widehat{\bf{H}}_l^{j} \}  \}$ of the received signal in \eqref{eq:uplink-signal-model}. 
Multiplying  $\Cyj^{-1/2}$ with $\vect{y}_j$ corresponds to whitening of the received signal such that $\mathbb{E} \{ \Cyj^{-1/2} \vect{y}_j (\Cyj^{-1/2} \vect{y}_j )^{\Htran} \,|\, \{\widehat{\bf{H}}_l^{j} \}  \}=\vect{I}_M$, meaning that the total variance of the signal, interference, plus noise is one in all directions.
After this change of basis, the desired signal from UE $k$ is received from the spatial direction $\Cyj^{-1/2} \hat{\vect{h}}_{jk}^{j}$ and the variance of the interference plus noise component must be at its weakest in that direction, since the total variance is always one.
Hence, MR combining maximizes both the desired signal and minimizes the interference plus noise after the whitening, leading to $(\Cyj^{-1/2} \hat{\vect{h}}_{jk}^{j})^{\Htran} \Cyj^{-1/2} \vect{y}_j  = (\hat{\vect{h}}_{jk}^{j})^{\Htran} \Cyj^{-1} \vect{y}_j$, from which the optimal receive combining $\Cyj^{-1}\hat{\vect{h}}_{jk}^{j}$ can be identified. In summary, M-MMSE combining is obtained by whitening followed by MR combining.  The whitening process is illustrated in Fig.~\ref{figure:MMSEprewhitening}. The M-MMSE vector is easily identified from the whitened signal (shown to the right), while this is not the case when inspecting the original signal (shown to the left). 


Table~\ref{tab:cost_linear_processing} summarizes the computational complexity of M-MMSE in \eqref{eq:MMSE-combining} (in terms of number of complex multiplications per coherence block), as obtained from \cite[Sec. 4.1.2]{massivemimobook} under the assumption that the statistical matrices $\{{\bf Z}_j:\forall j\}$ and $\{\vect{R}_{li}^{j} \Psiv_{li}^{j}:\forall j,l,i\}$ are precomputed and stored at the BSs, which requires approximately $\frac{M^3-M}{3}\tau_p+LKM^{3}$ complex multiplications. Note that the ``channel estimation'' column accounts for the computation of MMSE channel estimates from a BS to all UEs. Notice that alternative, but suboptimal, multicell schemes with lower complexity can be derived from M-MMSE. Two examples are the multicell ZF in \cite{BjornsonHS17} (with or without regularization) and the \emph{partial} M-MMSE that uses only inter-cell channels causing strong interference \cite{Pollin2019}.

The use of channel estimates to UEs in other cells in M-MMSE (or other suboptimal multicell solutions) 
is the key distinguishing factor from the basic combining schemes, such as MR and ZF. 
If we artificially assume that there is \emph{no pilot contamination} and restrict BS~$j$ to only compute and utilize  the knowledge of the channel estimates ${{\widehat{\bf H}}_j^{j}}$ of its own UEs for the design of combining vectors, the instantaneous effective SINR in Theorem \ref{theorem:uplink-SE} reduces to
\begin{align}
\gamma_{jk}^{\rm {ul}} 
= \frac{ |  \vect{v}_{jk}^{\Htran} \hat{\vect{h}}_{jk}^j |^2  }{ 
 \vect{v}_{jk}^{\Htran}  \left(  \sum\limits_{i=1,i\ne k}^K \hat{\vect{h}}_{jk}^j {(\hat{\vect{h}}_{jk}^j)}^{\Htran} +   \overline{\vect{Z}}_j\right) \vect{v}_{jk}  
}   \label{eq:uplink-instant-SINR_single-cell}
\end{align}
where $\overline{\vect{Z}}_j = \sum\nolimits_{i=1}^{K}\vect{R}_{ji}^j - \vect{\Phi}_{ji}^j + \sum\nolimits_{l=1,l \neq j}^{L}  \sum\nolimits_{i=1}^{K} \vect{R}_{li}^j 
+  \frac{\sigma_{\rm{ul}}^2}{\rho_{\rm{ul}}}    \vect{I}_{M}$. The above SINR has also the form of a generalized Rayleigh quotient and $\gamma_{j1}^{\rm {ul}},\ldots, \gamma_{jK}^{\rm {ul}} $ are jointly maximized by  
\begin{align} \label{eq:S-MMSE-combining}
{\vect{V}}_{j}^{\rm{S-MMSE}}=
\left(  \widehat{\vect{H}}_{j}^j{( \widehat{\vect{H}}_{j}^j)}^{\Htran} + \bar{\vect{Z}}_j\right)^{-1}  \!\!  \widehat{\vect{H}}_{j}^j.
\end{align}
This scheme has been called ``MMSE combining'' in \cite{Hoydis2013,Guo2014a,Krishnan2014a}, among others, since it minimizes the conditional MSE $\mathbb{E} \{ |s_{jk} - \vect{v}_{jk}^{\Htran}\vect{y}_j |^2\,|\, \widehat{\vect{H}}_{j}^{j} \}$ in the absence of pilot contamination. 
This demonstrates the fact that the MSE-minimizing scheme takes a different form depending on the underlying assumptions, which is a potential source of confusion. For example, ${\vect{V}}_{j}^{\rm{S-MMSE}}$ was heuristically applied to scenarios with pilot contamination in \cite{Hoydis2013,Krishnan2014a}, in which case the name becomes misleading since it does not minimize the MSE. Clearly, \eqref{eq:S-MMSE-combining} coincides with M-MMSE when only a single cell is active, thus we will call it the \emph{single-cell MMSE (S-MMSE) combining} as proposed by \cite{EmilEURASIP17}.

The main difference from M-MMSE in \eqref{eq:MMSE-combining} is that only the intra-cell estimates $\widehat{\vect{H}}_{j}^j$ are used in \eqref{eq:S-MMSE-combining}, while 
$\widehat{\vect{H}}_{l}^j {(\widehat{\vect{H}}_{l}^j)}^{\Htran} +  \sum\nolimits_{i=1}^{K} (\vect{R}_{li}^{j} - \vect{\Phi}_{li}^j)$ is replaced with its average $\sum\nolimits_{i=1}^{K} \vect{R}_{li}^{j}$ for all cells~$l \neq j$. Since there is no need to compute estimates of the channels to UEs in other cells, S-MMSE has a lower computational complexity, which is quantified in Table~\ref{tab:cost_linear_processing}.
The price to pay is the weaker interference suppression: if a UE at the cell edge switches to another BS, S-MMSE immediately stops suppressing its interference while M-MMSE continues as if nothing happened. This is an undesired feature of S-MMSE since the UE might still have a channel to its old BS that is better than many of the UEs that reside in the old cell.


The MR and ZF schemes in \eqref{eq:combining-schemes} can be derived as simplifications of \eqref{eq:S-MMSE-combining}  in the low and high SNR regimes, respectively \cite[Sec. 4.1.1]{massivemimobook}. The computational complexities of MR and ZF are also quantified in Table~\ref{tab:cost_linear_processing}, for comparative purposes.


\subsection{Downlink spectral efficiency and transmit precoding design}

The BS in cell~$l$ transmits the DL signal $
\vect{x}_l = \sum_{i=1}^{K} \vect{w}_{li} \varsigma_{li}$
where  $\varsigma_{li} \sim \CN(0,\rho_{\rm {dl}})$ is the DL data signal intended for UE~$i$ in cell $l$, assigned to a precoding vector $ \vect{w}_{li} \in \mathbb{C}^{M}$ that determines the spatial directivity of the transmission and satisfies $ \|  \vect{w}_{li} \|^2  =1$ so that $\rho_{\rm {dl}}$ represents the transmit power. The received signal $y_{jk} \in \mathbb{C}$ at UE~$k$ in cell~$j$ is given by
\begin{align} \notag
y_{jk} =\underbrace{   \vphantom{\sum_{i=1,i\ne k}^{K} }   ( \vect{h}_{jk}^{j})^{\Htran} \vect{w}_{jk} \varsigma_{jk}}_{\textrm{Desired signal}} &+ \underbrace{\sum_{i=1,i\ne k}^{K}  (\vect{h}_{jk}^{j})^{\Htran} \vect{w}_{ji} \varsigma_{ji}}_{\textrm{Intra-cell interference}} \\&+ \underbrace{\sum_{l=1, l\neq j}^{L} \sum_{i=1}^{K}  (\vect{h}_{jk}^{l})^{\Htran} \vect{w}_{li} \varsigma_{li}}_{\textrm{Inter-cell interference}} + \underbrace{ \vphantom{\sum_{i=1,i\ne k}^{K} }n_{jk}}_{\textrm{Noise}}\label{eq:downlink-signal-model}
\end{align}
where $n_{jk} \sim \CN(0,\sigma_{\rm{dl}}^2)$ is the receiver noise. 
Characterizing the capacity is harder in the DL than in the UL since it is unclear how the UE should best estimate the effective precoded channel $( \vect{h}_{jk}^{j})^{\Htran} \vect{w}_{jk}$ that it needs for decoding. 
An achievable SE that has received great attention in Massive MIMO can be computed by using the following \emph{hardening bound}. 
\begin{theorem} \label{theorem:downlink-capacity-forgetbound} 
The DL ergodic channel capacity of UE~$k$ in cell~$j$ is lower bounded by 
\begin{equation} \label{eq:downlink-SE-expression-forgetbound}
\underline{\mathacr{SE}}^{\mathrm{dl}}_{jk} = \frac{\tau_d}{\tau_c} \log_2   (  1 +
\underline{\mathacr{SINR}}^{\mathrm{dl}}_{jk} ) \quad \textnormal{[bit/s/Hz]} 
\end{equation}
with 
\begin{equation} \label{eq:downlink-SINR-expression-forgetbound}
\begin{split}
\underline{\mathacr{SINR}}^{\mathrm{dl}}_{jk} =  \frac{ | \mathbb{E}\{ \vect{w}_{jk}^{\Htran} \vect{h}_{jk}^{j} \} |^2  }{ 
\sum\limits_{l=1}^{L} \sum\limits_{i=1}^{K}  \mathbb{E} \{  | \vect{w}_{li}^{\Htran} \vect{h}_{jk}^{l}  |^2 \}
- | \mathbb{E}\{ \vect{w}_{jk}^{\Htran} \vect{h}_{jk}^{j} \} |^2 + \frac{\sigma_{\rm{dl}}^2}{\rho_{\rm {dl}}}  } 
\end{split}
\end{equation}
where the expectations are computed with respect to the channel realizations. The pre-log factor $\frac{\tau_d}{\tau_c}$ accounts for the fraction of samples per
coherence block that are used for DL data.
\end{theorem}

The above lower bound has dominated since the early articles on Massive MIMO \cite{marzetta2010noncooperative,Hoydis2013,ngo2013energy} and is achieved when the UE treats the mean of its precoded channel as the true one.
This is a reasonable assumption for channels that exhibit \emph{channel hardening} \cite[Sec. 2.5]{massivemimobook}, but the bound can be rather loose for channels
with little or no hardening. An alternative approach consists in estimating the precoded
channels in a better way, either explicitly as in \cite{Ngo2017a} or implicitly as in \cite{Caire17}. The use of DL pilots is unnecessary since $( \hat{\vect{h}}_{jk}^{j})^{\Htran} \vect{w}_{jk}$ is a positive scalar for all precoding schemes of interest, thus the phase is known a-priori and the magnitude can be estimated by only analyzing the received signals; this becomes easier the larger $\tau_d$ is. None of these refined bounds are considered in this tutorial since \eqref{eq:downlink-SE-expression-forgetbound} is sufficient to demonstrate the impact of spatial correlation.

The DL SE in \eqref{eq:downlink-SINR-expression-forgetbound} of UE $k$ in cell $j$ depends on the precoding vectors $\{\vect{w}_{li}\}$ of all UEs in the entire network. This stands in contrast to the UL SEs in Theorems \ref{theorem:uplink-capacity-forgetbound} and \ref{theorem:uplink-SE} that only depend the UE's own combining vector ${\bf v}_{jk}$. This makes optimal precoding design challenging in practice. A common heuristic approach relies on the UL-DL duality \cite{Bjornson2016a}, which holds between the UatF capacity bound in the UL  and hardening bound in the DL. The duality states that the SE achieved in the UL can be achieved also in the DL, if the UL combining vectors are used as DL precoding vectors and the DL transmit power is allocated properly. The UL-DL duality motivates to select the DL precoding vectors based on the UL combining vectors as
\begin{equation} \label{eq:precoding-based-on-combining}
\vect{w}_{jk} = \frac{ \vect{v}_{jk} }{ \|  \vect{v}_{jk} \| } 
\end{equation}
where $\left[ \vect{v}_{j 1} \, \ldots \, \vect{v}_{j K}  \right] = \vect{V}_j$. By selecting $\vect{V}_j$ according to one of the UL combining schemes described earlier, the corresponding precoding scheme is obtained; that is, $\vect{V}_j = \vect{V}_j^{\textrm{M-MMSE}}$ yields M-MMSE precoding, and so forth.
One key advantage is that the computation of the precoding vectors at BS~$j$ requires only $MK$ complex multiplications, which are needed to compute $\|  \vect{v}_{jk} \|$ in \eqref{eq:precoding-based-on-combining} for every UE. 


\begin{figure}[t!]
\begin{center}
\subfloat[UL sum SE with different combining schemes.]{\label{figure:ul_se_vs_antennas_CorrelatedChannels}
\begin{overpic}[unit=1mm,width=1\columnwidth]{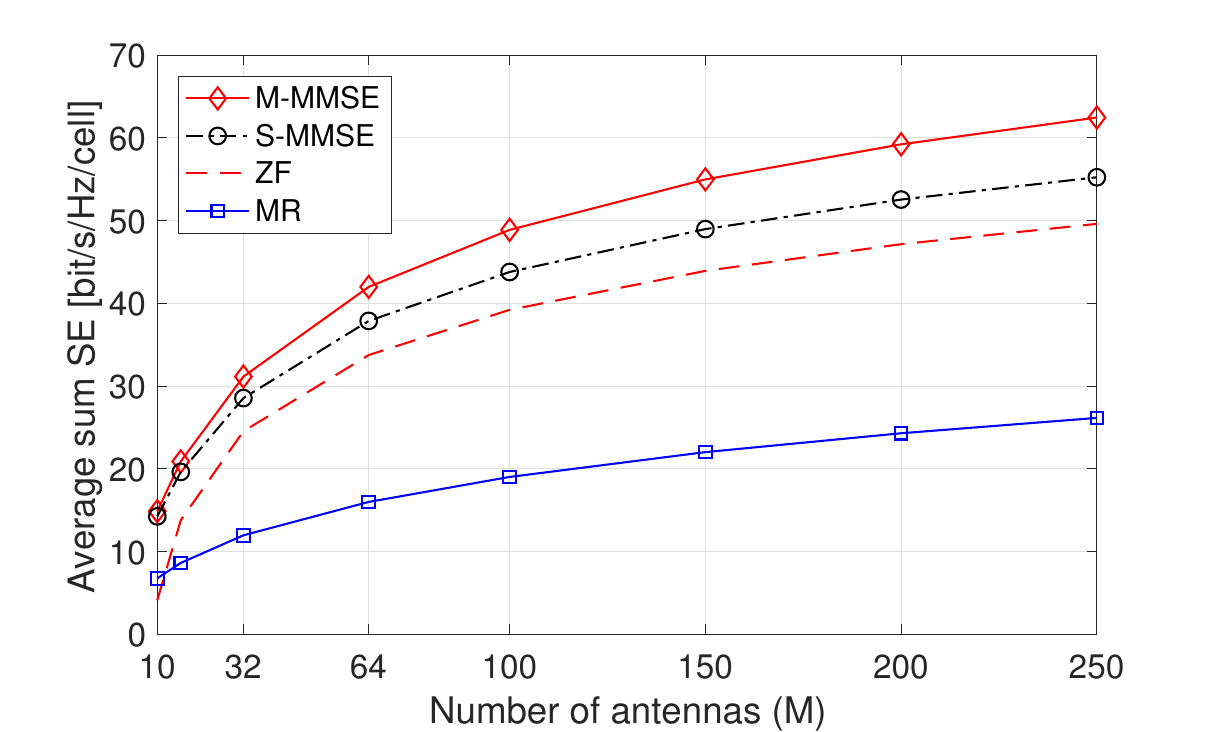}
\end{overpic}} \\
\subfloat[DL sum SE with different combining schemes.]{\label{figure:dl_se_vs_antennas_CorrelatedChannels}
\begin{overpic}[unit=1mm,width=1\columnwidth]{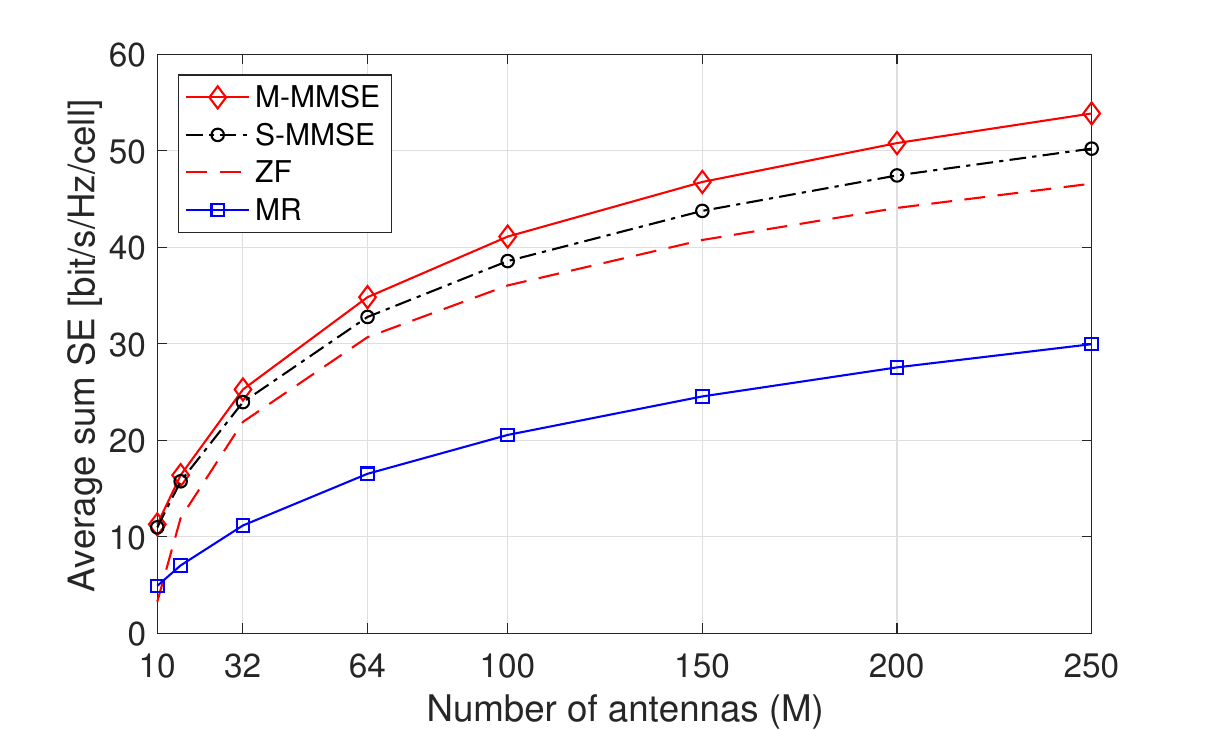}
\end{overpic}} 
\end{center} 
\caption{Average UL and DL sum SE as a function of $M$ with different combining and precoding schemes with the correlated Rayleigh fading model in \eqref{correllatedChannelModel}, by using the network setup in Fig.~\ref{figure_network_layout} with pilot reuse factor $f=1$.}\label{se_vs_antennas_CorrelatedChannels}
\end{figure}
\section{Performance Benefits of Optimized Signal Processing in Massive MIMO 2.0}
To quantify the  SE that can be achieved in Massive MIMO with optimized signal processing and spatial channel correlation, we now compare the different UL and DL schemes by using the network setup in Fig.~\ref{figure_network_layout} and the system parameters in Table~\ref{table:system_parameters_running_example}. Each BS is equipped with a uniform linear array with half-wavelength antenna spacing.  Each channel consists of $S= 6$ scattering clusters, which are modeled by the Gaussian local scattering model \cite[Sec. 2.6]{massivemimobook}. Hence, the $(m_1,m_2)$th element of $\vect{R}_{li}^j$ is given by
\begin{align}\notag
&\left[ \vect{R}_{li}^j \right]_{m_1,m_2} =\beta_{li}^j 10^{\frac{f_{m_1} +f_{m_2}}{10}}\times\\&\frac{1}{S} \sum_{s=1}^S e^{\mathsf{j}\pi (m_1-m_2) \sin({\varphi}_{li,s}^j) }  e^{-\frac{\sigma_{\varphi}^2}{2}\left(\pi (m_1-m_2) \cos({\varphi_{li,s}^j}) \right)^2}\label{correllatedChannelModel}
\end{align}
where $\beta_{li}^j$ is the large-scale fading coefficient (reported in Table~\ref{table:system_parameters_running_example}) and $f_m \sim \mathcal{N} (0, \sigma_f^2)$ represents i.i.d.~log-normal channel gain variations with $\sigma_f=2$, which model the gain variations observed from measurements in Fig.~\ref{figure:measurements_fhh}b. Let $\varphi_{li}^j$ be the geographical angle to UE $i$ in cell $l$ as seen from BS~$j$. Cluster $s$ is characterized by the randomly generated nominal
angle-of-arrival $\varphi_{li,s}^j\sim \mathcal{U}[\varphi_{li}^j-40^\circ, \varphi_{li}^j+40^\circ]$ and the angles of the multipath components are Gaussian distributed around the nominal angle with standard deviation $\sigma_{\varphi}^2= 5^\circ$.


\begin{table}[t!]
         \caption{Average UL and DL sum SE [bit/s/Hz/cell] with correlated Rayleigh fading for $M=100$ and different pilot reuse factors $f$. } 
\begin{center} 
    \begin{tabular}{|c||c|c|c||c|c|c|c|}
    \hline
    &\multicolumn{3}{|c||}{Uplink}& \multicolumn{3}{|c|}{Downlink}\\ \hline
    Scheme & $f=1$ & $f=2$ & $f=4$ &$f=1$ & $f=2$ & $f=4$\\ \hline \hline
    M-MMSE  & 48.71  & 54.91 & 56.09 &39.89&42.94&43.28\\ \hline
    S-MMSE  & 43.69  & 43.81 & 41.36&37.24&37.17&34.98\\ \hline
    ZF  & 39.26  & 39.82 & 36.91 &34.65&34.55&32.63\\ \hline
    MR & 18.94   & 18.42  &  16.55 &20.11&20.08&18.27\\
    \hline
    \end{tabular}\label{table:combiningschemeswithCorrelated:SE} 
\end{center}  
\end{table}

\begin{figure}[t!]
\begin{center}
\includegraphics[width=1\columnwidth]{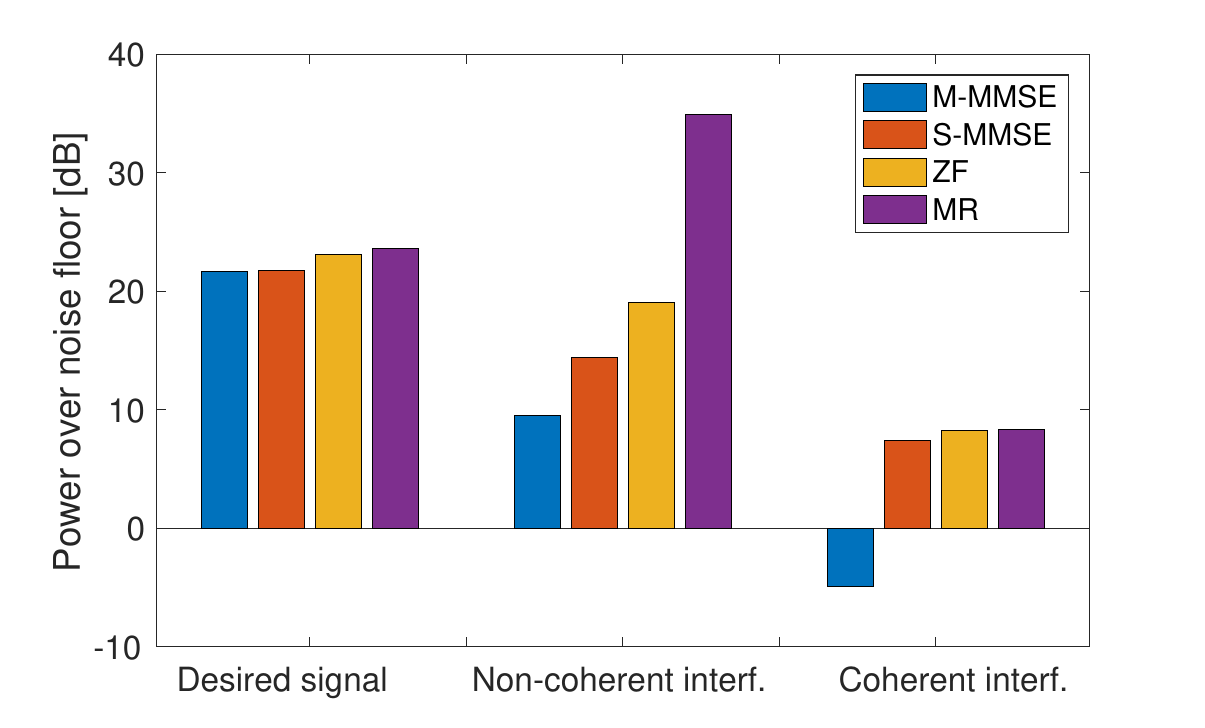}
\end{center} 
\caption{Average UL power of the desired signal, non-coherent interference, and coherent interference for the weakest UE in the cell with $M=200$ and $f=2$. Correlated Rayleigh fading with M-MMSE, S-MMSE, ZF and MR is considered.}\label{figure:ul_average_power_weakest_correlated} 
\end{figure}

Fig.~\ref{figure:ul_se_vs_antennas_CorrelatedChannels} shows the average UL sum SE for the pilot reuse factor $f=1$. M-MMSE provides the highest SE, which  passes from $14.89$ bit/s/Hz to $62.47$ bit/s/Hz as $M$ increases. The suboptimal schemes are quite competitive when $M$ is small, but in the Massive MIMO regime of $M\ge 64$, the losses are noticeable. S-MMSE loses more than $9\%$ in SE and ZF loses more than $18\%$. 
Interestingly, the SE of M-MMSE is $2.5$ times higher than with MR. 
The superior SE of M-MMSE for any value of $M$ is due to the fact that it finds the optimal tradeoff between interference suppression and coherent combining of the desired signal, as shown in Fig.~\ref{figure:MMSEprewhitening}. 

Fig.~\ref{figure:dl_se_vs_antennas_CorrelatedChannels} shows the average DL sum SE in the same setup with $f=1$. The precoding schemes behave in a similar way as their UL counterparts. M-MMSE provides the highest SE, followed by S-MMSE and ZF. MR provides only 45-60\% of the SE provided by M-MMSE. Since the trends are the same, we will for brevity focus on the UL.

From this example, it might seem that M-MMSE can only provide a 10\% SE gain, but there is much more than that to be harvested. Table~\ref{table:combiningschemeswithCorrelated:SE} reports the UL SEs for $M=100$ and different pilot reuse factors. M-MMSE benefits particularly much from having $f>1$. Thanks to the improved channel estimation quality, it can now better suppress the interference from UEs in the surrounding cells and the SE gain is 20\%. 
Since the other schemes do not suppress interference from other cells, their SE reduces when $f$ is increased. Similar trends are observed in Table~\ref{table:combiningschemeswithCorrelated:SE} for the DL. 
To showcase how M-MMSE operates, Fig.~\ref{figure:ul_average_power_weakest_correlated} shows the average power (normalized by the noise power) after receive combining for the weakest UE in the cell when $M=200$ and $f=2$. The power is divided into desired signal, non-coherent interference, and coherent interference. The figure shows that M-MMSE substantially reduces the coherent interference as compared to the other scheme that can only reduce the non-coherent interference.


\begin{figure}[t!]
\begin{center}
\begin{overpic}[unit=1mm,width=1\columnwidth]{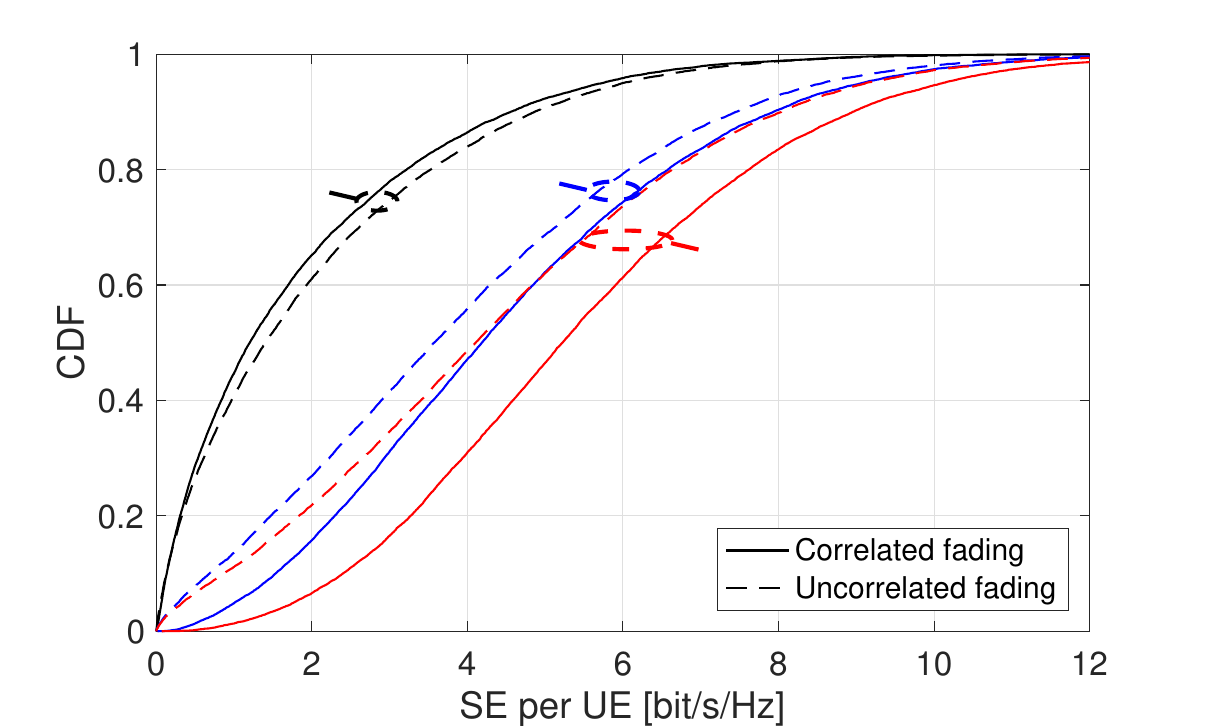}
\put(22,44){\scriptsize MR}
\put(35,44){{\color{blue}\scriptsize S-MMSE}}
\put(59,38){{\color{red}\scriptsize M-MMSE}}
\end{overpic}
\end{center}
\caption{CDF of the UL SE per UE with different combining schemes when $M=100$ and $f=2$. Uncorrelated and correlated Rayleigh fading are considered for the multicell setup in Fig.~\ref{figure_network_layout}.} \label{figureCDF-UL} 
\end{figure}

The simulation results have thus far focused on the average sum SE. To study how the spatial correlation and choice of combining scheme affect UEs in different parts of the cells, Fig.~\ref{figureCDF-UL} shows the cumulative distribution function (CDF) of the UL SE for an arbitrary UE in the network when $M=100$ and $f=2$. The randomness is due to random UE locations and log-normal fading realizations. Uncorrelated and correlated Rayleigh fading channels are compared. Spatial correlation is detrimental for MR, but for all other combining schemes the CDF curves are shifted to the right by the correlation. The highest benefit from having spatial correlation is achieved with M-MMSE. This does not mean that spatial correlation is always beneficial. A UE at a given location might achieve higher SE with uncorrelated fading, which is an effect that cannot be seen from CDFs. 
However, as a UE moves around in the network the probability of achieving a particular SE is consistently higher under correlated fading.

\subsection{Pilot contamination is not a fundamental asymptotic limitation}
We now study the asymptotic SE with MR and M-MMSE combining/precoding under spatially correlated channels in the ``Marzetta limit'', when $M\to \infty$. With MR combining, the following result can be proved from Corollary \ref{cor:uplink-capacity-forgetbound-MR}.

\begin{corollary} 
\label{cor:uplink-capacity-forgetbound-MR-asymptotics} 
With correlated Rayleigh fading, the SINR $\underline{\gamma}_{jk}^{\mathrm{UL}}$ of UE~$k$ in cell $j$ behaves 
 \begin{align}\label{eq:uplink-rate-expression-forgetbound-MR-correlated}
\underline{\gamma}_{jk}^{\mathrm{UL} \infty} =  \frac{  \tr \! \left( \vect{R}_{jk}^{j} {(\Psiv_{jk}^{j})}^{-1} \vect{R}_{jk}^{j} \right)  }{ \underbrace{\sum\limits_{l \in \Pu_{j}  \setminus \{j\} }  \frac{ \left| \tr  \left( \vect{R}_{lk}^{j} {(\Psiv_{jk}^{j})}^{-1}  \vect{R}_{jk}^{j} \right)  \right|^2  }{\tr  \left( \vect{R}_{jk}^{j} {(\Psiv_{jk}^{j})}^{-1} \vect{R}_{jk}^{j} \right) }}_{\textrm{Coherent interference}}  } .
\end{align}
when $M \to \infty $, in the sense that $\underline{\gamma}_{jk}^{\mathrm{UL} } -\underline{\gamma}_{jk}^{\mathrm{UL} \infty} \to 0$.
\end{corollary}

The asymptotic value of  $\underline{\gamma}_{jk}^{\mathrm{UL} \infty}$ lacks a simple expression, but it demonstrates that only coherent interference limits the SE. Unlike the case of uncorrelated fading, this interference is determined by how different $\vect{R}_{jk}^{j}$ is from the pilot-contaminating UEs' correlation matrices $\vect{R}_{lk}^{j}$ with $l\in \mathcal{P}_l\setminus \{j \}$.  If $\vect{R}_{jk}^{j}$ is asymptotically spatially orthogonal to all of them, i.e., $ \frac{1}{M} \tr \big( \vect{R}_{jk}^{j} \vect{R}_{lk}^{j} \big) \to 0$, then the coherent interference vanishes and thus the SINR grows unboundedly as $M \to \infty $. As discussed earlier, this is unlikely to happen in practice, therefore, the SE achieved by MR combining typically converges to a finite limit as $M\to \infty$, even with spatially correlated channels. 
A similar result can be obtained in the DL \cite[Sec.~4.4]{massivemimobook}, with the only difference that the set of correlation matrices that are involved in the coherent interference terms are different; UL interference comes through the UEs' channels, while DL interference comes through the BSs' channels. Unlike the case of uncorrelated fading, a UE can be subject to strong coherent interference in the UL but not in the DL, or vice versa. Consequently, the system should consider both the UL and DL SEs when assigning pilots to UEs.

Based on the asymptotic results with MR combining/precoding and the fact that similar conclusions hold for S-MMSE \cite{Hoydis2013}, one may suspect that the SE with spatially correlated channels has always a finite limit due to pilot contamination (except in extreme cases with asymptotic spatial orthogonality). This suspicion was recently shown to be wrong in \cite{BjornsonHS17}.

\begin{theorem}
\label{theorem:M-MMSE-UL-asymptotics}
With correlated Rayleigh fading, the UL (DL) SE of a given UE with M-MMSE combining (precoding) grows without bound as $\log_2(M)$ when $M\rightarrow \infty$, if the correlation matrix of the UE is asymptotically linearly independent of the set of correlation matrices of its pilot-contaminating UEs.
\end{theorem}

The above theorem proves that Massive MIMO is not asymptotically limited by pilot contamination when taking optimized signal processing and spatially correlated channels into consideration. Consequently, neither MR and S-MMSE, nor any other scheme that only suppresses intra-cell interference can be asymptotically optimal; even if this has been claimed in the literature since the beginning of Massive MIMO. 
In other words, the deceivingly simple conclusions that can be drawn from analyzing uncorrelated fading channels---particularly regarding coherent interference and pilot contamination---do not extend to practical channels, which are spatially correlated.

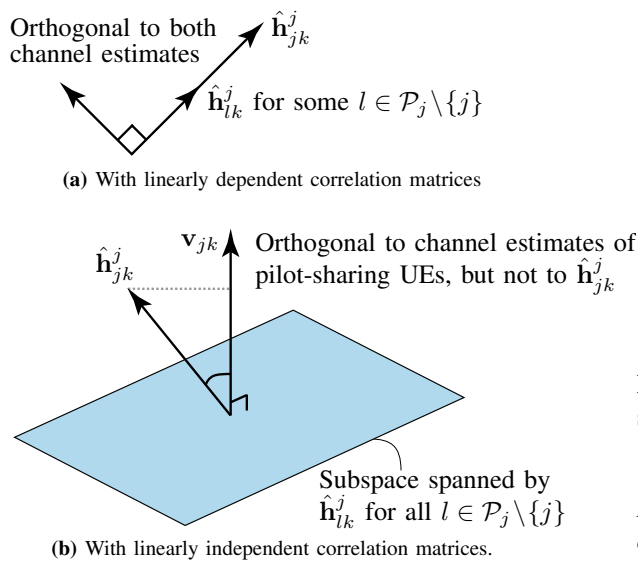
\begin{figure}[t!]
\begin{center}
\subfloat[With linearly dependent correlation matrices]{\label{figure_basic_pilotcontrejection2}
\begin{overpic}[unit=1mm,width=.8\columnwidth]{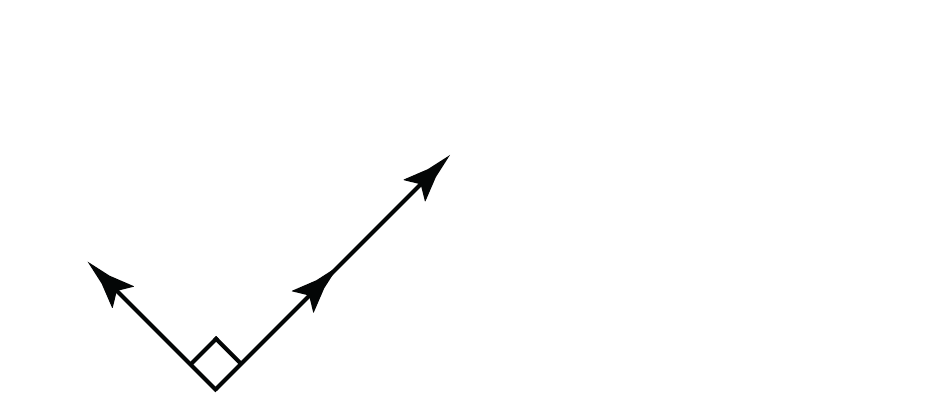}
\put(0,23){Orthogonal to both}
\put(0,18){channel estimates}
\put(49,23){$\hat{\vect{h}}_{jk}^j$}
\put(37,9){$\hat{\vect{h}}_{lk}^j$ for some $l \in \mathcal{P}_{j}\!\setminus\!\{j\}$}
\vspace{-9mm}
\end{overpic}} \\ 
\subfloat[With linearly independent correlation matrices.]{\label{figure_basic_pilotcontrejection}
\begin{overpic}[unit=1mm,width=.8\columnwidth]{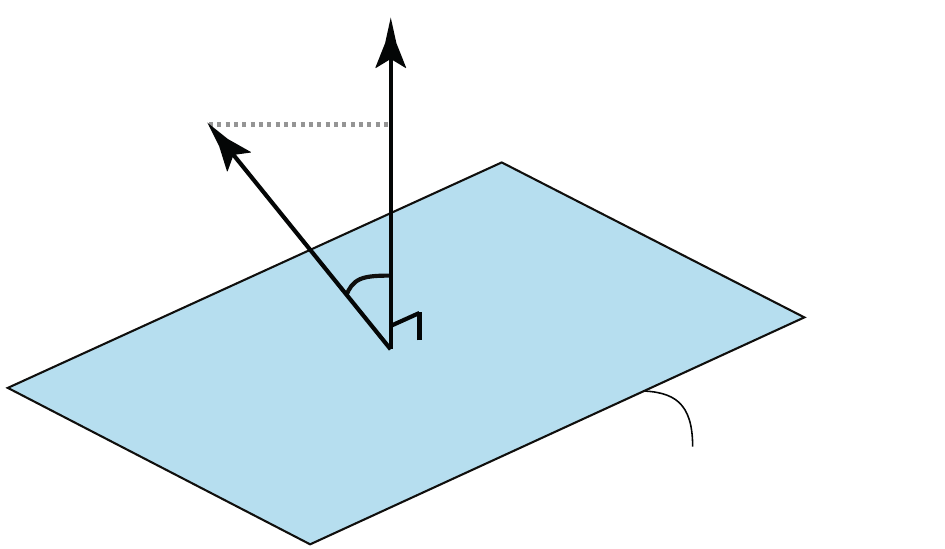}
\put(16,48){$\hat{\vect{h}}_{jk}^j$}
\put(32,53){$\vect{v}_{jk}$}
\put(46,52){Orthogonal to channel estimates of}
\put(46,46){pilot-sharing UEs, but not to $\hat{\vect{h}}_{jk}^j$}
\put(58,7.5){Subspace spanned by}
\put(58,1.5){$\hat{\vect{h}}_{lk}^j$ for all $l \in \mathcal{P}_{j}\!\setminus\!\{j\}$}
\end{overpic}} 
\end{center} 
\caption{Geometric illustration of how channel estimates are related for UEs that reuse the same pilot. The channel estimates are parallel when the spatial correlation matrices are linearly dependent, while they are linearly independent when the spatial correlation matrices are linearly independent. In the latter case, the illustrated combining vector $\vect{v}_{jk}$ rejects the coherent interference from the pilot-sharing UEs, while a non-zero part of the desired signal remains.}\label{basic_pilotcontrejection}
\end{figure}

These results can be surprising at a first sight but a closer look reveals that they are rather intuitive. Consider BS~$j$ and let us compare the estimate $\hat{\vect{h}}_{jk}^{j}$ of an intra-cell UE with the estimate $\hat{\vect{h}}_{lk}^{j}$ of a pilot-sharing UE in another cell, i.e., $l \in \mathcal P_j$. In this case, $ \Psiv_{lk}^{j} = \Psiv_{jk}^{j}$ and it follows from \eqref{eq:MMSEestimator_h_jli} that
\begin{align} 
\hat{\vect{h}}_{jk}^{j}    =  \vect{R}_{jk}^{j} \vect{z}_{jk}^{p} \quad\textnormal{and}\quad
\hat{\vect{h}}_{lk}^{j}    =  \vect{R}_{lk}^{j} \vect{z}_{jk}^{p}
\end{align}
where $\vect{z}_{jk}^{p} = \frac{1}{\tau_p\sqrt{\rho_{\rm{ul}}}}{(\Psiv_{jk}^{j})}^{-1}  \left(\vect{Y}_j^{p}\bphiu_{jk}\right)$ is the same for both UEs. These channel estimates are strongly correlated, but the difference 
\begin{equation} 
\hat{\vect{h}}_{jk}^{j}  - c \hat{\vect{h}}_{lk}^{j}    =  \bigg(\vect{R}_{jk}^{j} - c \vect{R}_{lk}^{j} \bigg)\vect{z}_{jk}^{p}
\end{equation}
is only zero for some $c \in \mathbb{R}$ if $\vect{R}_{jk}^{j}$ and $\vect{R}_{li}^{j}$ are linearly dependent, which is very unlikely as demonstrated in Example~\ref{example:linear}. This principle is illustrated geometrically in Fig.~\ref{basic_pilotcontrejection}. The key insight is that, for linearly independent channel estimates (or, equivalently, correlation matrices), it is possible to find a combining vector $\vect{v}_{jk}$ that is orthogonal to $\hat{\vect{h}}_{lk}^{j}$ (i.e., $\vect{v}_{jk}^{\Htran} \hat{\vect{h}}_{lk}^{j}=0$) and has a non-zero inner product $\vect{v}_{jk}^{\Htran} \hat{\vect{h}}_{jk}^{j}$ with the channel estimate of the desired UE. 
This heuristic combiner is shown in Fig.~\ref{basic_pilotcontrejection}(b) and will make the SINR grow unboundedly with $M$ if the correlation matrices are also asymptotically linearly independent.
Since the optimal M-MMSE combining (and M-MMSE precoding) provides a higher SINR than any other scheme, including the heuristic scheme just described, it also rejects the coherent interference while retaining an array gain that is proportional to  $M$. This is why Theorem~\ref{theorem:M-MMSE-UL-asymptotics} requires asymptotically linearly independent correlation matrices. In the special case of spatially uncorrelated channels, $\hat{\vect{h}}_{jk}^{j}$ and $\hat{\vect{h}}_{lk}^{j}$ become linearly dependent, i.e., $\hat{\vect{h}}_{lk}^{j}    =  \beta_{lk}^{j}/\beta_{jk}^{j}\hat{\vect{h}}_{jk}^{j}$ as also shown in Fig.~\ref{basic_pilotcontrejection}(a). This makes it impossible for BS~$j$ to find a vector $\vect{v}_{jk}$ that is orthogonal to $\hat{\vect{h}}_{lk}^{j} $ while being non-orthogonal to $\hat{\vect{h}}_{jk}^{j}$. Hence, the reason for the asymptotic limit with uncorrelated  fading is that we cannot suppress interference from pilot-sharing UEs without also removing an equal fraction of the desired signal.

\subsection{Numerical validation of asymptotic analysis with spatially correlated channels}

\begin{figure}[t!]
\begin{center}
\includegraphics[width=1\columnwidth]{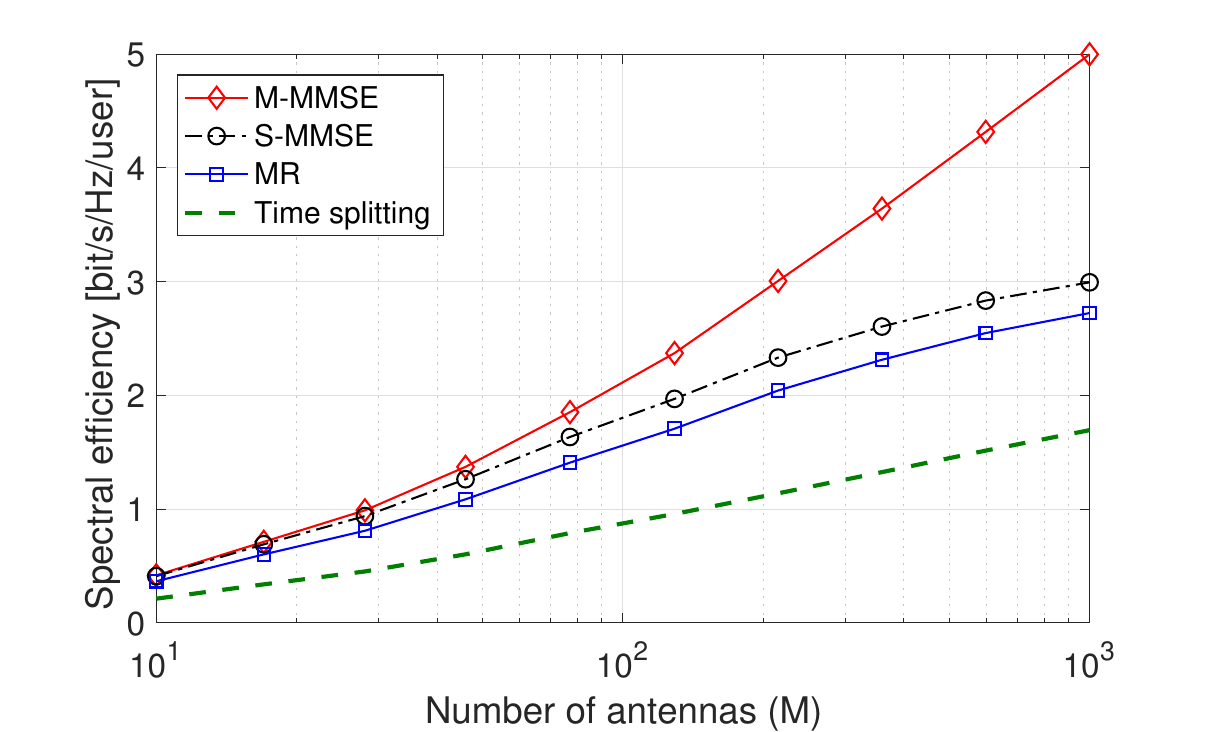}
\end{center}
\caption{UL SE per UE as a function of $M$ for the multicell setup in \cite[Fig.~3]{BjornsonHS17} with $L=4$ and $K=2$.} \label{figureAsymptotics}
\end{figure}

To illustrate the asymptotic behavior of M-MMSE, we consider the challenging symmetric setup in \cite[Fig.~3]{BjornsonHS17} with $L=4$ cells and $K=2$ UEs per cell. The BSs are located at the four corners of the coverage area and the UEs are located pairwise close to each other.
The star-marked UEs share a pilot, while the plus-marked UEs share another one. 
The asymptotic behavior of the UL SE per UE is shown in Fig.~\ref{figureAsymptotics} with M-MMSE, S-MMSE, and MR.  Both S-MMSE and MR converge to asymptotic limits of around 3\,bit/s/Hz per UE as $M$ grows. In line with Theorem~\ref{theorem:M-MMSE-UL-asymptotics}, M-MMSE provides an SE that grows without bound as $\log_2(M)$ (the horizontal scale is logarithmic). Hence, the 10-20\% SE gains, observed in previous simulations, only represent the gap for a given number of antennas, while the asymptotic gap grows without bound.

Since any practical system will have a finite number of cells, we could alternatively alleviate the pilot contamination problem by dividing the coherence blocks orthogonally between the cells. The curve ``Time splitting'' considers this case in which the $L=4$ cells are active in different coherence blocks, and M-MMSE is used for receive combining (which  coincides with S-MMSE in this case).
The SE with time splitting also grows without bound, but with a smaller slope than with M-MMSE, due to the additional pre-log factor of $1/L = 1/4$.
Hence, even in a small system with $L=4$, it is inefficient to avoid pilot contamination by time splitting; even MR provides higher SE than time splitting in the considered range of antennas.

The lack of an asymptotic limit does not imply that pilot contamination disappears entirely by using M-MMSE. There is still a substantial SE loss caused by the estimation errors and a fraction of the signal power is sacrificed in interference suppression, but there is no convergence to a fundamental capacity limit.

\subsection{Practical relevance of asymptotics in Massive MIMO 2.0}\label{ref:Asymptotic_discussion}

Asymptotic analyses are frequent in statistical signal processing (e.g., infinite SNR or number of samples) and in information theory (e.g., infinitely long code blocks). While some people question their usefulness, others argue that they are useful to understand the ultimate performance or to obtain tight low-complexity performance approximations. One can easily identify several compelling reasons to disregard the ``Marzetta limit''; that is, the asymptotic limit where $M\to\infty$. Firstly, deploying wireless networks with a nearly infinite $M$ is impossible in a finite-sized world. Secondly, the conventional channel models break down when $M\to\infty$, since one would eventually receive more power than transmitted. Thirdly, the technology will neither be cost nor energy efficient, since the cost and energy grow linearly with $M$, while the SE only grows logarithmically with $M$. Fourthly, the sum capacity is often maximized when $M/K$ is fairly small \cite{Bjornson2016a,massivemimobook}, which implies that the traditional large-system limit where $M,K\to \infty$ with a fixed ratio is of more practical interest \cite{SanguinettiSPAWC2019}. In light of these shortcomings: \emph{What is the purpose of the ``Marzetta limit'' in Massive MIMO?}

The answer is that the ``Marzetta limit'' tells us how to design Massive MIMO 2.0; a future-proof technology capable of delivering performance far beyond what Marzetta predicted in his seminal paper. The performance difference between the various combining/precoding schemes is fairly small when considering only 64 antennas, but the benefit of M-MMSE will grow larger over time, as antennas become a commodity and are deployed everywhere. 

New deployment concepts with very large surfaces of antennas have recently emerged \cite{Hu2018a,Amiri2018a,Bjornson2019a}, for which $M$ can be extremely large, while $K$ remains to be limited by the constraint $K\leq \tau_c$. Distributed deployments and higher carrier frequencies are two other ways to accommodate more antennas per km$^2$. These concepts are further discussed in Section~\ref{sec:future-directions}.
With this in mind, consider a wireless network with any finite number, $K$, of UEs that each have finite-valued data rate requirements in UL and DL. Theorem \ref{theorem:M-MMSE-UL-asymptotics} proves that we can always satisfy these requirements by deploying sufficiently many antennas, even in the presence of pilot contamination. In fact, the theorem ensures that there exists a finite number of antennas, $M$, that allows to deliver the required SE with only $\tau_c=2$ samples per coherence block: one for pilots and one for data. This is made possible by exploiting the spatial correlation that appears naturally in wireless channels.



\section{Acquiring Spatial Correlation Knowledge}\label{SectionVI}

The key message from the above results is that the existence and exploitation of spatial channel correlation is a game changer for Massive MIMO, particularly in future systems with hundreds or thousands of antennas. In the analysis, we have assumed that all deterministic quantities are perfectly known, which is the common practice in communication theory.
Specifically, if BS~$j$ wants to implement the M-MMSE scheme in \eqref{eq:MMSE-combining}, it needs to know the MMSE estimates $\{\hat{\vect{h}}_{li}^{j}:\forall l,i\}$ and also $\vect{Z}_j = \sum\nolimits_{l=1}^{L} \sum\nolimits_{i=1}^{K} (\vect{R}_{li}^{j} - \vect{\Phi}_{li}^j) +  \frac{\sigma_{\rm{ul}}^2}{\rho_{\rm{ul}}}  \vect{I}_{M}$. The computation of $\hat{\vect{h}}_{li}^{j}$ in \eqref{eq:MMSEestimator_h_jli} requires knowledge of the following deterministic quantities:
\begin{itemize}
\item The correlation matrix $\vect{R}_{li}^{j} = \mathbb{E}\{\vect{h}_{li}^{j}({\vect{h}_{li}^{j}})^{\Htran}\}$ of $\vect{h}_{li}^{j}$;
\item The sum  $\Psiv_{li}^{j}= \tau_p\mathbb{E}\{(\vect{Y}_j^{p}\bphiu_{li})(\vect{Y}_j^{p}\bphiu_{li})^{\Htran}\} = \sum_{m \in \Pu_{l} }\vect{R}_{mi}^{j}  +  \frac{1}{\tau_p}\frac{\sigma_{\rm{ul}}^2}{\rho_{\rm{ul}}}  \vect{I}_{M}$ of correlation matrices of the pilot-sharing UEs. 
\end{itemize}
This knowledge is also sufficient to compute $\vect{R}_{li}^{j} - \vect{\Phi}_{li}^j$ in $\vect{Z}_j$ since $\vect{\Phi}_{li}^j =\vect{R}_{li}^{j}
{(\Psiv_{li}^{j})}^{-1}  \vect{R}_{li}^{j}$. 
Any deterministic quantity of fixed dimension can be estimated to any given accuracy using a finite amount of pilot resources, which thus has a negligible impact on the system as the available communication resources (e.g., time) go to infinity.
There are nevertheless reasons to question the practicality of knowing the channel statistics perfectly, particularly since  all the correlation matrices have dimensions that grow with $M$. The channel statistics can change when a UE moves, the UE will not be active forever, and the set of interfering UEs will also changes due to data traffic variations, thus the resources that can be spent on learning the channel statistics are often limited in practice.
Therefore, in this section, we first show that the key results are also valid if only partial knowledge of the correlation matrices is available. Then, we discuss methods to acquire full or partial knowledge of these matrices when $M$ is large.

\subsection{Spectral efficiency with partial knowledge of channel statistics}

Consider BS~$j$ and assume that it wants to estimate $\vect{h}_{li}^{j}$.
To dispense with full knowledge of the correlation matrices, the BS can simply ignore the 
correlation between the elements in $\vect{h}_{li}^{j}$ by estimating each element  separately and using only the signals received on the corresponding antenna. This yields what we call the element-wise MMSE (EW-MMSE) estimator that uses only the main diagonals of $\{\vect{R}_{li}^{j}: l = 1,\ldots,L\}$ \cite{massivemimobook}. 
\begin{lemma}\label{lemma:EW-MMSE}
The EW-MMSE estimate of $\vect{h}_{li}^{j}$ based on the observation $\vect{Y}_j^{p}\bphiu_{li}$ is
\begin{equation} \label{eq:EWMMSEestimator_h_jli}
\hat{\vect{h}}_{li}^{j,{\rm{EW}}}  = \vect{R}_{li}^{j,{\rm{diag}}} \big(\Psiv_{li}^{j,{\rm{diag}}}\big)^{-1}  \left(\frac{1}{\tau_p\sqrt{\rho_{\rm{ul}}}}\vect{Y}_j^{p}\bphiu_{li}\right)
\end{equation}
where $\vect{R}_{li}^{j,{\rm{diag}}}$ and $\Psiv_{li}^{j,{\rm{diag}}}$ are diagonal matrices with elements $\big\{[ \vect{R}_{li}^j ]_{nn}\big\}$ and $\big\{ \sum_{l'\in \mathcal{P}_l} [\vect{R}_{l'i}^j ]_{nn} +  \frac{\sigma_{\rm{ul}}^2}{\tau_p\rho_{\rm{ul}}}\big\}$, respectively.
\end{lemma}
EW-MMSE and MMSE estimation are equivalent when the channel vectors have independent elements, but they are generally different. 
Each BS only needs to know the diagonal of the spatial correlation matrices to implement EW-MMSE estimation, which are easier to acquire in practice. 

Another important difference from the MMSE estimator is that the estimate $\hat{\vect{h}}_{li}^{j,{\rm{EW}}} $ and its estimation error ${\vect{h}}_{li}^{j}  - \hat{\vect{h}}_{li}^{j,{\rm{EW}}} $ are statistically correlated. Therefore, to design the combining vectors $\{{\bf v}_{jk}: k=1,\ldots,K\}$ and quantify the UL SE with the EW-MMSE estimator, we cannot use the SE expression provided in Theorem \ref{theorem:uplink-SE} but need to use the UatF bound in Theorem \ref{theorem:uplink-capacity-forgetbound}, which can be applied with any channel estimator.  Inspired by the optimality of M-MMSE combining, a heuristic choice for $\vect{v}_{jk}$ in \eqref{eq:uplink-SINR-expression-forgetbound} with the EW-MMSE estimator is
\begin{equation} \label{eq:EW_MMSE-combining}
\vect{V}_{j}^{\rm{M-MMSE-EW}} \triangleq  \Bigg(  \sum\limits_{l=1}^L \widehat{\vect{H}}_{l}^{j,\rm{EW}}{(\widehat{\vect{H}}_{l}^{j,\rm{EW}})}^{\Htran} + \vect{S}_j  \Bigg)^{\!-1} \!\!\!\!   \widehat{\vect{H}}_{jk}^{j,\rm{EW}}
\end{equation} 
where 
\begin{equation}
\vect{S}_j =  \sum_{l=1}^{L}\sum_{i=1}^{K} \Big(\vect{R}_{li}^{j,{\rm{diag}}} - \vect{R}_{li}^{j,{\rm{diag}}}\big(\Psiv_{li}^{j,{\rm{diag}}}\big)^{-1}\vect{R}_{li}^{j,{\rm{diag}}}\Big) +  \frac{\sigma_{\rm{ul}}^2}{\rho_{\rm{ul}}}  \vect{I}_{M}.
\end{equation} 
Specifically, we have taken M-MMSE combining from \eqref{eq:MMSE-combining} and replaced all the correlation matrices with their diagonal counterparts. Clearly, \eqref{eq:EW_MMSE-combining} coincides with M-MMSE when all the correlation matrices $\{\vect{R}_{li}^{j}\}$ are truly diagonal. 
The computational complexity of M-MMSE-EW is reported in Table~\ref{tab:cost_linear_processing}. The EW-MMSE estimator has a much lower computational complexity than the MMSE estimator with M-MMSE since the inverse in \eqref{eq:EWMMSEestimator_h_jli} only involves a diagonal matrix.

 \begin{figure}[t!]
\begin{center}
\includegraphics[width=1\columnwidth]{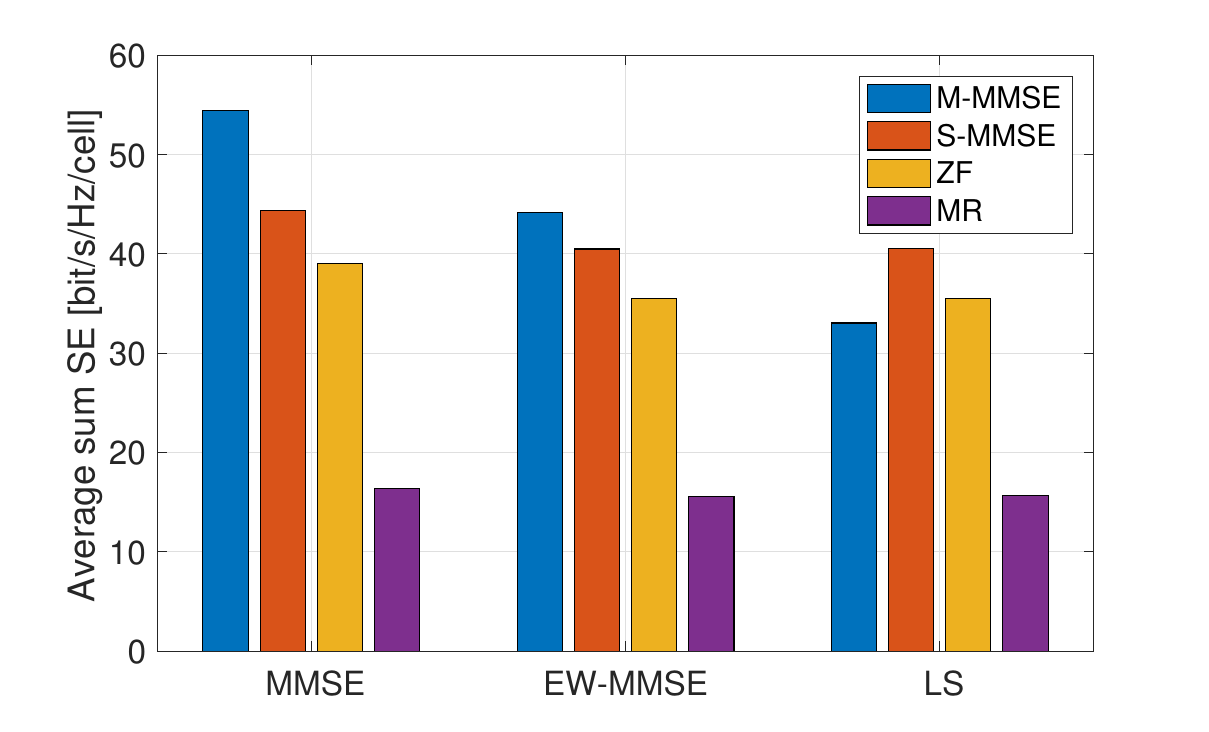}
\end{center}
\caption{Average UL sum SE when using MMSE, EW-MMSE, or LS channel estimators, for the network setup in Fig.~\ref{figure_network_layout} with $M = 100$ and $f=2$. Four different combining schemes are considered when using different channel estimator.} \label{fig:SE_with_diff_channel_estimators} 
\end{figure}

Fig.~\ref{fig:SE_with_diff_channel_estimators} shows a bar diagram of the average UL sum SE with M-MMSE, S-MMSE, ZF, and MR combining when either the MMSE or EW-MMSE estimators are used. 
We also consider the LS estimator, which requires no prior statistical information and computes the estimate of ${\vect{h}}_{li}^{j}$ as $\hat{\vect{h}}_{li}^{j,{\rm LS}}  = \frac{1}{\tau_p\sqrt{\rho_{\rm{ul}}}}\vect{Y}_j^{p}\bphiu_{li}$. For any choice of combining scheme, the MMSE estimator provides the highest SE. The SE loss incurred by using the EW-MMSE estimator is only 8\% with MR but increases to 23\% with M-MMSE. 
Generally speaking, M-MMSE is the scheme that is most sensitive to the choice of estimation scheme; it performs poorly with the LS estimator since the norms of the inter-cell channel estimates are greatly overestimated with that estimator. In contrast, S-MMSE, ZF, and MR performs almost equally well with the EW-MMSE and LS estimators.


The impression from Fig.~\ref{fig:SE_with_diff_channel_estimators} is that M-MMSE gives a noticeable gain over S-MMSE only when the MMSE estimator is used, while M-MMSE combining does not  perform particularly well with EW-MMSE. However, the figure only considers $M=100$ and not the asymptotic behavior as $M\to \infty$.
The following result is proved in \cite{BjornsonHS17}.

\begin{theorem} \label{theorem:approximate_M-MMSE}
With M-MMSE-EW, the UL SE of UE $k$ in cell $j$ grows without bound as $\log_2(M)$ when $M\to \infty$, if the diagonal correlation matrices $\vect{R}_{lk}^{j,\rm{diag}}$ for $l \in \mathcal {P}_j$ are asymptotically linearly independent.
\end{theorem}

This theorem proves that M-MMSE-EW has the same scaling behavior as M-MMSE, if the diagonals of the correlation matrices are known and linearly independent between pilot-sharing UEs. Recall from Fig.~\ref{figure:measurements_fhh}b that diagonal elements of measured correlation matrices are non-uniform and UE-specific, thus the linear independence is likely satisfied in practice. A downlink counterpart to Theorem~\ref{theorem:approximate_M-MMSE} can also be obtained \cite{BjornsonHS17}.

\begin{figure}[t!]
\begin{center}
\begin{overpic}[unit=1mm,width=\columnwidth]{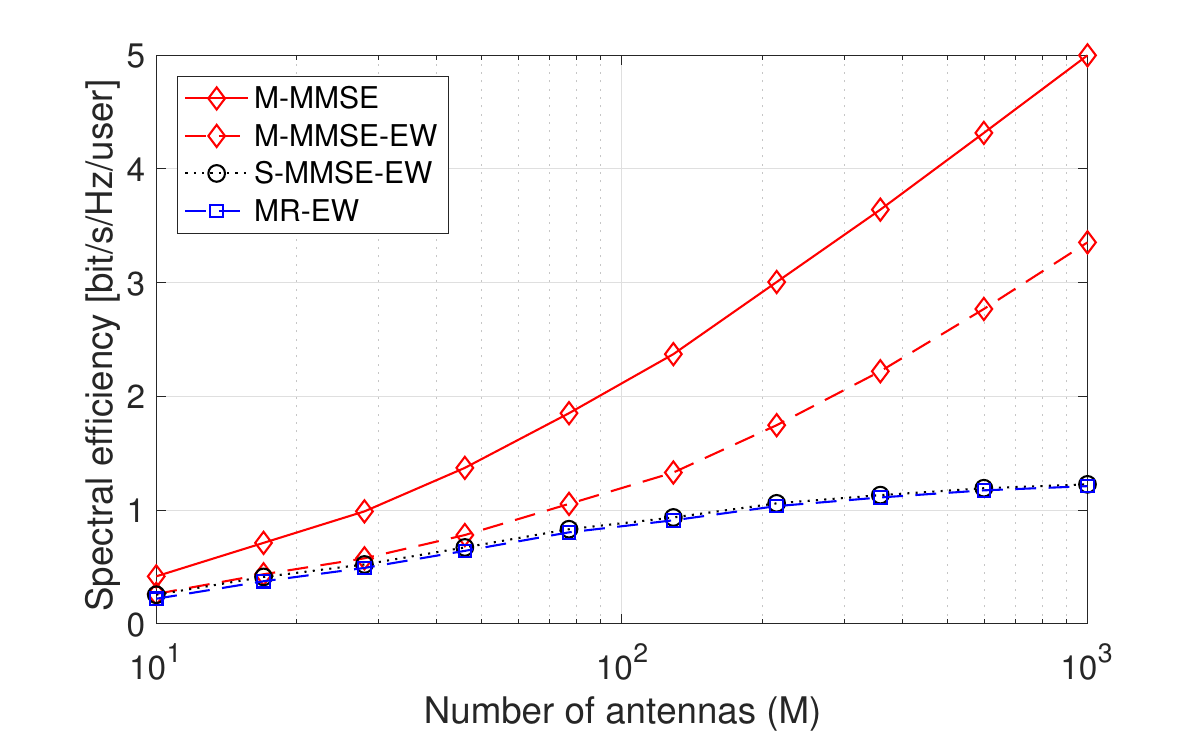}
\end{overpic}
\end{center}
\caption{UL SE per UE as a function of $M$ for the multicell setup in \cite[Fig.~3]{BjornsonHS17} when using the EW-MMSE estimator. Comparisons are made with the SE with M-MMSE as obtained in Fig.~\ref{figureAsymptotics}, by using the MMSE channel estimator.}  \label{figureAsymptotics_EW-MMSE} 
\end{figure}

Although the scaling behavior is the same, we have already observed in Fig.~\ref{fig:SE_with_diff_channel_estimators} that a performance loss occurs with M-MMSE-EW. This is further quantified in Fig.~\ref{figureAsymptotics_EW-MMSE} for the setup in \cite[Fig.~3]{BjornsonHS17}. The UL SE with M-MMSE-EW grows unboundedly, but there is an 32\% loss in SE compared to M-MMSE when $M=10^3$. All the combining schemes achieve lower SEs with the EW-MMSE estimator (compared to the MMSE estimator in Fig.~\ref{figureAsymptotics}) since it neglects the  correlation between the elements of channel vectors. For example, an asymptotic limit of  1.2 bit/s/Hz (instead of $3$ bit/s/Hz) per UE is achieved with S-MMSE-EW and MR-EW as $M$ grows.

\subsection{Sample correlation matrices and regularization}

Before looking into the estimation of the (full or partial) correlation matrices $\{\vect{R}_{li}^{j}\}$ of the individual UE channels, we exemplify how $\Psiv_{li}^{j}$ can be acquired at BS~$j$ by using standard methods, even when $M$ increases. 
Since $\Psiv_{li}^{j}$ depends on the correlation matrices of the pilot-sharing UEs, it is constant over the transmission bandwidth and evolves slowly in time compared to the fast variations of channel vectors. The measurements in \cite{Viering_2002} suggest roughly two orders of magnitude slower variations. Therefore, we may reasonably assume that $\Psiv_{li}^{j}$ does not change over $\tau_s$ coherence blocks, where $\tau_s$ can be at the order of thousands. In these circumstances, the classical approach  is to approximate $\Psiv_{li}^{j}$ with the sample correlation matrix. Suppose BS~$j$ has received the pilot vector $\vect{Y}_j^{p}$ in~\eqref{eq:uplink-pilot-model} in $N\le \tau_s$ coherence blocks. We denote these $N$ observations by $\vect{Y}_j^{p}[1],\ldots,\vect{Y}_j^{p}[N]$. The sample correlation matrix is then given by
\begin{align}\label{eq:Psiv_lij}
\widehat\Psiv_{li}^{j,\mathrm{sample}} = \frac{1}{N} \sum_{n=1}^N \big(\vect{Y}_j^{p}[n]\bphiu_{li}\big)\big(\vect{Y}_j^{p}[n]\bphiu_{li}\big)^{\Htran}.
\end{align}
Note that the uplink pilot signals that are transmitted to estimate the channel realizations are also used for estimating the correlation matrices, thus no extra signaling is required to obtain $\widehat\Psiv_{li}^{j,\mathrm{sample}}$.
For a particular diagonal element of $\widehat\Psiv_{li}^{j,\mathrm{sample}}$, the law-of-large numbers implies that the sample variance converges to the true variance as $N\to \infty$.
The standard deviation of the sample variance decays as $1/\sqrt{N}$ and is independent of $M$, thus relatively small values of $N$ are sufficient to obtain a fairly good variance estimate. 
It is more challenging to obtain a sample
correlation matrix $\widehat\Psiv_{li}^{j,\mathrm{sample}}$ whose eigenvalues and eigenvectors are well
aligned with those of $\Psiv_{li}^{j}$ when the matrix is non-diagonal.
The reason is that the estimation errors in all the $M^2$ elements of $\widehat\Psiv_{li}^{j,\mathrm{sample}}$ affect the eigenstructure.
To partially overcome this issue, a classic approach is to regularize the matrix by computing the convex combination \cite{Ledoit2004a}
\begin{align}\label{eq:Qjjk_estimate}
\widehat\Psiv_{li}^{j}(\eta) = \eta\widehat\Psiv_{li}^{j,\,\mathrm{sample}} + (1-\eta) \widehat\Psiv_{li}^{j,\,\mathrm{diag}}, \quad  \eta \in [0,1]
\end{align}
where $\widehat\Psiv_{li}^{j,\,\mathrm{diag}}$ contains the main diagonal of $\widehat\Psiv_{li}^{j,\,\mathrm{sample}} $ and otherwise is zero. The diagonal elements of $\widehat\Psiv_{li}^{j}(\eta) $ are independent of $\eta$, while the magnitudes of the off-diagonal elements are $\eta$ times smaller than in the sample correlation matrix. 
The regularization makes $\widehat\Psiv_{li}^{j}(\eta)  $ a full-rank matrix for any $\eta \in [0,1)$ and $\eta$ can be tuned (for example by using numerical methods) to purposely underestimate the off-diagonal elements when these are considered unreliable.
There are other ways to compute correlation matrices when $N<M$, for which we refer the interested reader to \cite{Couillet2011a,Haghighatshoar2017a}.

\begin{figure}[t!] 
\begin{center}
\begin{overpic}[unit=1mm,width=1\columnwidth]{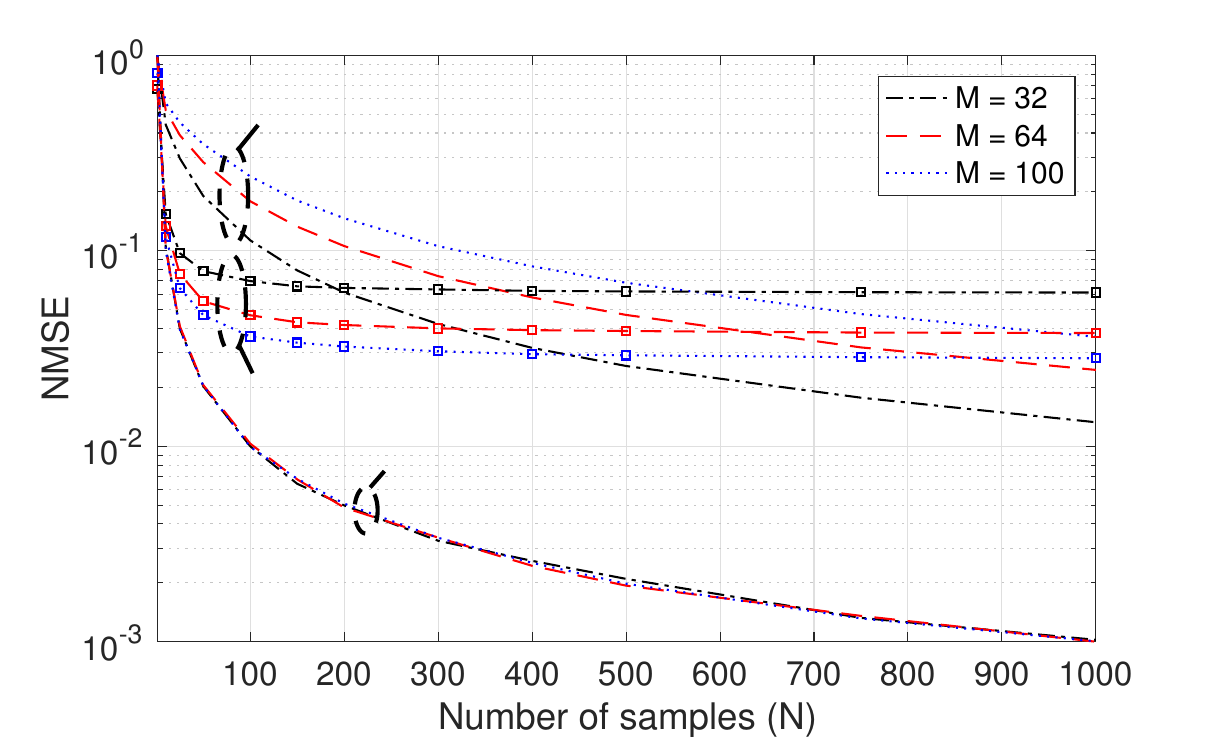}
\put(20,52){\footnotesize{Full matrix}}
\put(20,27){\footnotesize{DFT approx}}
\put(30,23){\footnotesize{Only diagonal elements}}
\end{overpic}
\end{center}
\caption{NMSE in the estimation of the full matrix $\Psiv_{li}^{j}$ by using the regularization approach in \eqref{eq:Qjjk_estimate} when $M=32, 64$ and $128$. The results are obtained for the multicell setup of Fig.~\ref{figure_network_layout} without i.i.d.~log-normal channel gain variations. The number of samples, $N$, refer to the number of independent observations of the pilot vector $\vect{Y}_j^{p}$ in~\eqref{eq:uplink-pilot-model}. The accuracy in the estimation of the diagonal elements only is also reported. As seen, they can be estimated efficiently using a small number of samples, that does not scale with $M$. We also report the accuracy in the estimation of the full matrix $\Psiv_{li}^{j}$ when the DFT approximation is used.} \label{figureNMSE} 
\end{figure}

We quantify the performance of the regularization approach by computing the NMSE defined as $
{\rm{NMSE}}(\eta) = {\mathbb{E}\big\{||\Psiv_{li}^{j} - \widehat\Psiv_{li}^{j}(\eta)  ||^2_F\big\}}/{|| \Psiv_{li}^{j}||^2_F}$
where the expectation is approximated by averaging over independent experiments. We consider the network setup in Fig.~\ref{figure_network_layout} without i.i.d.~log-normal channel gain variations. The NMSE is averaged over $10$ sets of random UE locations, and $\eta$ is numerically optimized for each $N$. Fig.~\ref{figureNMSE} reports the NMSE as a function of $N$ when $M\in \{32,64,100\}$. The NMSE reduces monotonically with $N$ and we need a few hundred samples, in the order of $3M$, to achieve an NMSE lower than $0.1$. We also report the accuracy in the estimation of only the diagonal elements of $\Psiv_{li}^{j}$. The results show that these  can be estimated very accurately using a small number of samples, which does not grow with $M$. Hence, the matrix $\Psiv_{li}^{j,{\rm{diag}}}$ that is needed in the EW-MMSE estimator is easy to estimate even for large values of $M$, while it is more challenging to estimate the full matrix $\Psiv_{li}^{j}$ that is needed by the MMSE estimator. 

Fig.~\ref{figureNMSE} reports also the NMSE achieved by assuming that the eigenvectors of all the correlation matrices in $\Psiv_{li}^{j} $ are well approximated by the DFT matrix ${\bf F}\in \mathbb{C}^{N\times N}$; recall Section~\ref{sec:ULA}. In this case, the estimate of $\Psiv_{li}^{j} $ is obtained as
\begin{align}
\widehat\Psiv_{li,\mathrm{dft}}^{j,\mathrm{sample}} = {\bf F}\widehat\Psiv_{li,\mathrm{dft}}^{j,\mathrm{diag}} {\bf F}^{\Htran}
\end{align}
where $\widehat\Psiv_{li,\mathrm{dft}}^{j,\mathrm{diag}}$ contains the main diagonal of the sample correlation matrix given by ${\bf F}^{\Htran}\widehat\Psiv_{li}^{j,\mathrm{sample}}{\bf F}$, which is obtained after multiplying $\widehat\Psiv_{li}^{j,\mathrm{sample}}$ in \eqref{eq:Psiv_lij} with the DFT and inverse DFT matrices from left and right, respectively. The results in Fig.~\ref{figureNMSE} show that the DFT approximation may improve the estimation accuracy in all investigated cases when $N$ is smaller than $200$ samples. As $N$ increases, the inaccuracy of the DFT approximation leads to an error floor, thus it eventually better to estimate the correlation matrices without imposing that structure. The switching point is at higher $N$ when $M$ is increased, which is expected since the DFT approximation becomes better as $M$ increases.

In principle, the individual correlation matrices $\vect{R}_{li}^{j}$, or their diagonals, can be estimated in the same way. But the issue is that we cannot use the existing pilot signaling because it is subject to pilot contamination; we somehow need to obtain observations of each UE's channel that are interference-free.
 While the research on this subject is in its infancy, a few methods already exist and are discussed next.


\subsection{Methods to estimate individual correlation matrices}

The simplest approach to obtain ``clean'' observations of $\vect{h}_{li}^{j}$ (without interference from pilot-sharing UEs)  is to use a specific phase for learning ${\vect{R}}_{li}^{j}$ where every UE in the network uses a unique orthogonal pilot \cite{Yin2013a}. If each UE repeats this pilot in $N_R$ different coherence blocks, 
we need $N_R KL$ extra pilots in total. These can be spread out over the $\tau_s$ coherence blocks where the channel statistics are assumed to be fixed. 
 This approach provides BS~$j$ with $N_R$ interference-free observations of $\vect{h}_{li}^{j}$ in noise, from which a sample correlation matrix can be formed, possibly by using regularization to get robustness. We call this method the ``\emph{R direct}" approach.

Alternatively, a two-stage estimation procedure can be used \cite{Bjornson2016c}, where each UE is still associated with a unique orthogonal pilot, but it is the pilot-sharing UEs that transmit it instead of the UE itself. In doing so, BS~$j$ obtains $N_R$ observations that can be used to form a sample correlation matrix $\widehat\Psiv_{li,-i}^{j,\mathrm{sample}}$ of $\Psiv_{li}^{j} - \vect{h}_{li}^{j} {(\vect{h}_{li}^{j})}^{\Htran}$, which includes all the pilot-contaminating UEs, but not the UE itself. As a second stage, if $\widehat\Psiv_{li}^{j,\mathrm{sample}}$ has been already computed, an estimate $\widehat{\vect{R}}_{li}^{j,\,\mathrm{sample}}$ of the sample correlation matrix can be obtained as follows:
\begin{align}\label{eq:11}
\widehat{\vect{R}}_{li}^{j,\,\mathrm{sample}}= \widehat\Psiv_{li}^{j,\,\mathrm{sample}} - \widehat\Psiv_{li,-i}^{j,\,\mathrm{sample}}.
\end{align}
We call this approach ``\emph{Via Q}'' and notice that $\widehat{\vect{R}}_{li}^{j,\,\mathrm{sample}} \to {\vect{R}}_{li}^{j}$ as $N,N_R \to \infty$. 
Whenever $N_R< N$, which is typically the case since it is expensive to allocate extra pilots for correlation matrix estimation,  the estimate $\widehat{\vect{R}}_{li}^{j,\,\mathrm{sample}}$ obtained by ``\emph{Via Q}''  contains more observations of $ \vect{h}_{li}^{j}$ than in the ``\emph{R direct}'' approach. However, it is also perturbed by the imperfect subtraction of the interfering UEs' correlation matrices. 
When the estimate is inaccurate, it can be regularized as described earlier.
We compare the two approaches numerically in the next section.

A different approach is explored in \cite{Vorobyov2018} in which each UE transmits two pilot sequences, where the second pilot is multiplied by a random phase shift. Specifically, the pilot sequence $\bphiu_{jk}[n]\in \mathbb{C}^{\taupu}$ used by UE~$k$ in cell~$j$ in the $n$-th coherence block is
\begin{align}
\bphiu_{jk}[n] =  
\left[\overline{\bphiu}_{jk}^{\,\Ttran} \,\,\, e^{{\mathsf{j}}\vartheta_l[n]}\overline{\bphiu}_{jk}^{\,\Ttran}\right]^{\Ttran}
\end{align}
where $\overline{\bphiu}_{jk}$ is taken from a pilot book $\overline{\mathbf{\Phi}} \in \mathbb{C}^{\taupu/2\times \taupu/2}$ of $\taupu/2$ (instead of $\tau_p$) mutually orthogonal pilot sequences, and $\{\vartheta_l[n]\}_{n=1}^N$ are independent realizations from a uniform distribution between $-\pi$ and $+\pi$, such that $\mathbb{E}\{ e^{{\mathsf{j}}\vartheta_l[n]}\} =0$.
These realizations are generated pseudo-randomly in a way that allows all the BSs to compute them locally. Let $\vect{Y}_j^{p,1}[n]$ and $\vect{Y}_j^{p,2}[n]$ be the pilot vectors received at BS~$j$ from the first and second pilot subsequences in coherence block $n$. An estimate of $\vect{R}_{li}^{j}$ is obtained as
\begin{align}
\!\!\widehat{\vect{R}}_{li}^{j,\mathrm{sample}} = \frac{1}{N} \sum_{n=1}^N \big(\vect{Y}_j^{p,1}[n]\overline{\bphiu}_{li}\big)\big(\vect{Y}_j^{p,2}[n]e^{-{\mathsf{j}}\vartheta_l[n]}\overline{\bphiu}_{li}\big)^{\Htran}
\end{align}
which converges to $\vect{R}_{li}^{j}$ as $N \to \infty$ since the pilot-sharing UEs' channels appear in $\vect{Y}_j^{p,1}[n]\overline{\bphiu}_{li}$ and $\vect{Y}_j^{p,2}[n]e^{-{\mathsf{j}}\vartheta_l[n]}\overline{\bphiu}_{li}$ with different random phase-shifts so that  these terms have zero mean.

Another approach is taken in \cite{Neumann-2017-A} where, instead of reserving a specific phase for learning ${\vect{R}}_{li}^{j}$, the pilot assignment is changed between different coherence blocks. Let ${\vect{\Pi}_{l}(t)}\in \{0,1\}^{K\times \tau_p}$ denotes the $t$-th pilot allocation to the UEs in cell $l$; that is, $[{\vect{\Pi}_l(t)}]_{ki}=1$ indicates that UE $k$ in cell $l$ transmits the UL pilot sequence $i$. 
Clearly, the pilot allocation matrix ${\vect{\Pi}(t)}
 = [\vect{\Pi}_1^{\Ttran}(t),\ldots,{\vect{\Pi}_L^{\Ttran}(t)}]^{\Ttran} \in \{0,1\}^{KL\times \tau_p}$ determines which UEs in the network contaminate each other. Assume that a set of $T$ pilot allocations is used and that all the sample correlation matrices $\{\widehat\Psiv_{li}^{j,\mathrm{sample}} (t): t =1,\ldots T\}$ of the contaminated observations have been already estimated. Since these matrices contain different linear combinations of the correlation matrices, the individual correlation matrices $\{{\vect{R}}_{li}^{j} :\forall l,i\}$ can be obtained as the solution of a linear system of equations,  where the coefficients are given by the joint allocation matrix ${\vect{\Pi}} = \left[{\vect{\Pi}(1)},\ldots,{\vect{\Pi}(T)}\right]\in \{0,1\}^{KL\times \tau_pT}$ and the constant terms depend on $\{\widehat\Psiv_{li}^{j,\mathrm{sample}} (t): t =1,\ldots T\}$. If $T \ge KL/\tau_p$ and the pilot allocation matrices are selected to give ${\vect{\Pi}}$  full row rank, the problem is well-posed and a unique solution exists \cite{Neumann-2017-A}. This approach can be also used to estimate the diagonal elements of $\{{\vect{R}}_{li}^{j} :\forall l,i\}$ by using $\{\widehat \Psiv_{li}^{j,\,\mathrm{diag}}(t): t = 1,\ldots, T\}$.

\begin{figure}[t!]
\begin{center}
\subfloat[Estimation of the full correlation matrices.]{\label{figure:section7_figure17_CovEstSEUL_2}
\begin{overpic}[unit=1mm,width=1\columnwidth]{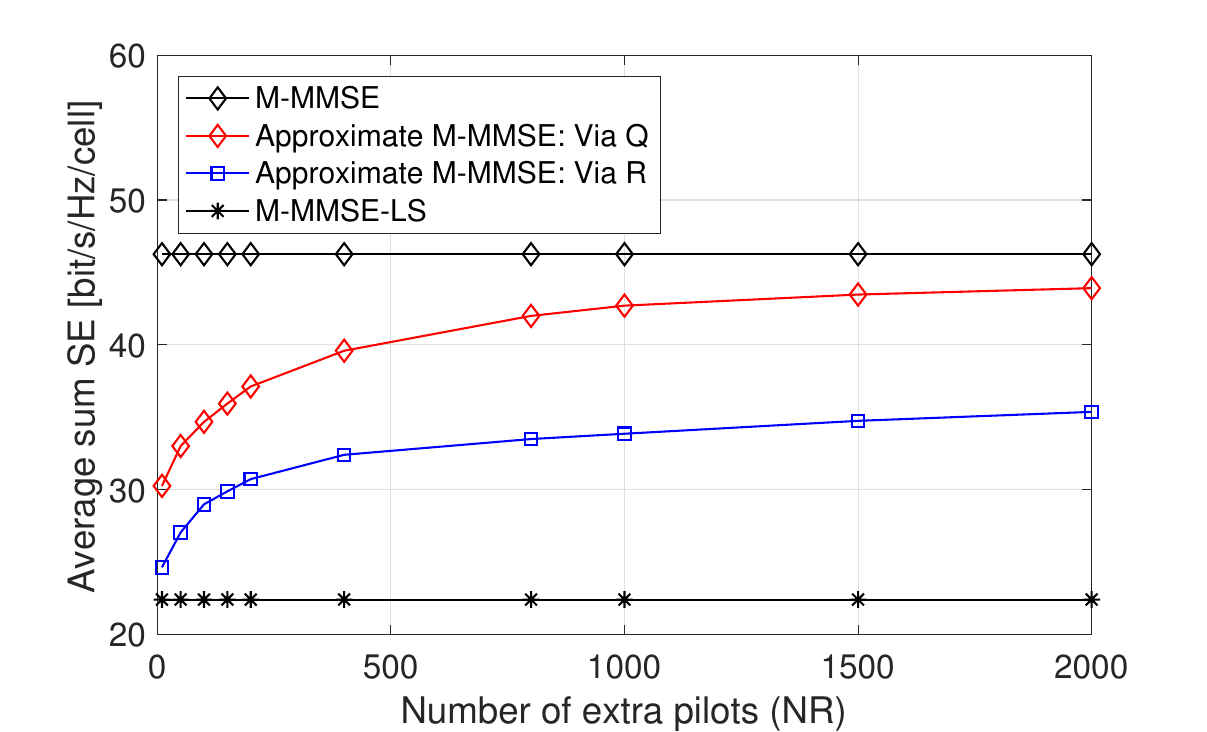}
\end{overpic}} \\
\subfloat[Estimation of only the diagonal elements of the correlation matrices.]{\label{figure:section7_figure17_CovEstSEUL_1}
\begin{overpic}[unit=1mm,width=1\columnwidth]{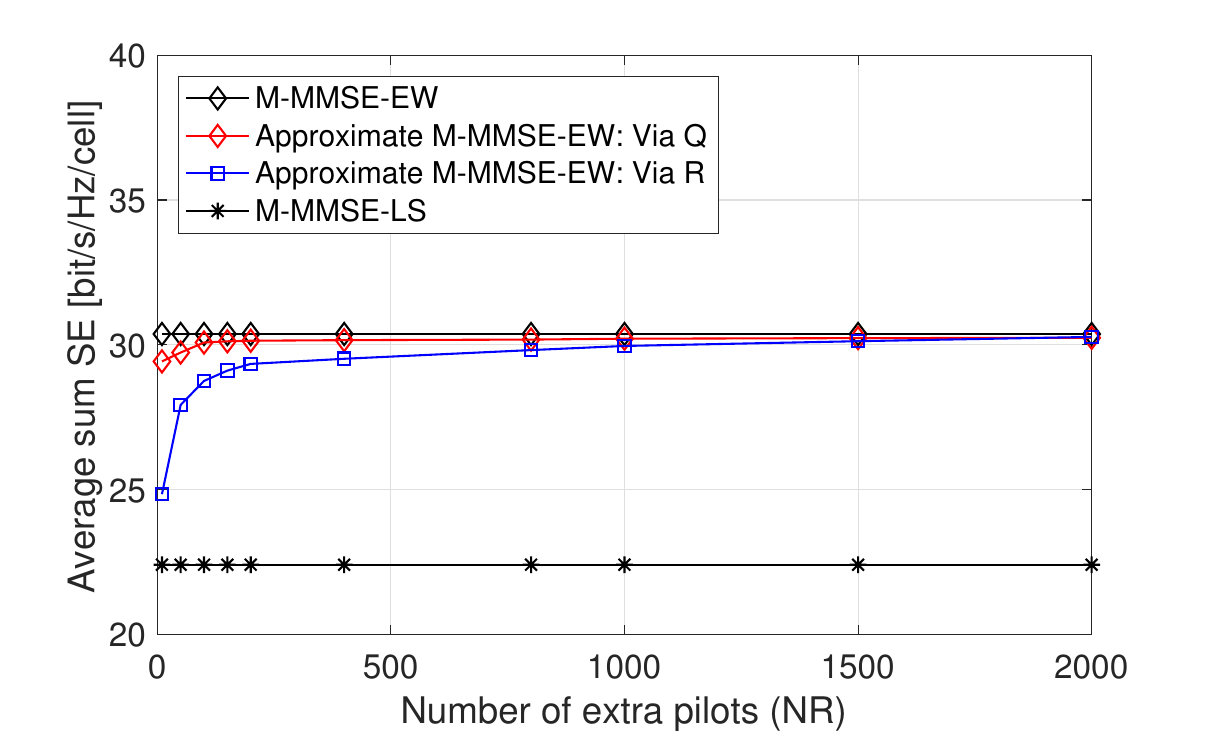}
\end{overpic}} 
\end{center}
\caption{UL SE achieved in a given cell with M-MMSE combining when the full correlation matrices or only its main diagonal elements are estimated with the 'Via Q' and 'R direct' methods. The network setup is that of Fig.~\ref{figure_network_layout} with $M=100$, $K=10$ and pilot reuse factor $f=1$.}\label{SE_with_imperfect_statistics}
\end{figure}

\subsection{Spectral efficiency evaluation with imperfect statistical knowledge}
To evaluate the impact that imperfect statistical knowledge has on the SE, we consider the network setup in Fig.~\ref{figure_network_layout} with $M = 100$ and $f = 1$. The ``R direct'' and ``Via Q'' methods are used for estimating the full or partial correlation matrices.{\footnote{A fair and comprehensive comparison of the different schemes discussed in Section VI.C requires extensive simulations, which are outside the scope this paper. Hence, we have only selected those schemes that fit well the simulated cases. We refer the interested reader to \cite{Vorobyov2018} and \cite{Neumann-2017-A} for a few comparisons.}} Channel statistics are assumed to be fixed over the system bandwidth $B$\,Hz and a time interval  $T_s$\,s. The number of coherence blocks contained into such a time-frequency block is
\begin{equation}
\tau_s = \frac{B}{B_c}\frac{T}{T_c} = \frac{BT_s}{\tau_c}.
\end{equation}
To quantify the value of $\tau_s$, we consider $B=20$ MHz and assume that $T_s= 0.5$ s, which corresponds to a medium/high mobility scenario in the sub-6\,GHz band. With $\tau_c = 200$, this means that the channel statistics are fixed for $\tau_s = 50000$ coherence blocks, thereby showing that one can easily put a few thousands of $N_R$ extra pilots for correlation matrix estimation and still keep the overhead low. 


Fig.~\ref{figure:section7_figure17_CovEstSEUL_2} shows the UL sum SE that is achieved by M-MMSE with the two correlation estimation methods as a function of $N_R$. 
The benchmarks are M-MMSE implemented using the MMSE estimator (with perfect correlation information) and the LS estimator (with no correlation information). We notice that a few tens of extra pilots are sufficient with both estimation methods to achieve higher SE than with LS. The ``Via Q'' approach outperforms the ``R direct'' method and requires a thousand extra pilots to achieve 92\% of the SE achieved when using the MMSE estimator. This means that the gain in estimation quality clearly outweighs the extra pilot overhead. However, the overhead increases if a larger number of antennas $M$ is considered (see Fig.~\ref{figureNMSE}) since the full correlation matrices are needed by M-MMSE. 

Fig.~\ref{figure:section7_figure17_CovEstSEUL_1} considers M-MMSE-EW combining, whose implementation requires only the estimation of the diagonal elements of the correlation matrices. In this case, only a few hundreds of extra pilots are required by both methods to achieve 98\% of the SE achieved by the MMSE estimator. This is a remarkable result because it does not need to grow with $M$, but the price to pay compared to Fig.~\ref{figure:section7_figure17_CovEstSEUL_2} is a loss in SE.

In summary, acquiring spatial correlation information and achieving an SE close to the ideal case is possible if the UEs are active for a sufficiently large number of coherence blocks, which may need to increase with the number of antennas. In practice, however, the active UEs in a cellular network can change quite rapidly due to changes in user behaviors and the bursty nature of packet transmission. The implementation of correlation estimation methods under these conditions is a challenge that requires further research activities.


%

\section{Asymptotically-optimal Signal Processing Schemes With Lower Complexity}

Although M-MMSE combining is optimal, it requires to compute the $M\times M$ matrix inverse in \eqref{eq:MMSE-combining} in every coherence block. Moreover, the channels estimates for all UEs in the network need to be computed. In practice, this may be too computationally demanding when $M$ increases and the network size is large. 
Therefore, it is of interest to look for alternative combining schemes (and their precoding counterparts) that have lower computational burden. Inspired by the structure of the M-MMSE combiner, given by $\vect{v}_{jk} = \Cyj^{-1}\hat{\vect{h}}_{jk}^{j}$, we now assume that  
\begin{equation} \label{eq:OBE-combining}
\vect{v}_{j k}=   \vect{W}_{jk}  \widehat{\vect{h}}_{jk}^j
\end{equation}
where $\vect{W}_{jk}\in \mathbb{C}^{M\times M}$ is an arbitrary matrix to be optimized only on the basis of channel statistics rather than of channel estimates, as with M-MMSE. Assuming that the statistics change slowly in time and thus $\vect{W}_{jk}$ can be precomputed and stored at BS~$j$, the computation of any $\vect{v}_{j k}$ in the form of \eqref{eq:OBE-combining} reduces to only one matrix-vector multiplication, which requires $M^2$ complex multiplications. This is substantially lower than the complexity of M-MMSE, which increases with $M^3$. Moreover, the computation of \eqref{eq:OBE-combining} only uses the channel estimates of the UEs in the own cell.

\begin{figure}[t!]
\begin{center}
\subfloat[Network setup of Fig.~\ref{figure_network_layout}.]{
\begin{overpic}[unit=1mm,width=1\columnwidth]{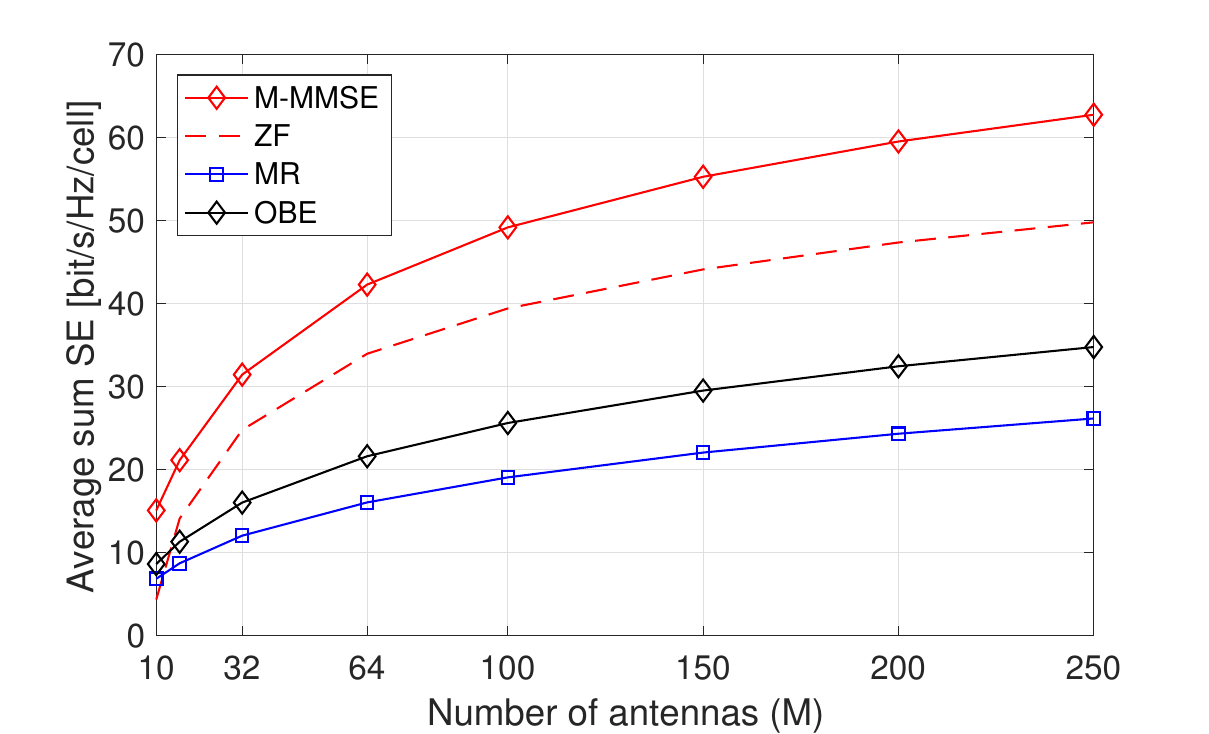}
\end{overpic}} \\
\subfloat[Network setup of {[17, Fig.~3]}.]{
\begin{overpic}[unit=1mm,width=1\columnwidth]{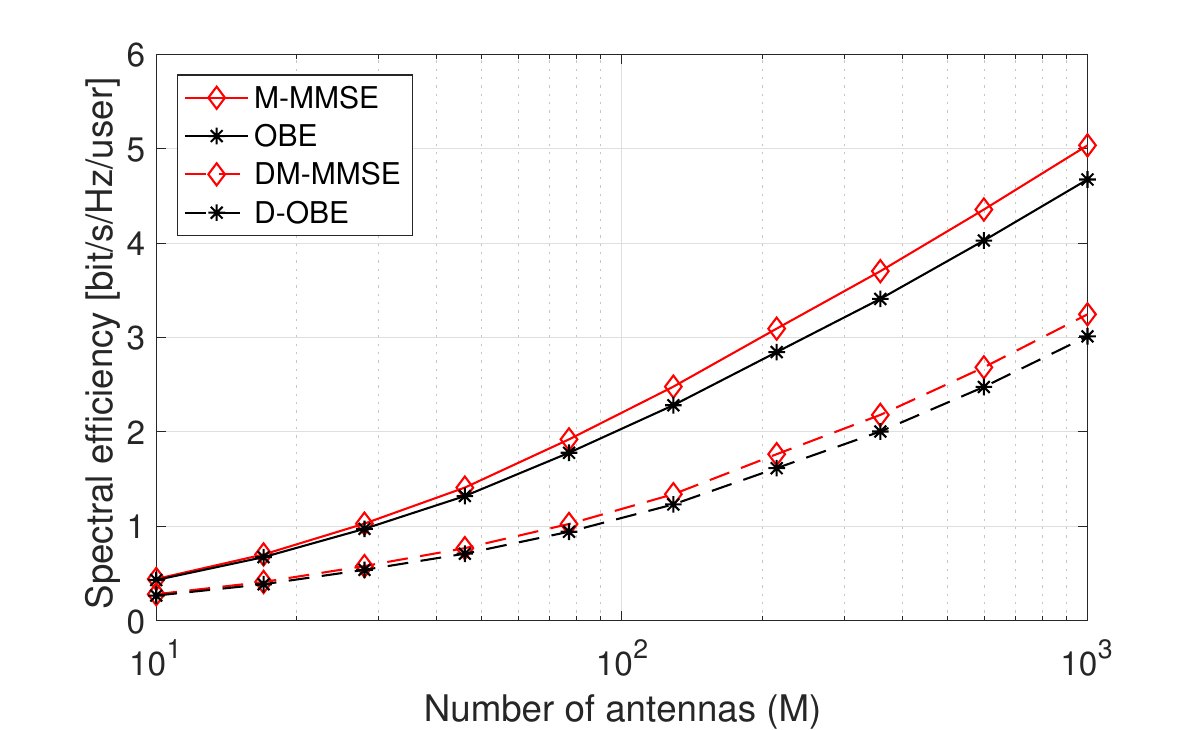}
\end{overpic}} 
\end{center}
\caption{UL SE as a function of the number of antennas with M-MMSE and OBE, and correlated Rayleigh fading for the network setups of Fig.~\ref{figure_network_layout} and \cite[Fig.~3]{BjornsonHS17}. In the former case, comparisons are also made with S-MMSE and MR.}\label{se_vs_antennas_OBE}
\end{figure}

The question is how to \emph{optimally} design $\vect{W}_{jk}$ in order to not incur a significant loss in SE when using \eqref{eq:OBE-combining}. Recently, \cite{Neumann-TSP17} proposed to select $\vect{W}_{jk}$ to maximize the UatF bound in Theorem \ref{theorem:uplink-capacity-forgetbound}. Plugging \eqref{eq:OBE-combining} into \eqref{eq:uplink-SINR-expression-forgetbound} yields
\begin{align} \label{eq:uplink-instant-SINR-combining}
\!\!\underline{\gamma}^{\mathrm{ul}}_{jk}  \!= \! \frac{\!\!\! \left| \tr \left( \vect{W}_{jk}^{\Htran} \vect{\Phi}_{jk}^j   \right)  \right|^2 }{\!\!\!\sum\limits_{  l\in \mathcal{P}_j } \!\!\!\left| \tr\left ( \big(\Psiv_{jk}^{j}\big)^{-1}\vect{R}_{jk}^j\vect{W}_{jk}^{\Htran} {\vect{R}}_{lk}^{j}   \right)  \right|^2 \!+\! \tr \left(\vect{W}_{jk} \vect{\Phi}_{jk}^j\vect{W}_{jk}^{\Htran} {{\bf U}}_{j} \right)    \!}\!\!
 \end{align}
where $\vect{\Phi}_{jk}^j =\vect{R}_{jk}^{j}
{(\Psiv_{jk}^{j})}^{-1}  \vect{R}_{jk}^{j}$ and ${{\bf U}}_{j} = \sum\limits_{l=1}^L\sum\limits_{i=1}^{K} {\vect{R}}_{li}^{j}+\frac{\sigma^2_{\rm {ul}}}{\rho_{\rm {ul}}}{\bf I}_M$. The ${\bf W}_{jk}$ that maximizes \eqref{eq:uplink-instant-SINR-combining} is \cite{Neumann-TSP17}
\begin{align}\label{eq:sectiond.7}
{\bf W}_{jk} & = {\big({{\bf U}}_{j}\big)}^{-1}\left(\sum_{l\in \mathcal{P}_j}\alpha_{lk}^j{\bf R}_{lk}^j\right)\big(\vect{R}_{jk}^j\big)^{-1}
\end{align}
where the scalar coefficients $\alpha_{lk}^j$ depend only
on the channel statistics and are obtained as $
\alpha_{lk}^j = [\boldsymbol{\alpha}_k^j]_l$ with $\boldsymbol{\alpha}_k^j = \big(\vect{\Gamma}_k^j+{\vect{I}_{|\mathcal{P}_j|}}\big)^{-1}\vect{e}_j$ where $ [\vect{\Gamma}_k^j]_{li} = \tr \big({\bf R}_{lk}^j (\Psiv_{jk}^{j})^{-1}   {\bf R}_{ik}^j({{\bf U}}_{j})^{-1}\big)$ $\forall l,i \in \mathcal{P}_j$ and $\vect{e}_j$ denotes the $j$th vector of the canonical basis. Substituting \eqref{eq:MMSEestimator_h_jli} and \eqref{eq:sectiond.7} into \eqref{eq:OBE-combining} yields \cite{SBH-GC-18}
\begin{align}\label{eq:OBE}
\vect{v}_{jk}^{\rm{OBE}} & = {\big({{\bf U}}_{j}\big)}^{-1} \left(\sum_{l\in \mathcal{P}_j}\alpha_{lk}^j\widehat{\bf {h}}_{lk}^j \right).
\end{align}
Although the BS only explicitly computes the channel estimates within the own cell, \eqref{eq:OBE} reveals that the combining vector of the form $\vect{W}_{jk}  \hat{\vect{h}}_{jk}^j$ that maximizes \eqref{eq:uplink-instant-SINR-combining} is implicitly created as a linear combination of the MMSE  estimates of the pilot-sharing UEs' channels, followed by a linear transformation with ${\big({{\bf U}}_{j}\big)}^{-1}$ that depends only on channel statistics. 
This is the reason why \eqref{eq:OBE} is called optimal bilinear equalizer (OBE) in \cite{Neumann-TSP17}. 
To quantify the computational complexity of OBE, we use  \eqref{eq:MMSEestimator_h_jli} to rewrite \eqref{eq:OBE} as
\begin{align}\label{eq:v_1_LS}
\!\!\!\!\!\vect{v}_{jk}^{\rm{OBE}}\!=\! \underbrace{ {\big({{\bf U}}_{j}\big)}^{-1}\!\!\left(\sum_{l\in \mathcal{P}_j}\alpha_{lk}^j{\bf R}_{lk}^j\right)\!\!\big(\Psiv_{jk}^{j}\big)^{-1}\!}_{{\bf \Sigma}_{jk} }  \underbrace{\!\left(\frac{1}{\tau_p\sqrt{\rho_{\rm{ul}}}}\vect{Y}_j^{p}\bphiu_{jk}\right)}_{{\textnormal{LS channel estimate}}}\!
\end{align}
which shows that, in practice, OBE only needs to compute the LS channel estimates
and then multiply them with the matrix ${\bf \Sigma}_{jk}$ that only depends on the channel statistics. Since the LS estimator has a complexity that scales linearly with $M$ (rather than quadratically as with the MMSE estimator), as reported in Table \ref{tab:cost_linear_processing}, this provides a further reduction in computational complexity compared to M-MMSE combining. For completeness, Table \ref{tab:cost_linear_processing} reports also the number of complex multiplications required for precomputing ${\bf \Sigma}_{jk}$. Specifically, the sum of the first two terms accounts for the complexity in computing ${\bf \Sigma}_{jk}$ for a given set of coefficients $\{\alpha_{lk}^j: \forall l \in \mathcal{P}_j\}$, while the sum of the last two terms quantifies the complexity of calculating $\{\alpha_{lk}^j: \forall l \in \mathcal{P}_j\}$. As is seen, the complexity of OBE is dominated by the precomputation of statistical parameters. Under the assumption that these are precomputed and stored, OBE has a lower computational complexity than M-MMSE. The price to pay is a significant loss in SE. For the network setup in Fig.~\ref{figure_network_layout}, this is quantified in Fig.~\ref{se_vs_antennas_OBE}(a) wherein the UL sum SE with OBE is $50\%$ lower than with M-MMSE. 
 OBE obtains a $40\%$ higher sum SE than MR at $M\ge 64$, but ZF is substantially better for all the considered values of $M$.
Interestingly, when the number of antennas grows unboundedly, the following result can be proved \cite[Th. 1]{Neumann-TSP17}.


\begin{theorem} \label{theorem:GMF_1}
If OBE combining is used with the matrix in \eqref{eq:sectiond.7} and the correlation matrices are asymptotically linearly independent, then $\underline{\mathacr{SE}}^{\mathrm{ul}}_{jk}$ increases logarithmically as $M\to \infty$.
\end{theorem}

The theorem implies that if ${\bf W}_{jk}$ in \eqref{eq:OBE-combining} is optimally designed, 
 then the same asymptotic scaling behavior as with M-MMSE  can be achieved. This is a remarkable result for a combining scheme that 
 has a much lower complexity than M-MMSE. The results of Theorem \ref{theorem:GMF_1} are exemplified in Fig.~\ref{se_vs_antennas_OBE}(b) for the scenario in \cite[Fig.~3]{BjornsonHS17}, where pilot contamination is particularly strong. M-MMSE provides the highest SE per UE since it is optimal. The loss incurred by using OBE is around 2--7\%. The SE per UE of both schemes grow unbounded as $M$ increases. If the diagonalized OBE (D-OBE) is considered, as obtained by using only the diagonals of correlation matrices, then similar asymptotic results are achieved under the assumption that the diagonals are known and linearly independent between pilot-sharing UEs~\cite[Th. 2]{Neumann-2017-A}. In this case, Fig.~\ref{se_vs_antennas_OBE}(b) shows that the SE with D-OBE is reduced by 5--7\% than with D-MMSE. Nevertheless, the performance achieved by OBE for the values of $M$ used in current networks is not so convincing. However, as the array sizes continue to grow, we encourage further research activities into finding alternative schemes that strike a good balance between complexity and SE.


\section{Concluding Remarks and Research Directions for Beyond 5G}
\label{sec:future-directions}

{This tutorial article showed that, by deploying Massive MIMO with more and more antennas and optimized signal processing, the capacity (spectral efficiency) of cellular networks can continue to grow for decades to come. The pilot contamination phenomenon, that was initially believed to create strict asymptotic capacity limits, can be resolved by exploiting the natural spatial correlation of propagation channels and by using  signal processing schemes that suppress (intra-cell and inter-cell) interference in the network. We refer to such technology evolution as \emph{Massive MIMO 2.0}. Its capacity is theoretically unlimited, but is practically limited when the number of antennas increases by the high computational complexity and the ability to learn the spatial channel correlation matrices.
These are two issues that deserve further research attention, but the previous two sections illustrated that good progress has recently been made in both directions.}
The described methods are model-based but can be complemented by data-driven solutions; for example, by learning the spatial correlation properties in different parts of a cell or reducing the complexity by learning suitable approximations.

We conclude this article by taking a wider look at what lies beyond 5G and discuss three new research directions in which Massive MIMO 2.0 will play an essential role.

\subsection{Large intelligent surfaces}

The number of antennas that fits into the conventional form factor of a BS site is fundamentally limited. Therefore, \emph{how can we approach the Marzetta limit $M \to \infty$ in practice?} One solution is to consider a very large electromagnetically active surface not deployed in a tower, but invisibly integrated into existing man-made structures, such as walls and windows. In this way, we can create arrays with huge aperture, using either a large number of discrete antenna elements or a continuous aperture with reconfigurable electromagnetic radiation properties. Different embodiments of this concept has recently appeared under the names large intelligent surfaces \cite{Hu2018a}, extremely large aperture arrays \cite{Amiri2018a}, and holographic Massive MIMO \cite{Bjornson2019a}. {Initial implementations have been made under the name holographic beamforming \cite{Black2017}.}

When using very large surfaces, most of the users will be in the radiative near-field \cite{Bjornson2019a}. Therefore, the standard far-field-based channel models must be revised. For example, the spatial channel correlation will not only carry angular information but also depth information, which is yet another spatial dimension that can be used to separate UEs by precoding and combining. This makes it an excellent scenario where the principles of Massive MIMO 2.0 will prevail; that is, utilize spatial correlation for interference suppression. The methodology for SE analysis can be readily applied to large surfaces, thus the main research challenges are related to resource allocation, hardware design, channel modeling, acquisition of spatial correlation, as well as experimental verification. In addition to providing communication services, the unprecedented spatial resolutions provided by large surfaces can be also used for positioning of objects and sensing \cite{Hu2018b}. 

A different but related concept is an intelligent reflecting surface \cite{Wu2018a}, which is not actively transmitting power but when illuminated by the signal from a BS or UE, it can reconfigure its reflective properties to form narrow beams with desired properties. 
The same concept has appeared under the names intelligent walls \cite{Subrt2012a}, {reconfigurable} reflectarrays \cite{Hum2014a}, and metasurfaces \cite{liaskos2018new}. 
{The general concept is supported by plenty of experiments \cite{Shaker2014a}, although there is a wide gap between the physics and communication literature that need to be bridged.}
The general idea is to gain control of the propagation environment instead of only controlling the transmitter and/or receiver. The obvious drawback 
is that channel estimation becomes very complicated; if the beam-search in hybrid beamforming is time-consuming, just imagine doing the same with much narrower beams and only having access to the concatenation of the channel between the UE and the surface and between the surface and the BS. Angle-domain representations of the channels combined with estimates of the UE position might be useful, if the approximation and estimation errors can be made sufficiently small. Machine learning tools may be promising to learn how to reflect the signals in different situations \cite{Bjornson2019a,Renzo2019a}. Since the reflective properties are frequency-independent, reflecting surfaces are mainly useful for frequency-flat channels, effectively limiting the use case to LoS scenarios. It also remains unclear in which use cases the ``passive'' reflecting surfaces are beneficial compared to ``active'' intelligent surfaces.

\subsection{Post-cellular network architectures}

This article has considered cellular networks with autonomous BSs and arbitrary geometry, as illustrated in Fig.~\ref{figure_network_layout}. However, the general theory behind Massive MIMO is not limited to that case, but has recently been applied also to cell-free networks \cite{Ngo2017b,Nayebi2017a,Bjornson2019c}. Attempts to remove cell boundaries, by introducing co-processing of communication signals at multiple BSs, has been going on for two decades \cite{Shamai2001a,Gesbert2010a}.
Although the theoretical gains of such cell-free operation are large, the achievable gains in 4G were marginal \cite{Fantini2016a}. This is why cellular Massive MIMO with autonomous BSs was adopted in 5G instead. Nevertheless, the continued network densification will make the fundamental weaknesses of the cellular paradigm evident: the signals are easily blocked when coming from a single BS, most UEs are at the cell edges and experience medicore SNRs, and inter-cell interference is the limiting factor unless every BS is equipped with Massive MIMO 2.0, which is not practical for very dense networks.

Suppose all the BS antennas are distributed over the coverage area instead of co-located in arrays at a few elevated locations, so that the UEs are surrounded by antennas instead of having a few BSs surrounded by UEs. How can we operate such a network? The ideal solution is to let each UE be served by coherent transmission and reception from all the antennas that can make a non-negligible impact on the UE's performance \cite{Bjornson2013d}. That effectively leads to a user-centric post-cellular network architecture. The phase-coherency is key to achieve array gains from distributed antennas and effectively focusing the signals onto only the spatial points where the UEs are. Recent papers have demonstrated substantial performance gains for cell-free operation compared to small-cell networks \cite{Ngo2017b,Nayebi2017a} (where each distributed antenna operates autonomously) and cellular Massive MIMO \cite{Bjornson2019c}. {There are also experimental verifications \cite{Shepard2018a} and attempts to commercialize the core concepts \cite{Perlman2015a}.}

The SE of a cell-free network can be easily evaluated by realizing that it can be modeled as a single-cell Massive MIMO system with strongly spatially correlated fading and some peculiar properties: 1) each pilot is reused by several UEs served by (partially) the same antennas; 2) there are widely different power allocations between antennas; and 3) each UE is only served by a subset of the antennas. The signal processing related to a particular UE can be either implemented locally (in the hardware connected to each antenna), at an edge-cloud processor, or divided between these entities \cite{Bjornson2019c}. Although the achievable SE follows from the Massive MIMO methodology, the post-cellular perspective opens a Pandora's box of new issues \cite{Interdonato2018}: how to implement initial access, soft handover, resource allocation, adaptive modulation and coding, MMSE-like interference suppression, or synchronization in a distributed and scalable manner? These operations were previously implemented on a per-cell basis, but now require a complete overhaul. The fundamental performance limits are also unknown: is there a maximum SE, measured in bit/s per m$^3$, that we can achieve when densifying the network? Can the asymptotic analysis that lead the way to Massive MIMO 2.0 be applied here as well \cite{SBH-GC-18}?

\subsection{Beyond 100\,GHz: Sub-THz communications}

In the quest for ever-increasing data rates, it is natural to continue looking for more bandwidth, which in turn pushes the operation towards higher frequencies than in the past. 5G is envisioned to operate in bands up to 86 GHz. However, there is at least 50\,GHz of suitable spectrum in the range 90--200\,GHz \cite{Saad2018a} and another 100\,GHz in the range 220--320\,GHz \cite{Shams2016a}. These bands have started be known as the sub-THz bands. When combined with MIMO and spatial multiplexing, an SE sufficient for breaking the 1\,Tbit/s barrier is within practical reach \cite{Faisal2019a}, but new waveforms and modulation schemes are needed to ease the implementation \cite{Akyildiz2016a}. To keep the SNR constant when increasing the carrier frequency, we need to increase the number of fixed-gain antennas that are used. The area that fits one antenna at 1\,GHz carrier frequency gives room for 100 antennas at 10\,GHz and 10000 antennas at 100\,GHz. Hence, we will reach the large-$M$ regime where the performance gains of Massive MIMO 2.0 are huge. It is also important to bear in mind that a $10\times$ bandwidth extension leads to 10\,dB loss in SNR, which must be compensated for using even more antennas and/or limiting the operational range to a few meters. Hence, it is likely that the sub-THz bands will be only used for outdoor backhaul links using very directive antennas  \cite{Edstam2017a} and short-range communications \cite{Faisal2019a}.

As in the case of 5G mmWave, the first implementations in sub-THz bands will look for the right balance between analog and digital processing \cite{Akyildiz2016a,Faisal2019a}, but digital solutions will probably prevail in the long run. Although the fundamental properties developed in the Massive MIMO literature are frequency independent \cite{massivemimobook}, every aspect of the signal processing need to be revisited when designing systems for higher frequencies. For example, the interconnect of thousands of antennas is hard, and calls for distributed or hierarchical signal processing implementations. Can methods like M-MMSE or OBE be implemented like that? Another fundamental question is how to perform effective channel estimation; the SNR per antenna will be low and the UEs will have many antennas so their pilots must be narrowly beamformed. The utilization of spatial correlation might hold the key to the solution, since only the dominant eigendirection needs to be estimated with high accuracy. Angle-domain representations might aid in the acquisition of spatial correlation but, as demonstrated in Fig.~\ref{figureNMSE}, it is an approximation and must be treated that way when evaluating algorithms---the influence of modeling inaccuracies probably grows with the carrier frequency. Signal blockage is another major issue in sub-THz bands but the risk that many spatially distributed antennas are simultaneously blocked is substantially lower. Therefore, distributed arrays will be necessary to achieve macro diversity against blockage. For this reason, the main use case will be in LoS-like scenarios with larger eigenvalue variations in the correlation matrices than at sub-6\,GHz frequencies. This might make ZF, and variations thereof, competitive compared to M-MMSE, and give room to developing other low-complexity algorithms that can cancel intercell interference.  Much work on channel modeling and characterization remains to confirm these predictions \cite{Akyildiz2016a,Ju2019}. Finally, hardware impairments might be the core limiting factor at these frequencies because of the less developed technology, stronger mutual coupling (crosstalk) in compact circuits, and stronger impact of phase noise and synchronization errors \cite{Bjornson2014a,Akyildiz2016a}.

\section*{Acknowledgments}

The authors would like to thank Lars Thiele, Moritz Lossow, Thomas Wirth, Martin Kurras, Leszek Raschkowski and Stephan Jaeckel, at Fraunhofer Heinrich Hertz Institute for kindly providing their measurement results.
\bibliographystyle{IEEEtran}
\bibliography{IEEEabrv,ref,ref_book}

\end{document}